\documentclass[aps,prd,superscriptaddress,nofootinbib,tighten,preprint]{revtex4}
\pdfoutput=1
\usepackage[english]{babel}
\usepackage{gensymb}
\usepackage{amsmath}
\usepackage{amssymb}
\usepackage{color}
\usepackage{comment}
\usepackage{footnote}
\usepackage[caption=false]{subfig}
\usepackage[mathscr]{euscript}
\usepackage{hyperref}
\usepackage{amsfonts}
\usepackage{amssymb}
\usepackage{multirow}
\usepackage{graphicx}
\usepackage{slashed}
\usepackage{float}
\usepackage{pstricks}
\usepackage{mathrsfs}
\usepackage[toc,page]{appendix}

\makeatletter
\newcommand{\xRightarrow}[2][]{\ext@arrow 0359\Rightarrowfill@{#1}{#2}}
\makeatother
\newcommand{\be}{\begin{equation}}
\newcommand{\ee}{\end{equation}}
\newcommand{\bea}{\begin{eqnarray}}
\newcommand{\eea}{\end{eqnarray}}

\newcommand{\zmt}{Z_{\mu \tau}}

\newcommand{\nn}{\nonumber}

\catcode`@=12
\def\la{\mathrel{\mathchoice {\vcenter{\offinterlineskip\halign{\hfil
$\displaystyle##$\hfil\cr<\cr\sim\cr}}}
{\vcenter{\offinterlineskip\halign{\hfil$\textstyle##$\hfil\cr<\cr\sim\cr}}}
{\vcenter{\offinterlineskip\halign{\hfil$\scriptstyle##$\hfil\cr<\cr\sim\cr}}}
{\vcenter{\offinterlineskip\halign{\hfil$\scriptscriptstyle##$\hfil\cr<\cr\sim
\cr}}}}}
\def\ga{\mathrel{\mathchoice {\vcenter{\offinterlineskip\halign{\hfil
$\displaystyle##$\hfil\cr>\cr\sim\cr}}}
{\vcenter{\offinterlineskip\halign{\hfil$\textstyle##$\hfil\cr>\cr\sim\cr}}}
{\vcenter{\offinterlineskip\halign{\hfil$\scriptstyle##$\hfil\cr>\cr\sim\cr}}}
{\vcenter{\offinterlineskip\halign{\hfil$\scriptscriptstyle##$\hfil\cr>\cr\sim
\cr}}}}}

\begin{document}  
\title{Reconciling dark matter, \boldmath$R_{K^{(*)}}$ anomalies and
\boldmath$(g-2)_{\mu}$ in an \boldmath${L_{\mu}-L_{\tau}}$ scenario} 

\author{Anirban Biswas}
\email{anirban.biswas.sinp@gmail.com}
\affiliation{School of Physical Sciences, Indian Association for the Cultivation of Science,
2A $\&$ 2B Raja S.C. Mullick Road, Kolkata 700032, India}

\author{Avirup Shaw}
\email{avirup.cu@gmail.com}
\affiliation{Theoretical Physics, Physical Research Laboratory, Ahmedabad 380009, India}

\begin{abstract}  
\noindent 
We propose an anomaly free unified scenario by invocation of an
extra local ${\rm U(1)}_{L_{\mu}-L_{\tau}}$ gauge symmetry. This
scenario simultaneously resolves the $R_{K^{(*)}}$ anomalies, the dark
matter puzzle and the long-standing discrepancy in muon's anomalous magnetic moment.
A complex scalar ($\eta$) having nonzero ${L_{\mu}-L_{\tau}}$ charge has been
introduced to break this new U(1) symmetry spontaneously. Moreover, for
the purpose of studying dark matter phenomenology and $R_{K^{(*)}}$
anomalies in a correlated manner, we introduce an inert ${\rm SU(2)}_L$
scalar doublet ($\Phi$), a $\mathbb{Z}_2$-odd real singlet scalar ($S$)
and a $\mathbb{Z}_2$-odd coloured fermion ($\chi$) which transforms vectorially under
the ${\rm U(1)}_{L_{\mu}-L_{\tau}}$ symmetry. This extra
gauge symmetry provides a new gauge boson $Z_{\mu\tau}$ which not only gives additional
contribution to both $b\to s \ell\ell$ transition and $(g-2)_{\mu}$ but also provides a
crucial annihilation channel for dark matter candidate $\rho_1$ of the present scenario. This $\rho_1$ is an admixture of CP-even neutral component
of $\Phi$ and $S$. Our analysis shows that the low mass dark
matter regime ($M_{\rho_1}\la 60$ GeV) is still allowed by the experiments
like XENON1T, LHC (via Higgs invisible branching) and Fermi-LAT,
making the dark matter phenomenology drastically different from the standard Inert Doublet
and the Scalar Singlet models. Furthermore, the present model is also fairly consistent
with the observed branching ratio of $B\to X_s\gamma$ in $3\sigma$ range
and is quite capable of explaining neutrino masses and mixings via Type-I seesaw mechanism
if we add three right handed neutrinos in the particle spectrum. Finally,
we use the latest ATLAS data of non-observation of a resonant $\ell^+\ell^-$ signal at
the 13 TeV LHC to constrain the mass-coupling plane of $Z_{\mu\tau}$. 
\vskip 5pt \noindent  
\end{abstract}

\maketitle

\section{Introduction}\label{intro}
With the discovery of the missing piece, the Higgs boson at the Large Hadron Collider (LHC)~\cite{Chatrchyan:2012xdj, Aad:2012tfa} at CERN the Standard Model (SM) of particle physics
has been turned into a complete theory. From the last several decades it has been a well known fact that most of the theoretical predictions of this theory are in good agreement with various experimental results. However, at the same time, different experimental results in various directions compelling us to formulate physics beyond the SM (BSM). For example, dark matter relic density has been measured with
a great precision from the temperature and polarization anisotropies of the cosmic
microwave background (CMB) radiation by experiments like WMAP \cite{Hinshaw:2012aka}
and Planck \cite{Ade:2015xua}. On top of that various indirect evidence such as
rotation curve \cite{Sofue:2000jx}, gravitational lensing of distant objects \cite{Bartelmann:1999yn},
collision between galaxy clusters (such as Bullet cluster \cite{Clowe:2003tk} etc.) etc. have
strongly support for the existence of dark matter. However, in the SM there is no such candidate for dark matter. On the other hand neutrino oscillation
experiments \cite{Fukuda:1998mi, Ahmad:2002jz, Araki:2004mb, Abe:2011sj}
firmly established massive nature of at least two neutrinos
and have accurately measured three intergenerational mixing angles, both of
which are missing in the SM due to non-existence of the right handed counterparts
of left handed neutrinos. Besides, the CP-violation in the quark sector is
not at all sufficient to explain the observed baryon asymmetry of the Universe
\cite{Tanabashi:2018oca}. Furthermore, there is an enduring
$\sim 3.5\sigma$ discrepancy \cite {Tanabashi:2018oca} between experimentally measured value of
the anomalous magnetic moment of muon [$(g-2)_\mu$] and its SM predictions, which strongly indicates
the presence of a new physics (NP) beyond the SM. 

Apart from the above mentioned facts, over the last few years different flavour physics experiments like LHCb, Belle and Babar have been consistently shown that experimental data for different observables are in significant disagreement with respect to the corresponding SM predictions. Indeed this situation demands the invocation of NP effects. Recently the LHCb collaboration has reported additional hints for violation of Lepton Flavour Universality (LFU) between $b \to s \mu^+ \mu^-$ and $b \to s e^+ e^-$ processes. The LFU violation\footnote{Evidences of LFUV via charge current semileptonic $b \to c \ell \nu$ transition processes have also been observed. For example experimental results show significant deviations for observables $R_{D^{(*)}}$~\cite{average} and $R_{J/\psi}$~\cite{Aaij:2017tyk} from the corresponding SM predictions.} (LFUV) can be measured with the help of following observables $R_K$ and $R_{K^*}$
\bea
R_{K^{(*)}} =\frac{{\rm Br} \left( B \to K^{(*)} \mu^+ \mu^-\right)}
{ {\rm Br} \left( B \to K^{(*)} e^+ e^-\right)} \,.
\eea
Summary of the corresponding experimental results with their SM predictions for different di-lepton invariant mass squared ($q^2$) ranges are given in Table~\ref{exp-data}.
\begin{table}[H]\label{exp-data}
\begin{center}
\begin{tabular}{|c|cr|cr|cr|} 
\hline 
Observable & ~~~~SM prediction &  & Measurement  &  &Deviations &\\
\hline 
$R_K : q^2 = [1.1,6] \, \text{GeV}^2$  & $1.00 \pm 0.01 $&  \cite{Descotes-Genon:2015uva,Bordone:2016gaq} &  $0.846^{+0.060+0.016}_{-0.054-0.014}$  & \cite{Aaij:2019wad} &  2.5$\sigma$ &\\
\hline
$R_{K^*} ^{\rm low}: q^2 = [0.045,1.1] \, \text{GeV}^2$  & $0.92 \pm 0.02$  &  \cite{Capdevila:2017bsm} &  $0.660^{+0.110}_{-0.070} \pm 0.024$  & 
\cite{Aaij:2017vbb} & $2.1\sigma-2.3\sigma$ &\\
\hline
$R_{K^*}^{\rm central} : q^2 = [1.1,6] \, \text{GeV}^2$  & $1.00 \pm 0.01 $&  \cite{Descotes-Genon:2015uva,Bordone:2016gaq} &  $0.685^{+0.113}_{-0.069} \pm 0.047$  &  \cite{Aaij:2017vbb} & $2.4\sigma-2.5\sigma$ &\\
\hline
%
\end{tabular}
\caption{The experimental values of the observables along with their SM predictions for different ranges of $q^2$.}
\end{center}
\end{table}
Deviations from the SM predictions shown in the Table~\ref{exp-data}\footnote{For $R_{K^*}$, new preliminary measurements have been given by Belle \cite{RKstar_Belle_update} for two $q^2$ ranges. For $q^2\in[0.1, 8]$ GeV$^2$ the value of $R_{K^*}$ is $0.90^{+0.27}_{-0.21}\pm 0.10$ while for $q^2\in[15, 19]$ GeV$^2$ the corresponding value is $1.18^{+0.52}_{-0.32}\pm 0.10$.} can be resolved by invoking additional
NP contributions to some of the Wilson Coefficients (WCs) which are involved in the effective Hamiltonian for 
$b\to s \ell \ell$ ($\ell\equiv$ charged lepton, i.e., electron (e) and muon ($\mu$)) transition. 
Furthermore, if these anomalies are associated with other observables for the rare processes $b\to s \mu \mu$ transitions, then it has been observed that a NP scenario with {\it additional} contribution to the WC $C^\mu_9$ (but not in
$C^e_9$ ) is more acceptable. The operator corresponding to the WC $C^\ell_9$ is $\mathcal{O}_9\equiv \frac{e^2}{16\pi^2}(\bar{s} \gamma_{\alpha} P_{L} b)(\bar{\ell} \gamma^\alpha \ell)$. From the Table~\ref{exp-data}, it is readily evident that NP interfere destructively with the SM, which ensures the sign of $C_9^{\text{NP},\mu}$
is negative. The best-fit value of $C_9^{\text{NP},\mu}$ is $\approx -1$ 
~\cite{Descotes-Genon:2013wba,Hiller:2014yaa,Ghosh:2014awa,Altmannshofer:2014rta,Descotes-Genon:2015uva,Hurth:2016fbr, Capdevila:2017bsm, Altmannshofer:2017yso, Aebischer:2019mlg}. Moreover, NP scenario with $C_9^{\text{NP},\mu} = -C_{10}^{\text{NP},\mu}$ (where the WC $C^\ell_{10}$ is associated with the operator $\mathcal{O}_{10}\equiv \frac{e^2}{16\pi^2}(\bar{s} \gamma_{\alpha} P_{L} b)(\bar{\ell} \gamma^\alpha\gamma_5\ell)$) is also a very appealing from the
model building point of view~\cite{Ghosh:2014awa,Altmannshofer:2014rta,Descotes-Genon:2015uva,Hurth:2016fbr,Capdevila:2017bsm, Altmannshofer:2017yso, Aebischer:2019mlg}.
Inspired by these results, several BSM scenarios using extra non-standard 
$Z$-boson\;\cite{Gauld:2013qba,Glashow:2014iga,Bhattacharya:2014wla, Crivellin:2015mga,
Crivellin:2015era,Celis:2015ara,Sierra:2015fma,Belanger:2015nma,Gripaios:2015gra,
Allanach:2015gkd,Fuyuto:2015gmk,Chiang:2016qov,Boucenna:2016wpr,Boucenna:2016qad,Celis:2016ayl,
Altmannshofer:2016jzy,Bhattacharya:2016mcc,Crivellin:2016ejn,Becirevic:2016zri,GarciaGarcia:2016nvr,Bhatia:2017tgo,Ko:2017yrd,Chen:2017usq,Baek:2017sew,Bonilla:2017lsq,Barman:2018jhz} and leptoquark~\cite{Hiller:2014yaa,Biswas:2014gga,Gripaios:2014tna,Sahoo:2015wya,Becirevic:2015asa,
Alonso:2015sja,Calibbi:2015kma,
Huang:2015vpt,Pas:2015hca,Bauer:2015knc,Fajfer:2015ycq,Barbieri:2015yvd,
Sahoo:2015pzk,
Dorsner:2016wpm,Sahoo:2016nvx,Das:2016vkr,Chen:2016dip,Becirevic:2016oho,Becirevic:2016yqi,Bhattacharya:2016mcc,Sahoo:2016pet,
Barbieri:2016las,Cox:2016epl, Alok:2017sui, Hati:2018fzc} 
have been demonstrated the viable interpretation of the anomalies.

In this article, we ameliorate some of these problems in a correlated manner within a single framework by introducing an extra local ${\rm U}(1)_{L_\mu-L_\tau}$ symmetry to the SM gauge symmetry,
where $L_{\mu}$ and $L_{\tau}$ indicate lepton numbers for the second and third
generations of charged leptons and their corresponding neutrinos. 
Apart from being an anomaly free gauged ${\rm U}(1)$ extension, the ${L_\mu-L_\tau}$
symmetry naturally violets the LFU between $e$ and $\mu$ because the ${L_\mu-L_\tau}$
charge of leptons are such that the corresponding new non-standard gauge boson 
couples only to $\mu (\tau)$ but not to $e$. This scenario was originally formulated
by Volkas et. al.\,\cite{He:1990pn,He:1991qd}. Thereafter, several variants of
${\rm U}(1)_{L_\mu-L_\tau}$ model have been studied in the context of different
phenomenological purposes: e.g.,\,\,contribution of the ${\rm U}(1)_{L_\mu-L_\tau}$
gauge boson to explain the $(g-2)_\mu$ anomaly~\cite{Ma:2001md, Baek:2001kca,
Heeck:2011wj, Harigaya:2013twa, Altmannshofer:2016brv, Biswas:2016yan, Biswas:2016yjr,
Banerjee:2018eaf}, dark matter phenomenology~\cite{Baek:2008nz, Das:2013jca, Patra:2016shz,
Biswas:2016yan, Biswas:2016yjr, Biswas:2017ait, Foldenauer:2018zrz},
generation of neutrino masses and mixing parameters~\cite{Ma:2001md, Choubey:2004hn,
Adhikary:2006rf, Baek:2015mna, Xing:2015fdg, Biswas:2016yan, Banerjee:2018eaf} etc.

For the purpose of explaining $b \to s \mu^+ \mu^-$ anomaly, this type of ${\rm U}(1)_{L_\mu-L_\tau}$ model has also been modified from its minimal version, {\it albeit} in a different approach \cite{Altmannshofer:2014cfa, Crivellin:2015mga,
Altmannshofer:2015mqa, Arnan:2016cpy, Altmannshofer:2016jzy, Chen:2017usq, Baek:2017sew,
Singirala:2018mio, Hutauruk:2019crc, Baek:2019qte}. In the present article, we introduce a $\mathbb{Z}_2$-odd bottom quark like non-standard fermion field
$\chi$ which is vectorial in nature under the ${\rm U}(1)_{L_\mu-L_\tau}$ symmetry.
It couples to all generations of the down-type SM quarks via Yukawa like interaction involving
a $\mathbb{Z}_2$-odd scalar doublet $\Phi$. Moreover, we 
introduce a $\mathbb{Z}_2$-odd singlet scalar $S$ which helps us to explain the flavour
anomaly, dark matter and $(g-2)_\mu$ anomaly simultaneously. A $\mathbb{Z}_2$-even complex
scalar singlet field $\eta$ with a nonzero $L_{\mu}-L_{\tau}$ charge 
has been introduced for the purpose of breaking of 
U(1)$_{L_\mu-L_\tau}$ symmetry spontaneously. Apart from these fields we have
the usual Higgs doublet field $H$ which breaks the  SU(2)$_{L}\times {\rm U}(1)_{Y}$ symmetry.
Therefore, in the broken phase of both electroweak (SU(2)$_{L}\times {\rm U}(1)_{Y}$)
and U(1)$_{L_\mu-L_\tau}$ symmetries, we have three physical $\mathbb{Z}_2$-odd
neutral scalars emerge from the mixing between $\Phi$ and $S$. The
lightest field among the three physical $\mathbb{Z}_2$-odd
neutral scalars can be considered as a potentially viable dark matter candidate. This is an admixture of both doublet and singlet
scalar representations and have distinct phenomenology compared
to the standard Inert Doublet \cite{Barbieri:2006dq, LopezHonorez:2006gr, Lundstrom:2008ai}
and the Scalar Singlet models \cite{McDonald:1993ex, Burgess:2000yq, Biswas:2011td, Cline:2013gha},
where the low mass dark matter regime is almost ruled by the latest bound on spin independent
scattering cross section from XENON1T \cite{Aprile:2018dbl} as well as by the upper limit on Higgs invisible
branching fraction from LHC \cite{Khachatryan:2016whc}. This is mainly due to the
fact that in these models in the low mass regime ($M_{\rm DM}\leq 62.5$ GeV),
dark matter candidate predominantly
annihilates into $b\bar{b}$ final state. 

On the contrary, in the present
scenario, the dark matter candidate in the low mass regime can annihilate 
into a pair of $L_{\mu}-L_{\tau}$ gauge boson $Z_{\mu\tau}$ and the branching fraction of
this annihilation channel is controlled by dark sector mixing angle
$\theta_D$. This actually makes the dark matter freeze-out process
extremely correlated with the flavour physics anomalies and $(g-2)_{\mu}$ anomaly,
where an $\mathcal{O}(\rm MeV)$ light $Z_{\mu\tau}$ plays a pivotal role.
Since, $Z_{\mu\tau}$ does not have direct couplings to the first generation
leptons and quarks, constraints from the LEP and more recently from the LHC
on the $g_{Z_{\mu\tau}}-M_{Z_{\mu\tau}}$ plane are relatively relaxed. Particularly,
light gauge boson with $M_{Z_{\mu\tau}}\la 100$ MeV and also with
moderate gauge coupling $g_{Z_{\mu\tau}}\la 10^{-3}$ is still allowed
from the experiments measuring neutrino trident processes namely CCFR \cite{Mishra:1991bv}
CHARM-II \cite{Geiregat:1990gz}. Moreover, apart from $(g-2)_{\mu}$ anomaly and
flavour physics related issues, such a light gauge boson has excellent
cosmological implication. The reason is that it can relax the $\sim3\sigma$ tension
between the measurements of Hubble constant ($H_0$) from two
different epochs\footnote{At two different redshifts ($z$), one is from 
the CMB experiment Planck \cite{Ade:2015xua} at high $z$ while another one
is from the local measurement using Hubble Space Telescope \cite{Riess:2016jrr}
at low $z$.}
by providing extra contribution to the
radiation energy density ($\Delta{N_{eff}}\sim 0.2-0.5$)
through the alteration of neutrino decoupling temperature \cite{Escudero:2019gzq}.       
In the present scenario the non-standard neutral gauge boson $Z_{\mu\tau}$ emerge from
all three neutral gauge bosons associated with ${\rm SU}(2)_{L}$,
${\rm U}(1)_{Y}$ and ${\rm U}(1)_{L_{\mu}-L_{\tau}}$ gauge groups by diagonalising a $3\times 3$ mixing matrix.
The additional contribution to the anomalous magnetic moment of muon
comes from an effective $\mu^{+} \mu^{-} \gamma$ vertex which has been generated from one loop penguin
diagram involving $Z_{\mu\tau}$. Moreover, we also have one
loop contribution from a diagram involving other BSM scalar (an orthogonal state of
the SM-like Higgs boson arises from the mixing between
$H$ and $\eta$ in the broken phase of the theory). However, its effect on $(g-2)_{\mu}$ is negligibly small.

To this end, we would like to mention another novel signature of the present scenario. 
The correlation between dark sector and flavour physics sector is
not only due to $L_{\mu}-L_{\tau}$ gauge boson but also due to all
the $\mathbb{Z}_2$-odd neutral particles (including dark matter candidate of the present scenario) along
with the coloured $\mathbb{Z}_2$-odd fermion $\chi$ generate non-standard one loop contributions
to produce $b \to s\mu^+\mu^-$ transition. In the present scenario, one can produce
non-standard contributions to both the WCs $C^\mu_9$ and $C^\mu_{10}$ respectively,
however, the contribution of the latter is insignificant and hence our analysis will
be furnished with $C_9^{\text{NP},\mu}$ only. The NP contribution to $C^\mu_9$
is obtained from non-standard penguin and self-energy diagrams and there is no
further NP contribution from box-diagram at one loop level. Moreover, we consider the constraint from the branching ratio of another flavour changing neutral current (FCNC) process $B \to X_s \gamma$.
Hence, we have computed the branching ratio of this decay in the present scenario. 
Further, neutrino masses and mixings can easily be addressed in these class
of $L_{\mu}-L_{\tau}$ models via Type-I seesaw mechanism by adding three right handed
neutrinos, which are singlet under the SM gauge groups and two of them have equal
and opposite $L_{\mu}-L_{\tau}$ charges for anomaly cancellation. Since a detailed
analysis on neutrino masses and mixings in the present scenario is beyond
the scope of this article, hence for the sake of completeness, we just have added three
right handed neutrinos in the Lagrangian and find the Majorana mass matrix for the
light neutrinos. A more comprehensive analysis on diagonalisation of the light neutrino
mass matrix and thereby finding the mass eigenvalues and mixing angles in
the $L_{\mu}-L_{\tau}$ scenario has already been done in \cite{Biswas:2016yan}. 

Finally, in order to impose the constraints on the parameter space of the
present scenario from the LHC experiment, we use the latest ATLAS data \cite{ATLAS:2019vcr} of
non-observation of a resonant $\ell^{+}\ell^{-}$ signal at the LHC running at 13 TeV
for the high mass range of $Z_{\mu\tau}$. Hence, we will estimate the cross section
for the process $pp \to Z_{\mu\tau} \to \ell^+ \ell^-$ at the 13 TeV LHC in the present
scenario. Consequently, it will be an interesting part of this exercise that, how
the LHC data can constrain the values of non-standard gauge coupling constant as well
as the $Z-Z_{\mu\tau}$ mixing angle.

The article is organised as follows. In Sec. \ref{model} we introduce the model with possible field content and interactions as well as we set our notations. Then in Sec. \ref{flav}, we show the calculational details of flavour physics observables and after that we will discuss $(g-2)_\mu$ anomaly in Sec. \ref{g2}. In Sec. \ref{dm}, we show the viability of our dark matter candidate of the present scenario considering all possible bounds from ongoing experiments and explain how can we correlate the dark matter with the flavour physics anomalies. We briefly discuss neutrino mass generation via Type-I seesaw mechanism
in Sec. \ref{neu}. Sec. \ref{cldr} deals with constraint that are obtained from non-observation
of a resonant $\ell^{+}\ell^{-}$ signal at the LHC running at 13 TeV. Finally, we summarize our
results and conclude in Sec. \ref{con}.
\section{The \boldmath${\rm U(1)}_{L_{\mu}-L_{\tau}}$ model}\label{model}
In order to facilitate our motivations (discussed in Section \ref{intro}), we propose
an anomaly free ${\rm U}(1)_{L_{\mu}-L_{\tau}}$ gauge extension of the SM.
This scenario is free from mixed gauge-gravitational and axial vector
gauge anomalies because these anomalies cancel between second and third
generations of charged leptons and their corresponding neutrinos
due to their equal and opposite ${L_{\mu}-L_{\tau}}$ charges. The
Lagrangian which remains invariant under the
${\rm SU}(3)_{C}\times {\rm SU}(2)_{L} \times {\rm U}(1)_{Y}
\times {\rm U}(1)_{L_\mu-L_\tau}\times \mathbb{Z}_2$ symmetry is given by,
\begin{eqnarray}\label{LagT}
\mathcal{L}&=&\mathcal{L}_{\rm SM} + \mathcal{L}_{N} + \mathcal{L}_{\chi}
+ (D_{\alpha}{\eta})^{\dagger} (D^{\alpha}{\eta}) + 
(D_{\alpha}{\Phi})^{\dagger} (D^{\alpha}{\Phi}) + \frac 12 \partial_{\alpha}S \partial^{\alpha}S\\ \nonumber 
&-&\frac{1}{4} \hat{B}_{\alpha \beta} \hat{B}^{\alpha \beta}
- \frac{1}{4} \hat{X}_{\alpha \beta} \hat{X}^{\alpha \beta} +
\frac{\epsilon}{2} \hat{X}_{\alpha \beta} \hat{B}^{\alpha \beta}-V(H, \eta, \Phi, S)\;,
\end{eqnarray}
where
\begin{eqnarray}
\hat{B}_{\alpha \beta} &=& \partial_{\alpha} \hat{B}_\beta - \partial_{\beta}
\hat{B}_{\alpha} \,\,\,\, {\rm and} \,\,\, \hat{X}_{\alpha \beta} = 
\partial_{\alpha} \hat{X}_\beta - \partial_{\beta} \hat{X}_{\alpha} \,\,,
\label{fieldtensor}
\end{eqnarray}
are field strength tensors for the two U(1) gauge fields\footnote{We are denoting the basis of gauge
fields having off-diagonal kinetic term by using a hat notation.} $\hat{B}_{\alpha}$
and $\hat{X}_{\alpha}$ respectively while the Lorentz indices $\alpha,\;\beta\equiv
0,1\ldots 3$. The term contains both field strength tensors is the
kinetic mixing term between $\hat{B}_{\alpha}$ and $\hat{X}_{\alpha}$,
which is not forbidden by any of the symmetries of the present model. Full list of particle contents
and their quantum numbers under various symmetry groups are given
in Table~\ref{gauqn}.

\begin{table}[h!]
	\centering
	\footnotesize
	\resizebox{16cm}{!}{
		\begin{tabular}{|c|c|c|c|c|c|c|c|c|c|c|c|c|c|c|c|c|c|}
			\hline\hline
			 \multicolumn{1}{|c|}{Gauge groups}& \multicolumn{13}{|c|}{Fermion fields}
			 & \multicolumn{4}{|c|}{Scalar fields} \\ 
\cline{2-18}
                         {} &\multicolumn{3}{|c|}{Quark fields}&\multicolumn{9}{|c|}{Lepton fields}
                         & \multicolumn{1}{|c|}{} & {}& {} & {}&\\
\cline{2-13}
                         {} &$Q_{Li}$ & $u_{Ri}$ & $d_{Ri}$ & $L_{Le}$ & $L_{L\mu}$ & 
                         $L_{L\tau}$ & $e_{R}$ & $\mu_{R}$ &  $\tau_{R}$ & $N_{eR}$ & 
                         $N_{\mu R}$ &  $N_{\tau R}$ &~$\chi$~&~$H$~&~$\eta$~&~$\Phi$~&$S$
                         \\ \hline 
			\multicolumn{1}{|c|}{${\rm SU}(3)_C$} & 3 & 3 & 3 & 1 & 1 &  1 & 1 & 1  & 1 
			& 1 & 1 & 1 & 3 & 1 & 1 & 1 & 1\\ \hline 
			\multicolumn{1}{|c|}{${\rm SU}(2)_L$} & 2 & 1 & 1 & 2 & 2 &  2 & 1 & 1  & 1
			 & 1 & 1 & 1 & 1 & 2 & 1 & 2 & 1\\ \hline 
			\multicolumn{1}{|c|}{${\rm U}(1)_Y$}  & $\frac 16$ & $\frac 23$ & -$\frac 13$ 
			& $\frac 12$ & $\frac 12$ & $\frac 12$ & -1 & -1  & -1 & 0 & 0 & 0 & -$\frac 13$ 
			& $\frac 12$ & 0 & $\frac 12$ & 0\\ \hline 
			\multicolumn{1}{|c|}{${\rm U(1)}_{L_{\mu}-L_{\tau}}$}  & 0 & 0 & 0 & 0 & 1 & -1 
			& 0 & 1  & -1 & 0 & 1 & -1 & -1 & 0 & -1 & 1 & 0\\ \hline\hline\hline
\multicolumn{1}{|c|}{$\mathbb{Z}_2$ symmetry}  & + & + & + & + & + & + & + & +  & + & + & + 
& + & - & + & + & - & -\\ \hline\hline
			\end{tabular}}
	\caption{Gauge quantum numbers and $\mathbb{Z}_2$ parity of different SM and BSM
	particles.}
 \label{gauqn}
\end{table}
As has been discussed in earlier that the $L_{\mu}-L_{\tau}$ extension
of the SM is anomaly free, however, for the purpose of neutrino
mass generation via Type-I seesaw mechanism we invoke three
SM gauge singlet right handed neutrinos ($N_{Ri}$) having
nonzero $L_{\mu}-L_{\tau}$ charge in such a manner so that their
inclusion does not introduce any further anomaly. The Lagrangian
of right handed neutrinos is denoted by $\mathcal{L}_{N}$ which contains
kinetic energy terms, mass terms and Yukawa terms associated
with the SM lepton doublets ($L_{Li}$) allowed by
the symmetries of the present model.
\begin{eqnarray}
\mathcal{L}_{N}&=&
\sum_{j=e,\,\mu,\,\tau}\frac{i}{2}\,\overline{N^j_{R}}\gamma^{\alpha}D_{\alpha} N^j_{R} 
-\dfrac{1}{2}\,M_{ee}\,\overline{(N^e_{R})^{c}}\,N^e_{R}
-\dfrac{M_{\mu \tau}}{2}(\overline{(N^\mu_{R})^{c}}\,N^\tau_{R}
+\overline{(N^\tau_{R})^{c}}\,N^\mu_{R})  \nn \\ &&
-\dfrac{y_{e \mu}}{2}(\overline{(N^e_{R})^{c}}\,N^\mu_{R} 
+ \overline{(N^\mu_{R})^{c}}\,N^e_{R})\,\eta
- \dfrac{y_{e \tau}}{2}(\overline{(N^e_{R})^{c}}N^\tau_{R} 
+ \overline{(N^\tau_{R})^{c}}\,N^e_{R})\,\eta^*
\nn \\ &&
-\sum_{i=e,\,\mu,\,\tau} y_{i}\,\overline{L^i}_{L}
\tilde {H} N^{i}_{R} + {\rm h.c.}\;,
\label{lagN}
\end{eqnarray}
where $\tilde {H}=i\,\sigma_2H^*$. $M_{ee}$, $M_{\mu \tau}$ are the bare mass parameters
while $y_{e\mu}$, $y_{e \tau}$ and $y_i$ are the dimensionless Yukawa couplings.
In order to generate $b \to s$ transition at one loop level
involving $\mathbb{Z}_2$-odd scalars a non-standard SU(2)$_{L}$ singlet
fermionic field $\chi$ with a colour charge has been
introduced in this scenario. This fermion is also $\mathbb{Z}_2$-odd
and has an electric charge identical to SM down-type quarks. Furthermore, both left and right chiral
parts of $\chi$ field have same $L_{\mu}-L_{\tau}$ charge making it
a vector like fermion under ${\rm U}(1)_{L_{\mu}-L_{\tau}}$ symmetry.
The Lagrangian of this field is given by 
\begin{eqnarray}
\mathcal{L}_{\chi}=i\,\bar{\chi}\gamma^\alpha D_\alpha\chi - M_{\chi}
\bar{\chi}\chi- \left(\sum_{j=1}^{3}\,f_j\, \overline{Q_{Lj}}\,\Phi\,\chi_R + {\rm h.c.}\right)\;, 
\label{lchi}
\end{eqnarray}
where $M_\chi$ is the bare mass parameter for the $\chi$ field and $f_j$s are
couplings of the Yukawa type interactions among the SM quark doublets
($Q_{Lj}$), $\mathbb{Z}_2$-odd scalar doublet $\Phi$ and the right chiral
part of $\chi$. The above Yukawa type interactions terms involving
$s$ and $b$ quarks have significant roles in $b\rightarrow s$ transition
and hence in the explanation of $R_{K^{(*)}}$ anomalies. The covariant derivative
$D_{\alpha}$ for the field $\chi$ is defined as
\begin{equation}
D_\alpha\chi\equiv \bigg(\partial_\alpha-i g_1 \frac 13 \hat{B}_\alpha +
i g_{Z_{\mu\tau}} n_\chi \hat{X}_\alpha + i g_3 \frac{\Lambda^a}{2} G^a_\alpha\bigg)\chi\;,
\label{Dchi}
\end{equation}
where $g_1$, $g_{Z_{\mu\tau}}$ and $g_3$ are the ${\rm U}(1)_Y$, ${\rm U}(1)_{L_\mu-L_\tau}$ and ${\rm SU}(3)_C$ gauge coupling constants  respectively. $n_\chi$ is the ${L_\mu-L_\tau}$ charge
of $\chi$. Further, $\Lambda^a$s ($a = 1,2 \ldots 8$) are the eight Gell-Mann matrices
representing the generators for SU(3)$_{C}$ while the
corresponding gauge fields are denoted by $G^a_\alpha$. 
The $4^{\rm th}$, $5^{\rm th}$ and $6^{\rm th}$
terms of the Eq.~(\ref{LagT}) represent the kinetic terms for
all the non-standard scalar representations ($\eta$, $\Phi$ and $S$)
introduced in the present model for specific purposes.
Particularly, the complex singlet (under the SM gauge
group) scalar $\eta$ is necessary to break the ${\rm U}(1)_{L_\mu-L_\tau}$
symmetry spontaneously as it is the only scalar field which has not only a 
${\rm U}(1)_{L_\mu-L_\tau}$ charge but also has a nonzero vacuum expectation value (VEV) $v_2$. Consequently, after ${L_\mu-L_\tau}$
symmetry breaking one obtains a massive non-standard neutral gauge boson. It has played
crucial roles in different aspects:\,\,e.g., $(g-2)_\mu$ anomaly explanation,
amelioration of the anomalies that are related to $b\to s \mu \mu$ transition and most importantly it provides
new annihilation channels for the dark matter candidate, which alters its
dynamics from the standard case. Moreover, a $\mathbb{Z}_2$-odd ${{\rm SU}(2)_{L}}$
scalar doublet $\Phi$ having both ${\rm U}(1)_Y$ and ${\rm U}(1)_{L_\mu-L_\tau}$
charges which are required to get the NP contribution to $b\rightarrow s$
transition via the Yukawa like interaction given in Eq.\,(\ref{lchi}). Although,
one of the neutral components of $\Phi$ (lightest one) is stable, but for
the simultaneous explanation of the dark matter enigma, $(g-2)_{\mu}$
anomaly and $R_{K^{(*)}}$ anomalies we include another real singlet scalar
field $S$ which is also odd under $\mathbb{Z}_2$ symmetry. Covariant derivatives for the scalar fields $\eta$ and $\Phi$
are given as follows
\begin{eqnarray}
D_\alpha \eta &\equiv& \bigg(\partial_\alpha+ i g_{Z_{\mu\tau}}
n_\eta \hat{X}_\alpha \bigg)\eta\;, 
\label{Deta} \\
D_\alpha\Phi &\equiv& \bigg(\partial_\alpha+i g_1 \frac 12
\hat{B}_\alpha + i g_{Z_{\mu\tau}} n_\Phi \hat{X}_\alpha + i g_2
\frac{\sigma^a}{2} W^a_\alpha\bigg)\Phi\;,
\label{Dphi}
\end{eqnarray}
where $\sigma^a$ are the three Pauli's spin matrices with $a$
runs from 1 to 3. $n_{X}$ denotes the ${L_\mu-L_\tau}$ charge
of the corresponding scalar fields $X=\Phi,\,\eta$.
Further, $g_2$ is the ${\rm SU}(2)_L$ gauge coupling constant
and $W^a_\alpha$s are the corresponding gauge bosons.

Finally, the scalar potential $V(H,\,\eta,\,\Phi,\,S)$ in Eq.\,(\ref{LagT})
contains those interactions terms among the scalar fields which
remain invariant under all the symmetries of the present model,
has the following form,
\begin{eqnarray}
V(H, \eta, \Phi, S)&=&-m^2_{H}(H^\dagger H)-m^2_\eta(\eta^\dagger \eta) + m^2_\Phi(\Phi^\dagger \Phi) + \frac{m^2_S}{2} S^2\\ \nonumber
&+& \lambda_H (H^\dagger H)^2 + \lambda_\eta (\eta^\dagger \eta)^2 + \lambda_\Phi (\Phi^\dagger \Phi)^2 + \frac{\lambda_S}{4} S^4\\ \nonumber
&+& \lambda_1(H^\dagger H)(\eta^\dagger \eta) + \lambda_2 (H^\dagger H)(\Phi^\dagger \Phi)
+ \lambda_3 (H^\dagger \Phi)(\Phi^\dagger H) \\ \nonumber
&+&  \lambda_4(\Phi^\dagger \Phi)(\eta^\dagger \eta)
+ \lambda_5(\Phi^\dagger \Phi)S^2 + \lambda_6(\eta^\dagger \eta)S^2
+ \lambda_7(H^\dagger H)S^2 \\ \nonumber
&+& \left[\lambda_8(H^\dagger \Phi)S\eta+{\rm h.c.}\right]\;,
\label{Vpot}
\end{eqnarray}
where $m_{H}$, $m_{\eta}$, $m_\Phi$ and $m_S$ are real parameters
having dimension of mass and $\lambda_i$s $(i= H, \eta, S, 1,2 \ldots 7)$
are dimension less, real quartic coupling constants because the corresponding
operators are self-conjugate in nature. However, the quartic coupling
$\lambda_8$ can in general be a complex parameter and thus can act as an
extra source of CP-violation. Since in this work we are not studying
any CP-violating effects, we have taken $\lambda_8$ as a real parameter
and this assumption will not alter our conclusions. Although, the term proportional to
$\lambda_8$ has important significance in this model as it
generates mixing between $\Phi$ and $S$. Later we will
discuss more elaborately on this issue. The component wise
structure of the scalar fields are given in the following
\begin{eqnarray}
H=
\begin{pmatrix}
h^+ \\
\dfrac{h_1+v_1+i z_1}{\sqrt{2}}
\end{pmatrix},
\,\,\,\,
\eta=
\begin{pmatrix}
\dfrac{h_2+ v_2+ i z_2}{\sqrt{2}}
\end{pmatrix},
\,\,\,\,
\Phi=
\begin{pmatrix}
\phi^+ \\
\dfrac{\phi^0+a^0}{\sqrt{2}}
\end{pmatrix},
\label{scalflds}
\end{eqnarray}
where $v_1$ and $v_2$ are the VEVs of the scalar fields\footnote{$H$ and $\eta$ are
even under $\mathbb{Z}_2$ symmetry and hence $\mathbb{Z}_2$ remains unbroken.} $H$ and
$\eta$ respectively. After breaking of both electroweak and $L_{\mu}-L_{\tau}$
symmetries by the respective VEVs $v_1$ and $v_2$, one can have mixing between the real components
$h_1$ and $h_2$ due to the presence of an interaction term proportional to
$\lambda_1$ in $V(H, \eta, \Phi, S)$. The mixing matrix in the basis
$\frac{1}{\sqrt{2}}(h_1\,\,\,h_2)^{T}$ has the following form,
\begin{eqnarray}
\mathcal{M}^2_{\rm scalar} = \left(\begin{array}{cc}
2\lambda_H v^2_1 ~~&~~ \lambda_1 v_1 v_2 \\
~~&~~\\
\lambda_1 v_1 v_2 ~~&~~ 2 \lambda_\eta v^2_2
\end{array}\right) \,\,.
\end{eqnarray}
Diagonalising the mass squared matrix by an orthogonal transformation, we obtain
two physical CP-even neutral scalars $H_1$ which has been considered as SM like Higgs of mass 125.5 GeV
and $H_2$. These fields are also even under $\mathbb{Z}_2$ symmetry similarly as
$h_1$ and $h_2$. The physical states $H_1$ and $H_2$ are related with previous
states $h_1$ and $h_2$ by the following relation,
\begin{eqnarray}
\left(\begin{array}{c} H_1 \\ H_2\end{array}\right)
=\left(\begin{array}{cc}\cos\theta_s ~-\sin\theta_s
\\ \sin\theta_s ~~~~\cos\theta_s \end{array}\right)
\left(\begin{array}{c} h_1 \\ h_2\end{array}\right) \,\,,
\label{CP-massmatrix}
\end{eqnarray}
where $\theta_s$ is the mixing angle which can
be expressed as, 
\begin{eqnarray}\label{CP-angle}
\theta_s &=& \frac{1}{2}~\tan^{-1}\left(\frac{\frac{\lambda_1}{\lambda_\eta}\frac{v_1}{v_2}}
{1 - \frac{\lambda_H}{\lambda_\eta}\frac{v^2_1}{v^2_2}}\right) \,\,.
\end{eqnarray}
Mass eigenvalues corresponding to the physical scalars $H_1$ and $H_2$
are given by,
\begin{eqnarray}
M_{H_1} &=& \sqrt{\lambda_H v^2_1 + \lambda_{\eta} v^2_2 + 
\sqrt{(\lambda_H v^2_1 - \lambda_\eta v^2_2)^2 + (\lambda_1 v_1 v_2)^2} }\ , \\
\label{CP-mass1}
M_{H_2} &=& \sqrt{\lambda_H v_1^2 + \lambda_{\eta} v^2_2 - 
\sqrt{(\lambda_H v^2_1 - \lambda_\eta v^2_2)^2 + (\lambda_1 v_1 v_2)^2} } \,\ .
\label{CP-mass2}
\end{eqnarray}
Furthermore, similar to the $\mathbb{Z}_2$-even sector, the $\mathbb{Z}_2$-odd
sector also exhibits mass mixing between $\phi^0$ and $S$. This also
happens when both $H$ and $\eta$ get nonzero VEVs and in this case the term proportional to
$\lambda_8$ in $V(H, \eta, \Phi, S)$ is solely responsible for such mixing. Therefore, 
the $\mathbb{Z}_2$-odd real singlet scalar $S$ mixes with CP-even component $\phi^0$
of the $\mathbb{Z}_2$-odd doublet $\Phi$. However, as there is no spontaneous CP-violation,
the CP-odd component $a^0$ remains decoupled from the CP-even fields and with respect
to the basis $\frac{1}{\sqrt{2}}(S\,\,\,\phi^0\,\,a^0)^{T}$, the $3\times 3$ odd-sector
mixing matrix has a block diagonal form,
\begin{eqnarray}
\mathcal{M}^2_{\rm DM} = \left(\begin{array}{ccc}
(m^2_S+v^2_1\lambda_7+v^2_2\lambda_6) & \frac{v_1v_2\lambda_8}{\sqrt{2}} & 0 \\
\frac{v_1v_2\lambda_8}{\sqrt{2}} & \frac 12\{2m^2_\Phi+v^2_1(\lambda_2+\lambda_3)+v^2_2\lambda_4\} & 0 \\
0  & 0 &  \frac 12\{2m^2_\Phi+v^2_1(\lambda_2+\lambda_3)+v^2_2\lambda_4\} \end{array}
\right)\,. \nonumber \\ 
\label{dm-mixing}
\end{eqnarray}
One can easily diagonalise this matrix using an orthogonal transformation
by an angle $\theta_D$ between $S$ and $\phi^0$. Therefore, after diagonalisation
we have three physical states $\rho_1$, $\rho_1$ and $\rho_3$, where
$\rho_1$ and $\rho_2$ are orthogonal linear combinations of $S$ and $\phi^0$
while the remaining physical scalar $\rho_3$ exactly coincides with $a^0$.
In matrix notation, the basis transformation can be shown as 
\begin{eqnarray}
\left(\begin{array}{c} \rho_1 \\ \rho_2 \\ \rho_3\end{array}\right)
=\left(\begin{array}{ccc}\cos\theta_D &-\sin\theta_D&0\\ 
\sin\theta_D& \cos\theta_D&0\\
0&0&1\end{array}\right)
\left(\begin{array}{c} S \\ \phi^0 \\ a^0\end{array}\right) \,\,,
\label{CP-massmatrix}
\end{eqnarray}
where the mixing angle $\theta_D$ can be expressed in terms
of parameters of the Lagrangian as, 
\begin{eqnarray}\label{dm-angle}
\theta_D &=& \frac{1}{2}~\tan^{-1}\left(\frac{2\sqrt{2}v_1 v_2\lambda_8}
{2m^2_\Phi-2m^2_S+v^2_1(\lambda_2+\lambda_3-2\lambda_7)+v^2_2(\lambda_4-2\lambda_6)}\right) \,\,.
\end{eqnarray}
Among the three states ($\rho_1$, $\rho_1$ and $\rho_3$), we choose $\rho_1$ as the lightest
odd particle (LOP) which is regarded as the stable dark matter candidate in this scenario. Thus,
the dark matter candidate in this scenario is an admixture of singlet and doublet states.
The expressions for the masses of these $\mathbb{Z}_2$-odd scalar fields are given below 
\begin{eqnarray}
M_{\rho_1}=\sqrt{(m^2_S+v^2_1\lambda_7+v^2_2\lambda_6)\cos^2\theta_D-\sqrt{2}v_1 v_2\lambda_8\cos\theta_D\sin\theta_D+M^2_{\rho_3}\sin^2\theta_D}\;, 
\label{mrho1} \\
M_{\rho_2}=\sqrt{(m^2_S+v^2_1\lambda_7+v^2_2\lambda_6)\sin^2\theta_D+\sqrt{2}v_1 v_2\lambda_8\cos\theta_D\sin\theta_D+M^2_{\rho_3}\cos^2\theta_D}\;,
\label{mrho2}
\end{eqnarray}
where
\begin{eqnarray}
M_{\rho_3}&=& \sqrt{m^2_\Phi+\frac 12 \left[ v^2_1(\lambda_2+\lambda_3)+v^2_2\lambda_4\right]}\,.
\label{mrho3}
\end{eqnarray}
Further using Eqs.\,(\ref{mrho1}-\ref{mrho3}), one can establish a
relation between $M_{\rho_1}$, $M_{\rho_2}$, $M_{\rho_3}$ and $\theta_D$, which has the
following form
\begin{eqnarray}
M^2_{\rho_3} = {M^2_{\rho_1}\sin^2\theta_D+M^2_{\rho_2}\cos^2\theta_D}\;.
\label{mho3-relation}
\end{eqnarray}
Therefore, the mass of the CP-odd scalar $\rho_3$ is not an independent
quantity in the present scenario and it becomes fixed though the above
relation once we know other parameters like $M_{\rho_1}$, $M_{\rho_2}$ and $\theta_D$.
This is a consequence of that, the $2\times2$ and $3\times3$ elements of
the dark sector mixing matrix $\mathcal{M}^2_{\rm DM}$ are identical. From the
symmetry argument this can be understood as follows. The splitting between the
coefficients of ${\phi^0}^2$ ($\varpropto$ $2\times 2$ element of
${\mathcal{M}^2_{\rm DM}}$) and ${a^0}^2$ ($\varpropto$ $3\times 3$
element of ${\mathcal{M}^2_{\rm DM}}$) of a $\mathbb{Z}_2$-odd doublet $\Phi$ is
obtained from a term like $(H^\dagger\Phi)^2$ (usual $\lambda_5$ term in the Inert Doublet
Model \cite{Barbieri:2006dq}), which is forbidden here by the ${\rm U(1)}_{L_{\mu}-L_{\tau}}$
symmetry invariance. Additionally, in the dark sector we also have a charged scalar
$\phi^\pm$ and its mass term is given by 
\begin{eqnarray}
M_{\phi^\pm}&=&\sqrt{M^2_{\rho_3}-\frac 12 v^2_1\lambda_3}\;.
\label{mphpm} 
\end{eqnarray}
Let us now find out the effects of the extra ${\rm U(1)}_{L_{\mu}-L_{\tau}}$
local gauge symmetry on the gauge sector and generate the physical states
of the gauge bosons with their proper mass terms. In the Eq.\,(\ref{fieldtensor}),
$\hat{B}_{\alpha}$ and $\hat{X}_{\alpha}$ are denoted as gauge fields corresponding
to gauge groups U(1)$_{Y}$ and U(1)$_{L_\mu-L_\tau}$ respectively. As mentioned
earlier, the kinetic terms for the two U(1) gauge fields with hat notation are not diagonal
and it is clearly evident from the presence of a mixing term
between two U(1) gauge fields proportional $\epsilon$. 
The kinetic mixing parameters is severely constrained
from the electroweak precision data (sensitive mainly in the low mass
regime of the extra gauge boson) \cite{Hook:2010tw, Cline:2014dwa}
and also from di-lepton searches at the LHC (for relatively high mass
regime i.e., few hundred GeV to few TeV range). Now, one can perform a basis transformation from ``hat'' states
to ``un-hat'' states, due to which the off-diagonal
kinetic term vanishes. This can be achieved by applying a
following transformation\footnote{This transformation matrix is
not a unique one. For a general $2\times 2$ real matrix,
we have four independent elements. However, using $c_1=c_2=1$ and $c_3$ = 0,
we have only three independent equations to solve for four variables. Here,
$c_1$, $c_2$ and $c_3$ are coefficients of $\frac{1}{4}B_{\mu\nu}B^{\mu\nu}$,
$\frac{1}{4}X_{\mu\nu}X^{\mu\nu}$ and $\frac{\epsilon}{2}B_{\mu\nu}X^{\mu\nu}$
respectively. Thus, one can express three elements in terms of the
fourth one and for each real value of that element, we will have a different
transformation matrix which eventually cancels the kinetic mixing term.
For the particular matrix that we have used here is obtained by setting
$2\times 1$ element of the transformation matrix equal to zero. Such a special
choice easily reproduces all the phenomena of electromagnetism.},
\begin{eqnarray}
\left(\begin{array}{c} B_{\alpha} \\ X_{\alpha}\end{array}\right) = \left(\begin{array}{cc}
1 &-\epsilon\\ 0 &\sqrt{1-\epsilon^2}\end{array}\right)\left(\begin{array}{c} \hat{B}_{\alpha} \\
\hat{X}_{\alpha}\end{array}\right)\,\,,
\label{hat-unhat-matrix}
\end{eqnarray}
and since experiment dictates $\epsilon \ll$ 1,
therefore using the approximation $\mathcal{O}(\epsilon^2)\approx 0$ we have 
\begin{eqnarray} 
\hat{B}_{\alpha} \simeq B_{\alpha} + \epsilon X_{\alpha} \,\,\,\,
{\rm and}\,\,\,
\hat{X}_{\alpha} \simeq  X_{\alpha} \,\, .
\label{hat-unhat-trans}
\end{eqnarray}
After the occurrence of both electroweak symmetry breaking (EWSB)\footnote
{In the present scenario, after EWSB one can readily determine
the mass of the $W^\pm$ gauge boson which is exactly equal to
that of the SM, i.e., $M_W = \frac 12 g_2v_1$.}
and ${L_\mu-L_\tau}$ breaking by the VEVs of the neutral components
of $H$ and $\eta$, we obtain a $3\times 3$ mass square matrix in the basis
of three neutral gauge bosons namely $W^\alpha_3$, $B^{\alpha}$, $X^{\alpha}$
using Eqs.\,(\ref{Deta}-\ref{Dphi}, \ref{hat-unhat-trans}), 
\begin{eqnarray}
\mathcal{M}^2_{\rm gauge} = \left(\begin{array}{ccc}
\frac{1}{4}g^2_2v^2_1 &-\frac{1}{4}g_2 g_1v^2_1 & -\frac{1}{4}g_2 g_1v^2_1\epsilon \\
-\frac{1}{4}g_2 g_1v^2_1 & \frac{1}{4}g^2_1v^2_1 & \frac{1}{4} g^2_1 v^2_1\epsilon \\
-\frac{1}{4}g_2 g_1v^2_1\epsilon  & \frac{1}{4} g^2_1 v^2_1\epsilon & g^2_{Z_{\mu\tau}}v^2_2 \end{array}\right)\,\, .
\label{gauge-mixing}
\end{eqnarray}
The above matrix has a special symmetry. If we rotate $W^{\alpha}_3$
and $B^{\alpha}$ by the Weinberg angle $\tan \theta_{\rm W} = \dfrac{g_1}{g_2}$,
the matrix $\mathcal{M}^2_{\rm gauge}$ reduces to a $2\times 2$ block diagonal
structure with respect to an intermediate state $\mathcal{Z}^\alpha \equiv \cos \theta_{\rm W}
W^\alpha_3 - \sin \theta_{\rm W} B^\alpha$ and $X^\alpha$ while the other
orthogonal state i.e. $A^\alpha = \sin \theta_{\rm W} W^\alpha_3 + \cos \theta_{\rm W} B^\alpha$
having zero mass eigenvalue becomes completely decoupled. This is possible
due to the special choice of the transformation matrix we have considered
in Eq.\,(\ref{hat-unhat-matrix}). Now, once we reduce a $3\times3$
matrix to a $2\times2$ block diagonal form, we already have made our life
very simple and next task is to perform another orthogonal transformation
between the states $\mathcal{Z}^{\alpha}$ and $X^{\alpha}$ to finally get
the physical $Z$ and $Z_{\mu\tau}$ bosons. This is mathematically demonstrated
below for both mass matrix as well as eigenstates,

\begin{eqnarray}
\mathcal{M}^2_{\rm gauge}\,
\xRightarrow{\mathcal{O}(\theta_{\rm W})}
\left(\begin{array}{ccc}
\frac{1}{4}(g_1^2+g_2^2)v_1^2 & 0 & -\frac{\epsilon}{4} g_1
\sqrt{g_1^2+g_2^2} v_1^2 \\
0 & 0 & 0 \\
-\frac{\epsilon}{4} g_1 \sqrt{g_1^2+g_2^2} v_1^2
& 0 & g^2_{Z_{\mu\tau}} v^2_2 
\end{array}
\right)\,
\xRightarrow{\mathcal{O}(\theta_{\mu\tau})} 
\left(\begin{array}{ccc}
M_Z & 0 & 0 \\
0 & 0 & 0 \\
0 & 0 & M_{Z_{\mu\tau}} 
\end{array}
\right)\,
\end{eqnarray}
and 
\begin{eqnarray}
\hspace{2cm}\left(\begin{array}{c} W^{\alpha}_3 \\B^{\alpha}
\\X^{\alpha}\end{array}\right)
\xRightarrow{\mathcal{O}(\theta_{\rm W})^{T}}
\left(\begin{array}{c} \mathcal{Z}^\alpha_3\\ A^{\alpha}\\ X^{\alpha}
\end{array}\right)
\xRightarrow{\mathcal{O}(\theta_{\mu\tau})^{T}}
\left(\begin{array}{c} Z^\alpha\\ A^{\alpha}\\ Z^{\alpha}_{\mu\tau}
\end{array}\right)\,,
\end{eqnarray}
where, the masses of two massive neutral gauge bosons ($Z$ and $Z_{\mu\tau}$)
are respectively given as
\begin{eqnarray}
M_{Z} &=& \sqrt{\frac{g^2_2(v^2_1+v^2_2)}{4}\cos^2\theta_{\mu\tau} + g^2_{Z_{\mu\tau}}v^2_2\sin^2\theta_{\mu\tau} + \frac{g_1\sqrt{(g^2_1+g^2_2)} v^2_1\epsilon}{4}\sin 2\theta_{\mu\tau}}\;, \\
\label{Zmass}
M_{Z_{\mu\tau}} &=& \sqrt{\frac{g^2_2(v^2_1+v^2_2)}{4}\sin^2\theta_{\mu\tau} + g^2_{Z_{\mu\tau}}v^2_2\cos^2\theta_{\mu\tau}- \frac{g_1\sqrt{(g^2_1+g^2_2)} v^2_1\epsilon}{4}\sin 2\theta_{\mu\tau}}\;,
\label{Z'mass}
\end{eqnarray}
and the two orthogonal transformation matrices are given by,

\begin{eqnarray}
\mathcal{O(\theta_{\rm W})} = \left(\begin{array}{ccc}
\cos \theta_{\rm  W} & \sin \theta_{\rm W} & 0\\
-\sin \theta_{\rm W} & \cos \theta_{\rm W} & 0 \\
0 & 0 & 1
\end{array}
\right)\,,\hskip 0.2in  
\mathcal{O(\theta_{\mu\tau})} = \left(\begin{array}{ccc}
\cos \theta_{\mu\tau} & 0 & \sin \theta_{\mu\tau}\\
0 & 1 & 0 \\
-\sin \theta_{\mu\tau} & 0 & \cos \theta_{\mu\tau}
\end{array}
\right)\,\,.
\end{eqnarray}

Finally, the gauge basis and the mass basis of the neutral gauge bosons
are related the following orthogonal transformation
\begin{eqnarray}
\hspace{2cm}\left(\begin{array}{c} Z^{\alpha} \\A^{\alpha} \\Z^{\alpha}_{\mu\tau}\end{array}\right)
&=& \mathcal{O}(\theta_{\rm W},\,\theta_{\mu\tau})^{T}
\left(\begin{array}{c} W^\alpha_3\\ B^{\alpha}\\ X^{\alpha}\end{array}\right)\,\, ,  
\end{eqnarray}
with
\begin{eqnarray}
\mathcal{O}(\theta_{\rm W},\,\theta_{\mu\tau})^{T} &=&
\mathcal{O}(\theta_{\mu\tau})^{T}\,\mathcal{O}(\theta_{\rm W})^{T} \nonumber \\ 
&=&
\left(\begin{array}{ccc}
\cos\theta_{\mu\tau} \cos\theta_{\rm W} &-\cos\theta_{\mu\tau}
\sin\theta_{\mu\tau}& -\sin\theta_{\mu\tau}\\
\sin\theta_{\rm W}&\cos\theta_{\rm W} & 0\\
\sin\theta_{\mu\tau} \cos\theta_{\rm W}&-\sin\theta_{\mu\tau}
\sin\theta_{\rm W}&\cos\theta_{\mu\tau}
\end{array}\right)\,\, ,
\label{u-matrix}
\end{eqnarray} 
where $\theta_{\rm W}$, as mentioned above, is the familiar Weinberg angle
and $\theta_{\mu\tau}$ is the mixing angle between two neutral gauge bosons
$Z$ and $Z_{\mu\tau}$. These mixing angles can be expressed in terms
gauge coupling constants, VEVs and the kinetic mixing parameters as follows,
\begin{eqnarray}
\theta_{\rm W} = \tan^{-1}\left(\dfrac{g_1}{g_2}\right)\ , \ \ \ \ \ 
\theta_{\mu\tau} = \frac{1}{2}\,\tan^{-1}\left(\dfrac{\dfrac{2\,\epsilon g_1}{\sqrt{g^2_1 + g^2_2}}}
{1 - \dfrac{4g^2_{Z_{\mu\tau}}}{g^2_1 + g^2_2} \dfrac{v^2_2}{v^2_1}}\right) \,\, .
\label{gauge-mix-angle}
\end{eqnarray}

Before we proceed any further, it is worthwhile to mention about the independent parameters.
In this model, in addition to the SM parameters, we have fourteen
new parameters in the scalar sector (excluding SM-Like Higgs boson mass
and VEV $v_1$), three additional Yukawa like coupling constants and
one mass term in the extended quark sector\footnote{Here, we are not considering
Yukawa like coupling constants and bare mass terms in the extended neutrino sector.}
and two more couplings in the gauge sector in the form of
new gauge coupling $g_{Z_{\mu\tau}}$ and kinetic mixing parameter $\epsilon$.
These twenty independent parameters are: $M_{H_2}$, $M_{\phi^{\pm}}$, $M_{\rho_1}$, $M_{\rho_2}$,
$M_{Z_{\mu\tau}}$, $\theta_D$, $\theta_s$, $\lambda_{\Phi}$, $\lambda_S$,
$\lambda_2$, $\lambda_4$, $\lambda_5$, $\lambda_6$, $\lambda_7$, $f_1$,
$f_2$, $f_3$, $M_{\chi}$, $g_{Z_{\mu\tau}}$ and $\theta_{\mu\tau}$.
In terms of these independent parameters the other parameters
appearing in the Lagrangian (Eq.\,(\ref{LagT})) can be
obtained using Eqs.\,(\ref{CP-angle}-\ref{CP-mass2}),
Eqs.\,(\ref{dm-angle},\,\ref{mrho1}), Eqs.\,(\ref{mrho3},
\ref{mphpm}) and Eqs.\,(\ref{Z'mass},\,\ref{gauge-mix-angle})\footnote{Additionally, one
needs to use minimization conditions of the scalar potential $V(H,\,\eta,\,\Phi,\,S)$.}.
\section{\boldmath${b \to s}$ flavour observables}\label{flav}
\subsection{\boldmath$R_{K^{(*)}}$ anomalies}\label{RKRKs}
In the present scenario the NP part of the effective Hamiltonian $\mathcal H_\text{eff}(\equiv \mathcal H_\text{eff}^\text{SM} +
\mathcal H_\text{eff}^\text{NP}$) that describes the $b \to s \ell\ell$ transitions is given by
\begin{equation}
\label{eq:HeffRK}
\mathcal{H}_\text{eff}^\text{NP} = - \frac{4\,G_F}{\sqrt{2}} V_{tb}V_{ts}^* \frac{e^2}{16\pi^2}
\sum_{\ell=e,\mu}
C^{\rm NP}_{9\;\ell} (\bar{s} \gamma_{\alpha} P_{L} b)(\bar{\ell} \gamma^\alpha \ell)  + 
C^{\rm NP}_{10\;\ell} (\bar{s} \gamma_{\alpha} P_{L} b)( \bar{\ell} \gamma^\alpha \gamma_5 \ell)+\text{h.c.} \,,
\end{equation}
where $G_F$ is the Fermi constant, $V_{ij}$ are the Cabibbo-Kobayashi-Maskawa (CKM) matrix elements.
Here we neglect other dimension-six operators for example, $C_7$ can not give significant contributions to the processes, because it corresponds
to the dipole operator that is strictly constrained by branching ratio of $B\to X_s \gamma$ \cite{Kawamura:2017ecz}. 
Also four-quark operators \cite{Jager:2017gal} cannot
play any significant role for the violation of LFU, hence they are irrelevant in this work.
Moreover, four-fermion contact interactions with scalar currents could be
a natural source of LFU violation, although they are highly constrained
by existing measurements of the $B_s \to \mu^+\mu^-$ and $B_s \to e^+e^-$
branching ratios~\cite{Aaij:2017vad,Aaltonen:2009vr}. The NP contribution to the WC $C^{\rm NP,\ell}_{9}=C^{\ell}_{9Z}+C^{\ell}_{9Z_{\mu\tau}}$ can be obtained from
\begin{eqnarray}\label{dc9}
C^{\ell}_{9Z(Z_{\mu\tau})}&=&-\frac{\sqrt{2}}{16 \pi \alpha_{\rm em}G_F V_{tb}V_{ts}^*}\frac{\mathscr{L}^9_{Z(Z_{\mu\tau})}}{M^2_{Z(Z_{\mu\tau})}}\bigg(-\frac{\mathcal{G}_{Z(Z_{\mu\tau})} f_2 f_3}{4 }\bigg[-2\ln(m^2_\chi)-1\\  \nonumber 
&&+h_q(x_1)(1-2x_1)\sin^2\theta_D+h_q(x_2)(1-2x_2)\cos^2\theta_D+h_q(x_3)(1-2x_3)\bigg] \\  \nonumber
&+&\frac{\mathcal{C}_{Z(Z_{\mu\tau})} f_2 f_3}{4 }\bigg[\{-\ln(M^2_{\rho_1})+h_w(x_1,r_1)\}\sin^2\theta_D+\{-\ln(M^2_{\rho_2})+h_w(x_2,r_2)\}\cos^2\theta_D\bigg]\\  \nonumber
&-&\frac{\mathcal{S}_{Z(Z_{\mu\tau})} f_2 f_3}{4 }\bigg[\{-\ln(M^2_{\rho_1})+h_s(x_1)\}\sin^2\theta_D+\{-\ln(M^2_{\rho_2})+h_s(x_2)\}\cos^2\theta_D\\  \nonumber
&&+\{-\ln(M^2_{\rho_3})+h_s(x_3)\}\bigg]\bigg)\;,
\end{eqnarray}
while the NP contribution to the WC
$C^{\rm NP,\ell}_{10}=C^{\ell}_{10Z}+C^{\ell}_{10Z_{\mu\tau}}$ is given by
\begin{eqnarray}\label{dc10}
C^{\ell}_{10Z(Z_{\mu\tau})}&=&-\frac{\sqrt{2}}{16 \pi \alpha_{\rm em} G_F V_{tb}V_{ts}^*}\frac{\mathscr{L}^{10}_{Z(Z_{\mu\tau})}}{M^2_{Z(Z_{\mu\tau})}}\bigg(-\frac{\mathcal{G}_{Z(Z_{\mu\tau})} f_2 f_3}{4 }\bigg[-2\ln(m^2_\chi)-1\\  \nonumber 
&&+h_q(x_1)(1-2x_1)\sin^2\theta_D+h_q(x_2)(1-2x_2)\cos^2\theta_D+h_q(x_3)(1-2x_3)\bigg] \\  \nonumber
&+&\frac{\mathcal{C}_{Z(Z_{\mu\tau})} f_2 f_3}{4 }\bigg[\{-\ln(M^2_{\rho_1})+h_w(x_1,r_1)\}\sin^2\theta_D+\{-\ln(M^2_{\rho_2})+h_w(x_2,r_2)\}\cos^2\theta_D\bigg]\\  \nonumber
&-&\frac{\mathcal{S}_{Z(Z_{\mu\tau})} f_2 f_3}{4 }\bigg[\{-\ln(M^2_{\rho_1})+h_s(x_1)\}\sin^2\theta_D+\{-\ln(M^2_{\rho_2})+h_s(x_2)\}\cos^2\theta_D\\  \nonumber
&&+\{-\ln(M^2_{\rho_3})+h_s(x_3)\}\bigg]\bigg)\;,
\end{eqnarray}
although we have found that the contribution of $C^{\rm NP,\ell}_{10}$ ($\ell\equiv\mu$) is insignificant\footnote{Due to this reason there is no significant NP contribution to the decay $B_s\to \mu^+\mu^-$. Therefore, there is no stringent constraint from the branching ratio of this process to our analysis.} and we will focus only on $C^{\ell}_{9Z(Z_{\mu\tau})}$ ($\ell\equiv\mu$) in rest of the analysis\footnote{Therefore, the present scenario can be considered as a typical scenario which can provide the NP contribution to $C^{\ell}_{9}$ ($\ell\equiv\mu$) only. Although, there is a NP contribution to $C^{e}_{9}$ but practically it has no significance due to very small mixing between $Z$ and $Z_{\mu\tau}$. Hence, the coupling between $Z_{\mu\tau}$ and $e^+e^-$ pair is effectively vanishing in nature.}. $\alpha_{\rm em}$ is the fine structure constant. Here $x_{1,2,3}=\frac{M^2_\chi}{M^2_{\rho_{1,2,3}}}$ and $r_{1,2}=\frac{M^2_{\rho_{3}}-M^2_{\rho_{1,2}}}{M^2_{\rho_{1,2}}}$. The expressions of the factors $g_{Z(Z_{\mu\tau})}$, $c_{Z(Z_{\mu\tau})}$, $s_{Z(Z_{\mu\tau})}$, $\mathscr{L}^9_{Z(Z_{\mu\tau})}$, $\mathscr{L}^{10}_{Z(Z_{\mu\tau})}$ and the functions $h_q(x)$, $h_w(x,r)$, $h_s(x)$ are given in the Appendix~\ref{flav_app}. In Fig.~\ref{newdia} we have shown relevant Feynman diagrams responsible for the additional contribution to the $b\to s \mu\mu$ transition. It is clearly evident from these Feynman diagrams that the NP contribution to the WC $C^{\rm NP,\ell}_{9}$ is provided by the non-standard bottom like fermion field $\chi$ and the dark matter candidate $\rho_1$ with its partners $\rho_2$ and $\rho_3$. Later we provide the dark matter phenomenology of a weakly interacting massive particle (WIMP) type dark matter candidate $\rho_1$ and related issues by considering the constraints of flavour physics observables that we have considered in this article. 
\begin{figure}[t]
\begin{center}
\subfloat[]{\label{fig1a}\includegraphics[scale=0.8,angle=0]{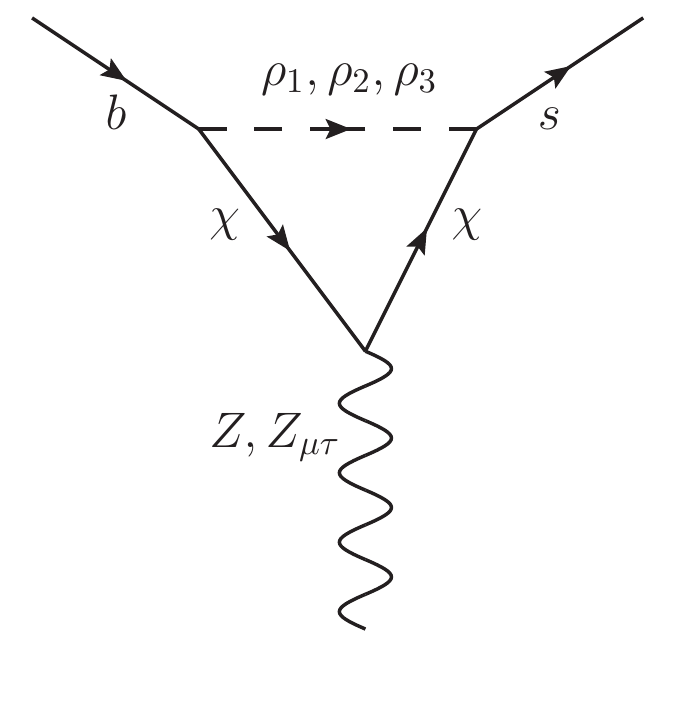}}
\subfloat[]{\label{fig1b}\includegraphics[scale=0.8,angle=0]{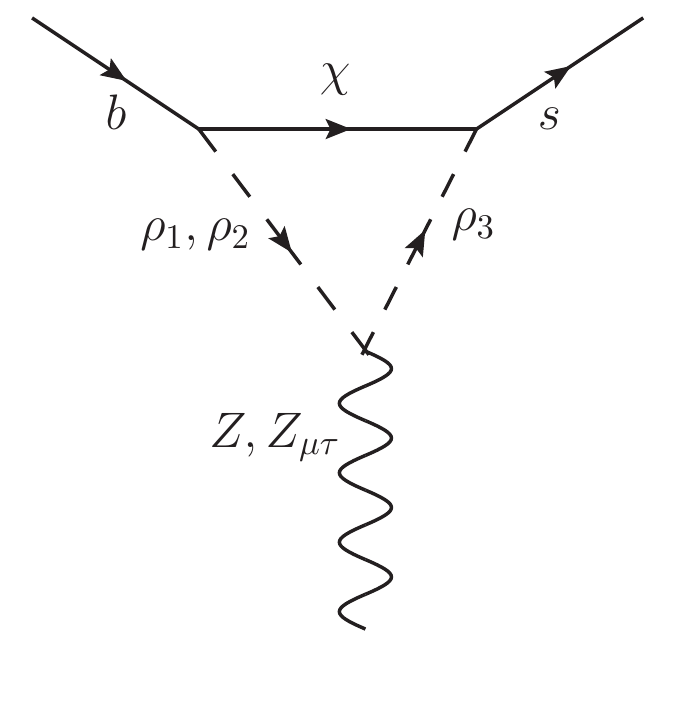}}
\subfloat[]{\label{fig1c}\includegraphics[scale=0.8,angle=0]{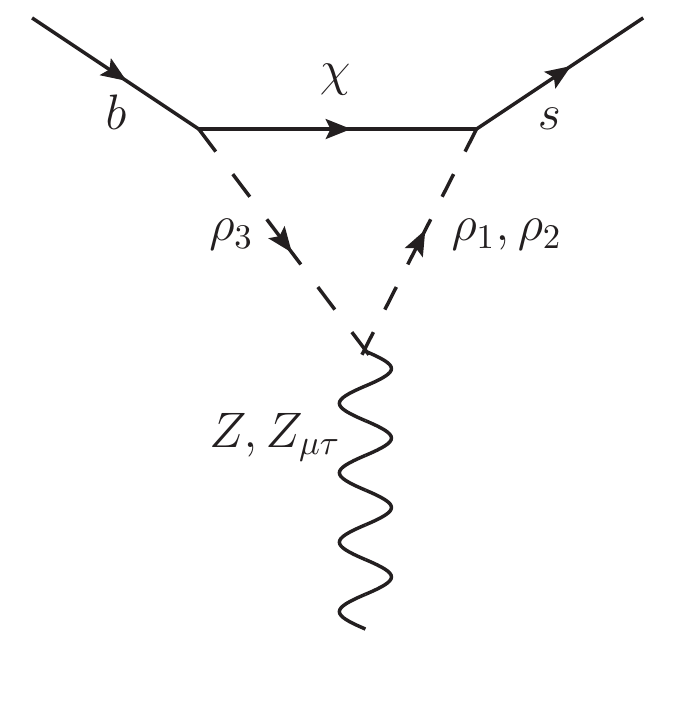}}\\
\subfloat[]{\label{fig1d}\includegraphics[scale=0.8,angle=0]{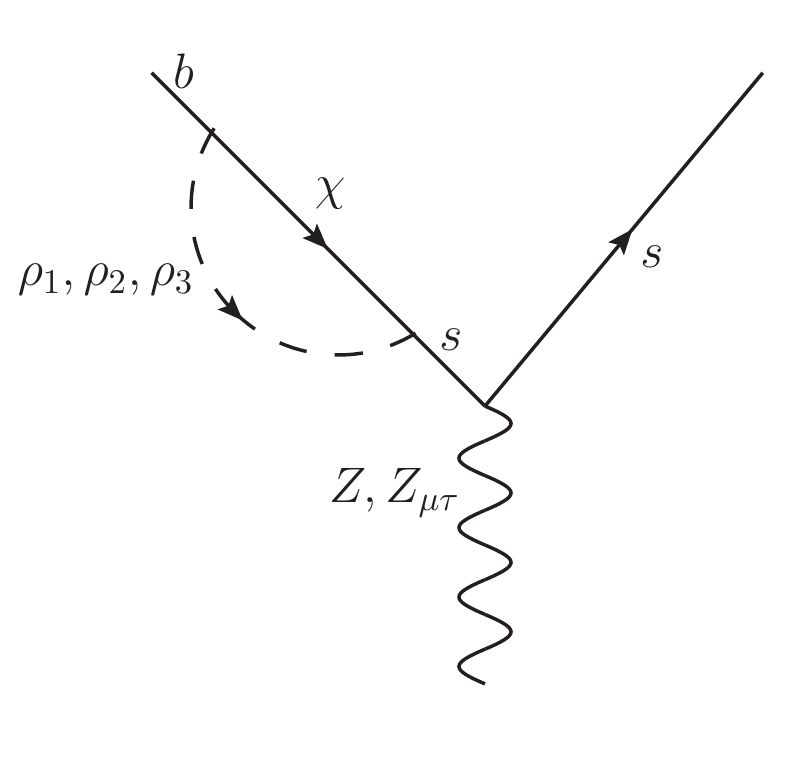}}
\subfloat[]{\label{fig1e}\includegraphics[scale=0.8,angle=0]{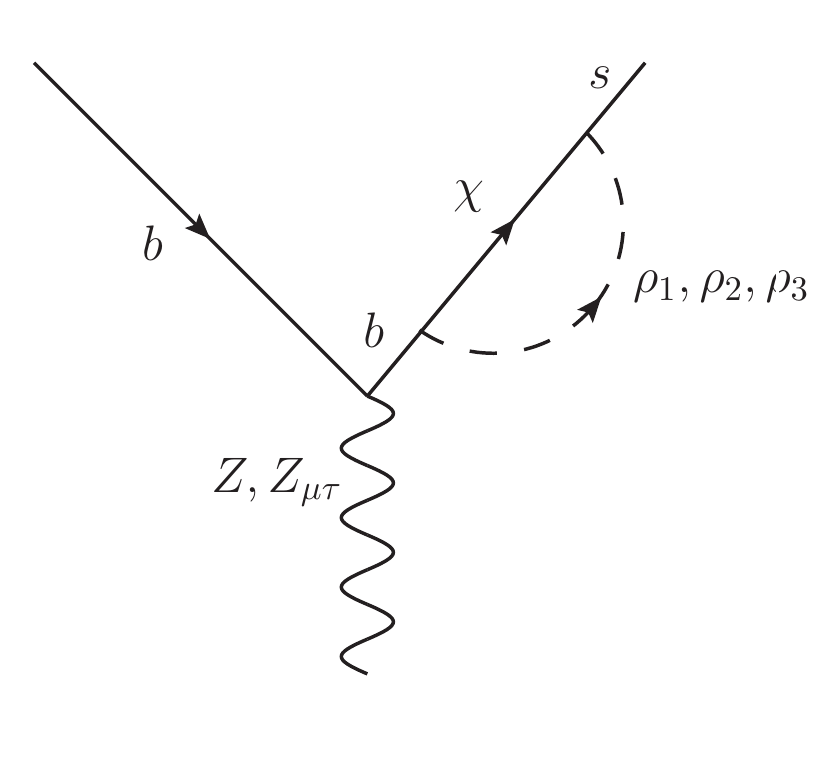}}
\caption{$Z$ and $Z_{\mu\tau}$-penguin and self-energy diagrams that
contribute to the decay of $b\to s \mu\mu$ in addition to SM contribution.}
\label{newdia}
\end{center}
\end{figure}
To ameliorate the tension between the SM prediction and experimental data
for $R_{K^{(*)}}$ we use $C^{\rm NP,\mu}_{9} \in [-1.26, -0.63]$ \cite{Aebischer:2019mlg}
in $2\sigma$ interval. For the purpose of notational simplicity, from now and onwards, we use $\Delta{C_9}$ for the total NP contributions to the WC $C_9$ for $\ell =\mu$, i.e., $C^{\rm NP,\mu}_{9} = C^{\mu}_{9Z}
+ C^{\mu}_{9Z_{\mu\tau}}=\Delta{C_9}$.   

\begin{figure}[h!]
\centering
\subfloat[Variation of $\Delta{C_9}$ with $M_{\rho_1}$
for $M_{\rho_2} = 506$ GeV, $M_{\chi}= 1300$ GeV,
$g_{Z_{\mu\tau}}=0.93\times 10^{-3}$, $M_{Z_{\mu\tau}}=0.076$ GeV and
$\theta_D = 0.095$ rad. \label{Fig:dc9-vs-mrho1}]{
\includegraphics[height=6cm,width=7.8cm,angle=0]{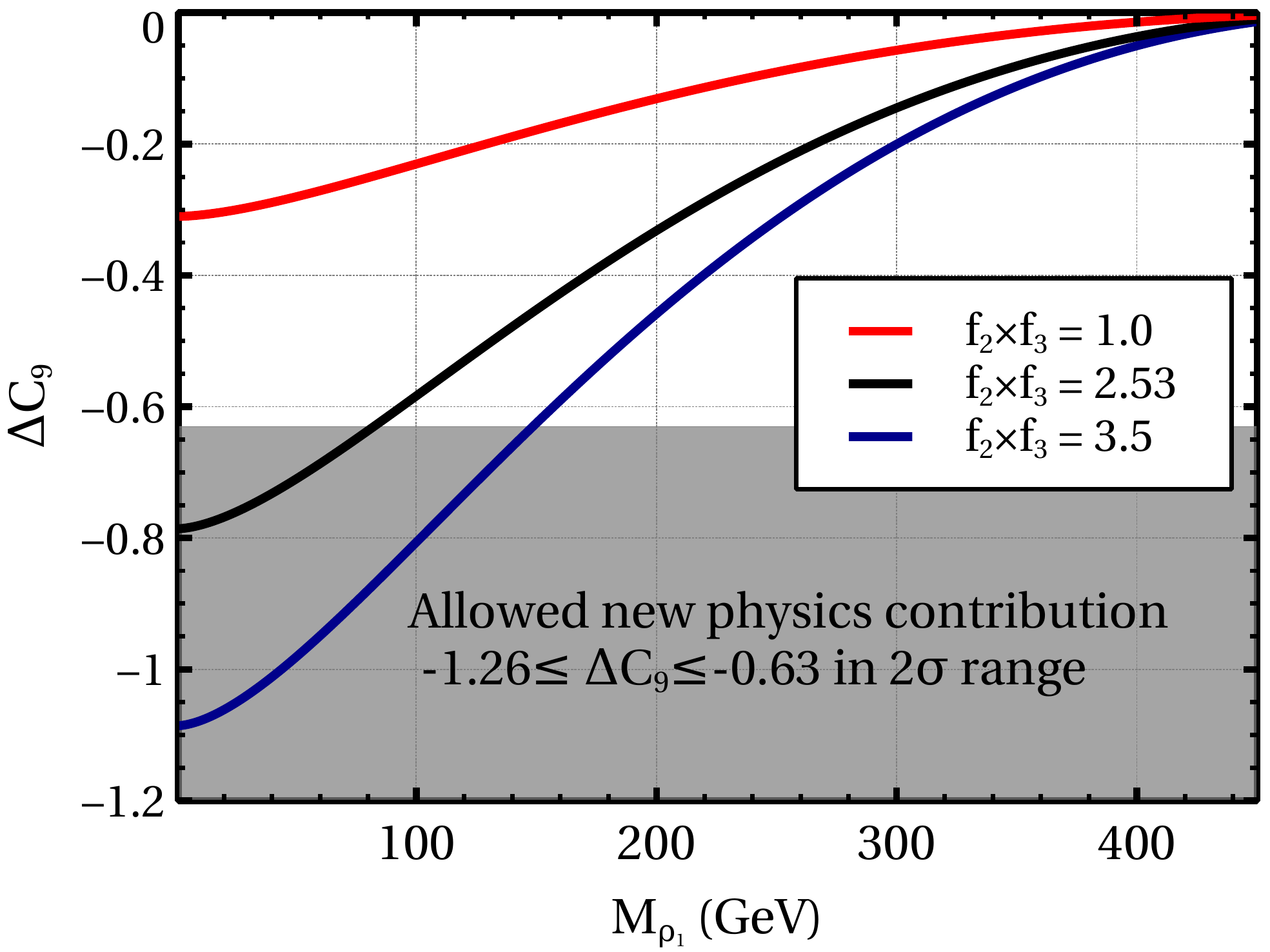}
}
\hskip 0.2in
\subfloat[Variation of $\Delta{C_9}$ with $M_{\chi}$
for $M_{\rho_2} = 506$ GeV, $M_{\rho_1}= 26.5$ GeV,
$g_{Z_{\mu\tau}}=0.93\times 10^{-3}$, $M_{Z_{\mu\tau}}=0.076$ GeV and
$\theta_D = 0.095$ rad. \label{Fig:dc9-vs-mchi}]{
\includegraphics[height=6cm,width=7.8cm,angle=0]{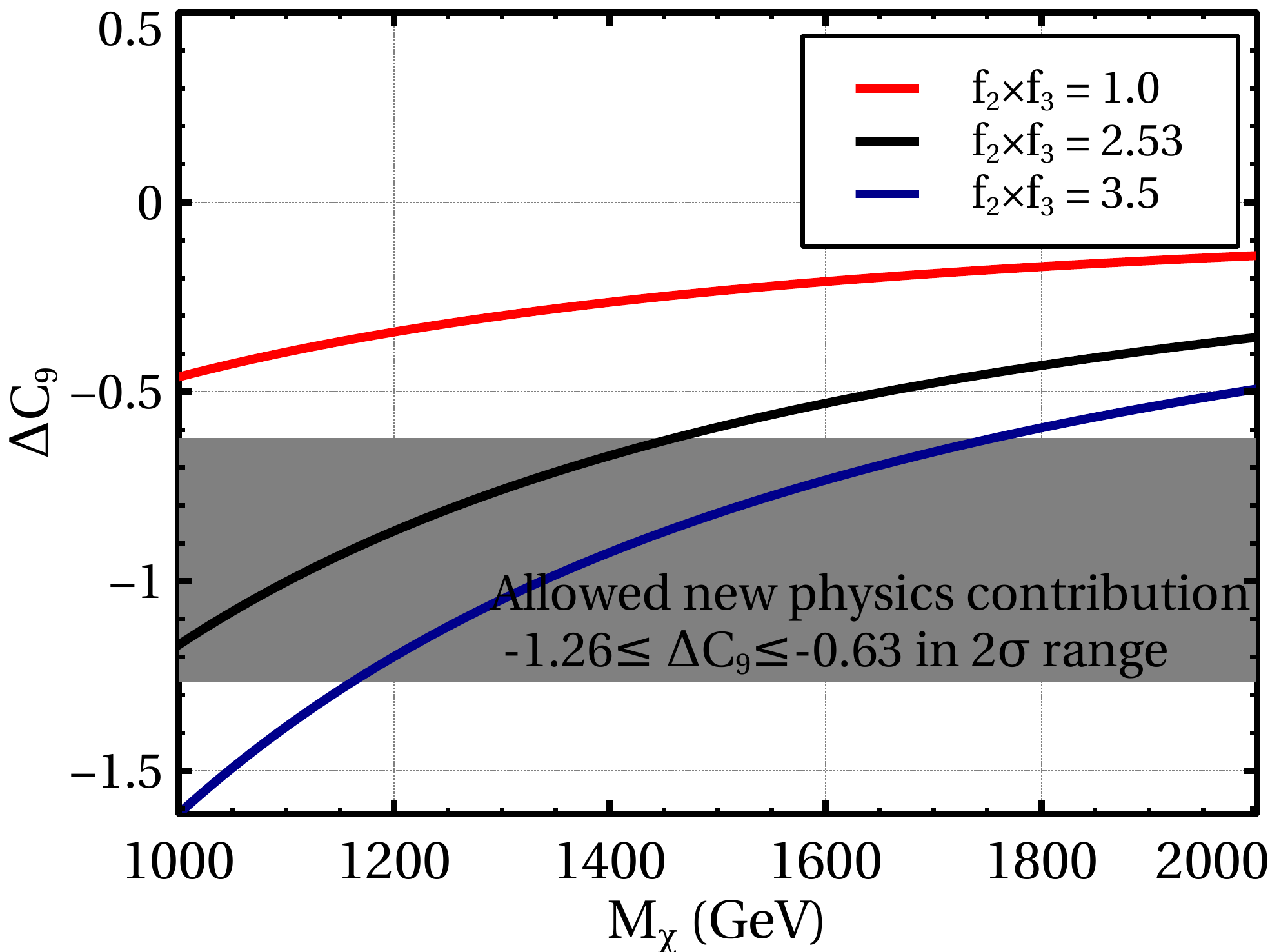}
}
\vskip 0.2in
\subfloat[Variation of $\Delta{C_9}$ with $M_{Z_{\mu\tau}}$
for $M_{\rho_1}=26.5$ GeV, $M_{\rho_2} = 506$ GeV, $M_{\chi}= 1300$ GeV,
$f_2\times f_3=2.53$ and $\theta_D = 0.095$ rad.\label{Fig:dc9-vs-mZp}]{
\includegraphics[height=6cm,width=7.8cm,angle=0]{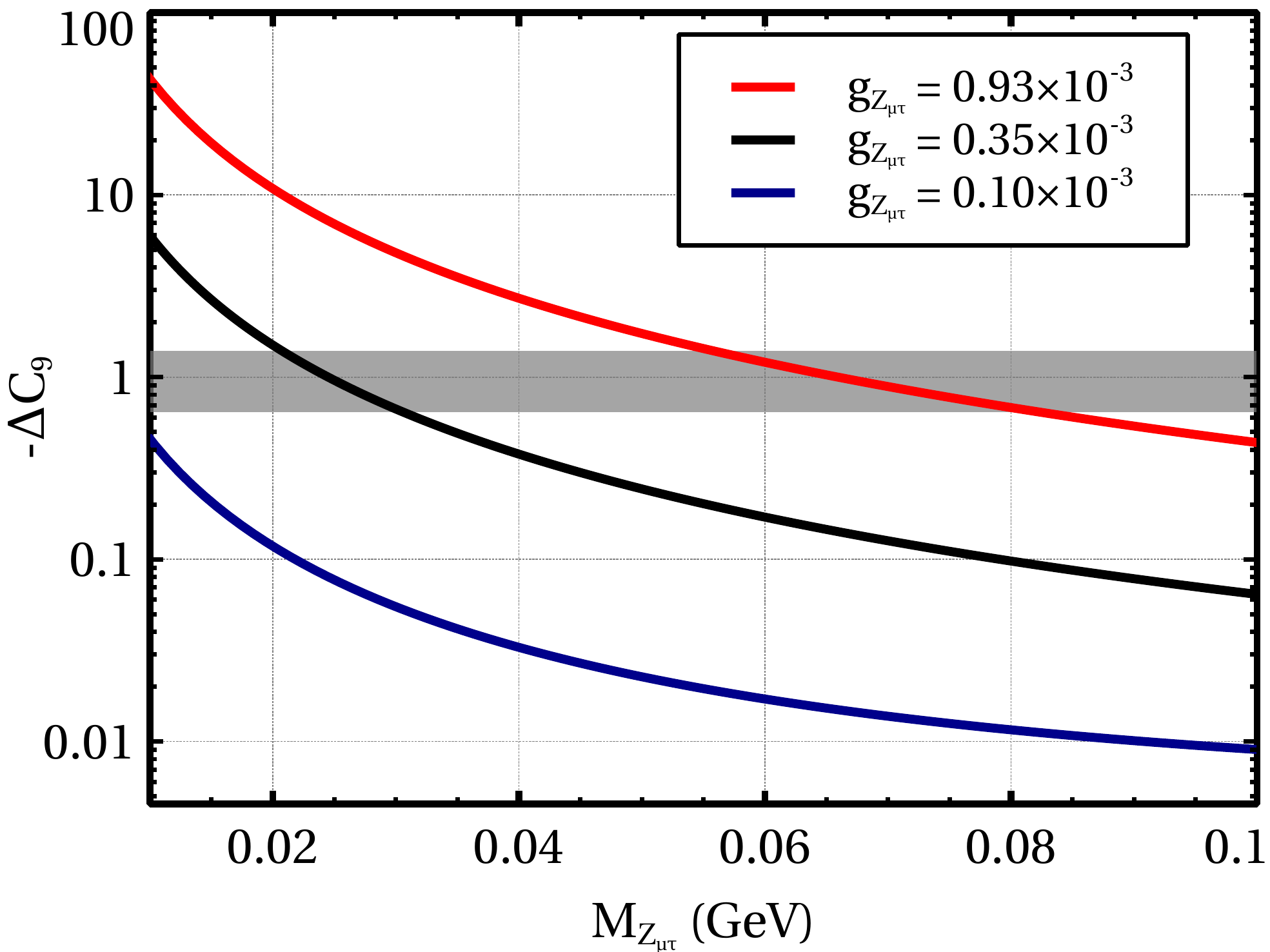}
}
\hskip 0.2in
\subfloat[Variation of $\Delta{C_9}$ with $\theta_D$ for
$M_{\rho_2} = 506$ GeV, $M_{\chi}= 1300$ GeV, $M_{Z_{\mu\tau}}=0.076$ GeV,
$g_{Z_{\mu\tau}}=0.93\times 10^{-3}$ and $f_2\times f_3=0.8$. \label{Fig:dc9-vs-thD}]{
\includegraphics[height=6cm,width=7.8cm,angle=0]{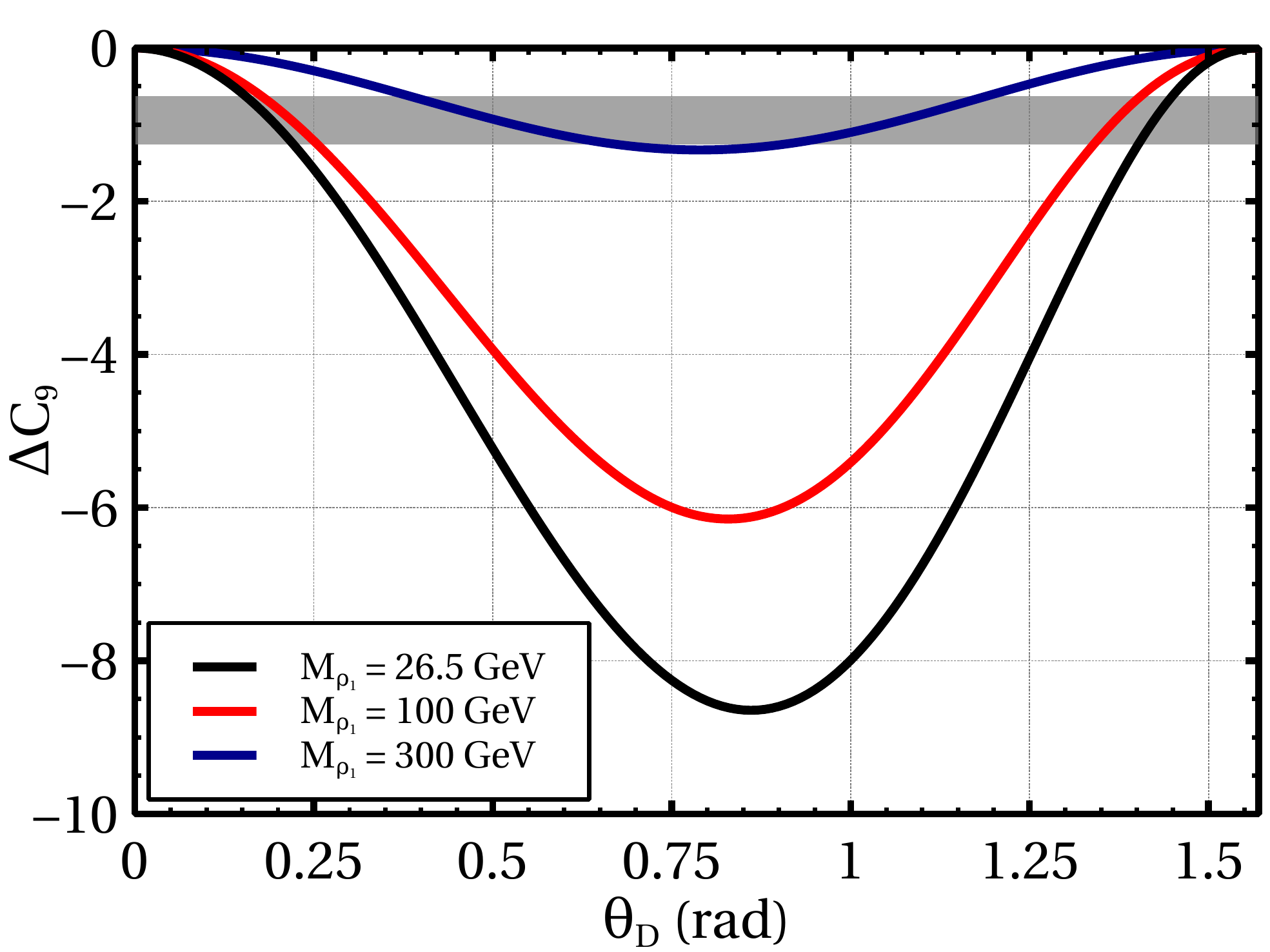}
}
\caption{Variation of $\Delta{C_9}$ with respect to different parameters.}
\label{Fig:dc9lineplot}
\end{figure}

In order to understand the dependence of $\Delta{C_9}$ on the model parameters 
we have shown the variation of $\Delta{C_9}$ in Fig.\,\ref{Fig:dc9lineplot}.  In this figure there are four panels 
which represent the variation of $\Delta{C_9}$ with respect to four important 
parameters namely $M_{\rho_1}$, $M_{\chi}$, $M_{Z_{\mu\tau}}$ and $\theta_D$. 
In Fig.\,\ref{Fig:dc9-vs-mrho1},
we have shown the variation of $\Delta{C_9}$ with mass of $\rho_1$ for three
different values of the product of Yukawa couplings $f_2$ and $f_3$. Here,
one can see that the magnitude of $\Delta{C_9}$ increases with decreasing values of 
mass of $\rho_1$ which enters into loop diagrams (see Feynman diagrams
shown in Fig.\,\ref{newdia}). Consequently, the loop functions are enhanced which in turn increase the magnitude of $\Delta{C_9}$.
Moreover, as the NP contributions to
the WC $C_9$ (Eq.\,(\ref{dc9})) is proportional to
Yukawa couplings $f_2$ and $f_3$, the magnitude of $\Delta{C_9}$
enhances with $f_2\times f_3$. This feature is also clearly demonstrated
in Fig.\,\ref{Fig:dc9-vs-mrho1}. Similar to this plot, in Fig.\,\ref{Fig:dc9-vs-mchi},
we have illustrated the effect of $M_{\chi}$ on $\Delta{C_9}$ for the 
same three different values of $f_2\times f_3$. Here also we have
found similar behaviour of $\Delta{C_9}$ with respect to
$M_{\chi}$ as we have observed for $M_{\rho_1}$. Further, we have also
displayed the effect of non-standard gauge boson mass $M_{Z_{\mu\tau}}$ on 
$\Delta{C_9}$ in Fig.\,\ref{Fig:dc9-vs-mZp} for three different values of gauge
coupling $g_{Z_{\mu\tau}}=0.93\times 10^{-3}$, $0.35\times 10^{-3}$
and $0.1\times 10^{-3}$ respectively. In this case, the magnitude of 
$\Delta{C_9}$ decreases caused by the propagator suppression for larger values of $M_{Z_{\mu\tau}}$.
It is clearly seen from Eq.\,(\ref{dc9}), where $\Delta{C_9}$ is inversely
proportional to $M^2_{Z_{\mu\tau}}$. On the other hand, in this plot $\Delta{C_9}$
increases significantly with the gauge coupling $g_{Z_{\mu\tau}}$ for
the considered mass range of $M_{Z_{\mu\tau}}$ ($0.01\leq
M_{Z_{\mu\tau}}\,({\rm GeV})\leq 0.1$). Finally, in Fig.\,\ref{Fig:dc9-vs-thD}
we have demonstrated the variation of $\Delta{C_9}$ with respect to the dark sector
mixing angle $\theta_D$ for three different choices of $M_{\rho_1}$. In this
plot, we have varied $\theta_D$ in range 0 to $\pi/2$. The oscillatory behaviour
of $\Delta{C_9}$ with respect to $\theta_D$ is due the combined effects of two
factors. One is the direct involvement of sine and cosine functions
within the expressions of $\Delta{C_9}$. Another one is the indirect effect
due to the change of $M_{\rho_3}$ with $\theta_D$, where the former undergoes
a full oscillation between $M_{\rho_2}$ to $M_{\rho_1}$ via Eq.\,(\ref{mho3-relation}) when
$\theta_D$ changes from $0$ to $\pi$. The morphology
of $\Delta{C_9}$ with respect to $\theta_D$ fits pretty well
with a function like $-A\sin^2\,(2\theta_D)$, where the exact
value of the normalisation constant $A$ depends on the values
of other parameters namely,  $M_{\rho_1}$, $M_{\rho_2}$, $g_{Z_{\mu\tau}}$,
$M_{Z_{\mu\tau}}$ and $M_{\chi}$. Moreover, the oscillatory behaviour
of $\Delta{C_9}$ vanishes if we set $M_{\rho_1}=M_{\rho_2}$. Under this condition,
the dependence of $\theta_D$ disappears from the expression of $M_{\rho_3}$
and consequently $\Delta{C_9}$ becomes independent of $\theta_D$. Furthermore,
in all the four plots of Fig.\,\ref{Fig:dc9lineplot},
the grey coloured band represents $2\sigma$ range allowed range of fit
value of $\Delta{C_9}$ for explaining $R_{K^{(*)}}$ anomalies \cite{Aebischer:2019mlg}. 

\subsection{\boldmath$B\to X_s\gamma$}\label{bsg}
The measurement of inclusive radiative $B$ decay process like $B\rightarrow X_s\gamma$ has also been shown deviation from the corresponding SM prediction. The world average experimental value of the branching ratio of this process is \cite{Amhis:2016xyh}
\begin{equation}\label{br_exp_bsg}
{\rm Br}^{\rm Exp}(B\rightarrow X_s\gamma)=(3.32\pm 0.16)\times10^{-4},
\end{equation}
for photon energy $E_{\gamma} >1.6$ GeV in the $B$-meson rest frame. Under the same conditions the corresponding SM prediction with higher order corrections is \cite{Misiak:2015xwa}
\begin{equation}\label{br_sm_bsg}
{\rm Br}^{\rm SM}(B\rightarrow X_s\gamma)=(3.36\pm 0.23)\times10^{-4}.
\end{equation}
It is quite evident that the theoretical prediction is in good agreement with the experimental value. Hence this small difference can tightly constrain any NP which contributes to this process. Keeping this in mind we have evaluated the NP contributions to this decay process in the present scenario. Consequently, we use the branching ratio of this process as one of the constraints in our analysis.

\begin{figure}[h!]
\begin{center}
\includegraphics[scale=1,angle=0]{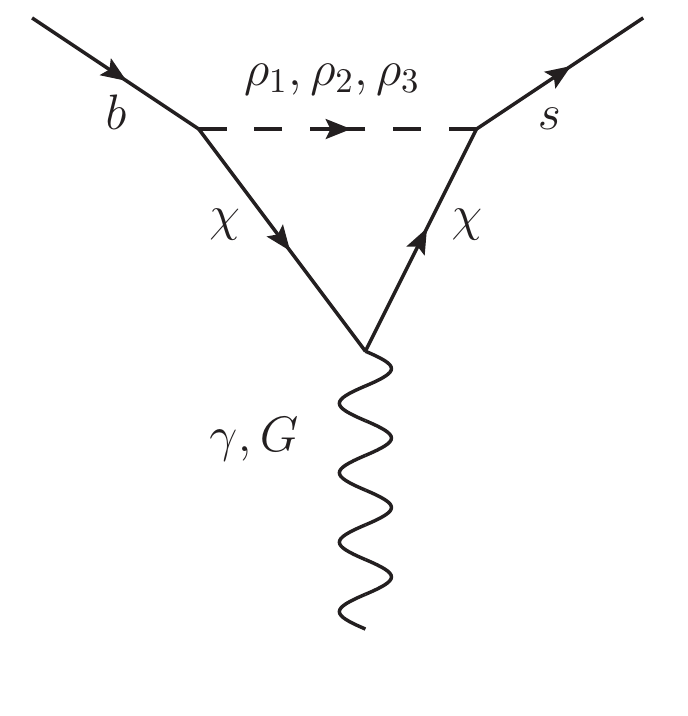}
\caption{The possible electromagnetic and chromomagnetic penguin
diagrams that are contributed to the decay $B\to X_s \gamma$ in
addition to the SM.}
\label{fig2}
\end{center}
\end{figure}

At quark level $B\to X_s\gamma$ decay is indicated by $b\to s\gamma$ transition. The effective Hamiltonian for this transition at the bottom quark mass ($\mu_b=m_b$) scale is given by (see ref.\;\cite{Buchalla:1995vs, Buras:1997fb})
\begin{equation} \label{Heff_at_mu}
{\cal H}_{\rm eff}(b\to s\gamma) = - \frac{G_{\rm F}}{\sqrt{2}} V_{ts}^* V_{tb}
\left[ \sum_{i=1}^6 C_i(\mu_b) \mathcal{O}_i + C_{7\gamma}(\mu_b) \mathcal{O}_{7\gamma}
+C_{8G}(\mu_b) \mathcal{O}_{8G} \right]\,.
\end{equation}
At first the WCs ($C_i$) have been calculated at electroweak scale ($\mu_W$) and using renormalisation group (RG) equations \cite{Buchalla:1995vs, Buras:1997fb, Buras:2003mk} they are evolved down to $\mu_b = m_b$ scale. The local operators $\mathcal{O}_1....\mathcal{O}_6$ represent four quark interactions and the explicit form of these operators can be found in \cite{Buras:1998raa}. The remaining operators $\mathcal{O}_{7\gamma}$ (electromagnetic dipole) and $\mathcal{O}_{8G}$ (chromomagnetic dipole)  which are the most important for this decay and the expressions for these operators at the leading order are given by 
\begin{equation}\label{O6B}
\mathcal{O}_{7\gamma}  =  \frac{e}{8\pi^2} m_b \bar{s}_{\alpha'} \sigma^{\alpha\beta}
          (1+\gamma_5) b^{\alpha'} F_{\alpha\beta},\qquad            
\mathcal{O}_{8G}     =  \frac{g_s}{8\pi^2} m_b \bar{s}^{\alpha'} \sigma^{\alpha\beta}
   (1+\gamma_5)\Lambda^a_{\alpha'\beta'} b^{\beta'} G^a_{\alpha\beta}\;, 
\end{equation}
with $\sigma^{\alpha\beta}=\frac{i}{2}[\gamma^\alpha, \gamma^\beta]$. The expressions of the WCs at $\mu_b$ scale is given by
\begin{eqnarray}
\label{C7eff}
C_{7\gamma}^{(0)eff}(\mu_b) & = & 
\eta^\frac{16}{23} C_{7\gamma}^{(0)}(\mu_W) + \frac{8}{3}
\left(\eta^\frac{14}{23} - \eta^\frac{16}{23}\right) C_{8G}^{(0)}(\mu_W) +
 C_2^{(0)}(\mu_W)\sum_{i=1}^8 h_i \eta^{a_i},
\\
\label{C8eff}
C_{8G}^{(0)eff}(\mu_b) & = & 
\eta^\frac{14}{23} C_{8G}^{(0)}(\mu_W) 
   + C_2^{(0)}(\mu_W) \sum_{i=1}^8 \bar h_i \eta^{a_i},
\end{eqnarray}
with
\begin{equation}
\eta  =  \frac{\alpha_s(\mu_W)}{\alpha_s(\mu_b)},~~~\alpha_s(\mu_b) = \frac{\alpha_s(M_Z)}{1 
- \beta_0 \frac{\alpha_s(M_z)}{2\pi} \, \ln(M_Z/\mu_b)}, \qquad 
\beta_0=\frac{23}{3}~,
\label{eq:asmumz}
\end{equation}
and 
\begin{eqnarray}\label{c2}
C^{(0)}_2(\mu_W) &=& 1,\\                             
C^{(0)}_{7\gamma} (\mu_W) &=& -\frac{1}{2} D'(x_t, x_1, x_2, x_3)=-\frac{1}{2}\{(D'_0(x_t)+ D'(x_1, x_2, x_3)\},\\ 
C^{(0)}_{8G}(\mu_W) &=& -\frac{1}{2} E'(x_t, x_1, x_2, x_3)=-\frac{1}{2}\{(E'_0(x_t)+ E'(x_1, x_2, x_3)\}.
\end{eqnarray}
Apart from these other WCs vanish at the electroweak scale $\mu_W$. The superscript ``0'' indicates the leading logarithmic (LO) approximation. The values of $a_i$, $h_i$ and $\bar h_i$ can be obtained from \cite{Buras:2003mk}. The total (SM+NP) contribution at the LO is represented by the functions $D'(x_t, x_1, x_2, x_3)$ and $E'(x_t, x_1, x_2, x_3)$ while the functions $D'_0(x_t)$ and $E'_0(x_t)$ are designated as the corresponding SM contributions at the electroweak scale \cite{Inami:1980fz}

\begin{equation}\label{dp0}
D'_0(x_t)= -{{(8x_t^3 + 5x_t^2 - 7x_t)}\over{12(1-x_t)^3}}+ 
          {{x_t^2(2-3x_t)}\over{2(1-x_t)^4}}\ln x_t~,
\end{equation}
\begin{equation}\label{ep0}
E'_0(x_t)=-{{(x_t^3-5x_t^2-2x_t)}\over{4(1-x_t)^3}} + {3\over2}
{{x_t^2}\over{(1 - x_t)^4}} \ln x_t~,
\end{equation}
with $x_t\equiv \frac{m^2_t}{M^2_W}$. The functions corresponding to electromagnetic and chromomagnetic dipole operators due to the NP particles (generated form Fig.~\ref{fig2}) are given in the following respectively
\begin{eqnarray}
D'(x_1, x_2, x_3)&=-&\frac{\sqrt{2}}{G_F V^\ast_{tb} V_{ts}}\frac{f_2f_3}{8}\frac 13 \left(\frac{\sin^2\theta_D}{M^2_{\rho_1}}h_b(x_1)+\frac{\cos^2\theta_D}{M^2_{\rho_2}}h_b(x_2)+\frac{1}{M^2_{\rho_3}}h_b(x_3)\right) \;,\\ 
\label{deldp}
E'(x_1, x_2, x_3)&=&\frac{\sqrt{2}}{G_F V^\ast_{tb} V_{ts}}\frac{f_2f_3}{8}\left(\frac{\sin^2\theta_D}{M^2_{\rho_1}}h_b(x_1)+\frac{\cos^2\theta_D}{M^2_{\rho_2}}h_b(x_2)+\frac{1}{M^2_{\rho_3}}h_b(x_3)\right) \;,\\ 
\label{delep}
\end{eqnarray}
while the function $h_b(x)$ is given in the Appendix~\ref{flav_app}.

In SM the branching ratio of $B\rightarrow X_s\gamma$ has been estimated at a very high level of accuracy including higher order QED and QCD corrections.  For example in refs.\;\cite{Chetyrkin:1996vx, Kagan:1998ym} one can find the full next-to leading order (NLO) QCD and QED corrections for this process in two different ways. The present precision level of experimental data requires that one should also include next-to-next-to leading order (NNLO) QCD corrections in this analysis. In this regard the first effort to measure NNLO QCD corrections for this process in SM was described in ref.\;\cite{Misiak:2006zs}. Finally in a recent article \cite{Misiak:2015xwa} one can find an updated and more complete NNLO QCD corrections to this process. Using the \cite{Misiak:2015xwa} one can calculate the branching ratio of $B\rightarrow X_s\gamma$ incorporating NNLO QCD corrections in NP scenario. Therefore, in the current article we also follow the same approach\footnote{This approach has also been used in the context of other BSM scenarios to measure the NP effects for this process: for example for nonminimal universal extra dimensional model 
\cite{Datta:2016flx} and for two higgs doublet model \cite{Arhrib:2017yby}.} (as given in \cite{Misiak:2015xwa}) to measure NP contribution for this process with NNLO QCD corrections
\begin{equation}\label{nnlo}
{\rm Br}^{\rm NNLO}(B\rightarrow X_s\gamma)\times10^{4}=(3.36\pm 0.23) -8.22\Delta C_7-1.99\Delta C_8.
\end{equation}
Here $\Delta C_7$ and $\Delta C_8$ represent for the NP contributions to WCs for electromagnetic and chromomagnetic 
dipole operators. In our convention, $\Delta C_7=-\frac12 D'(x_1,x_2,x_3)$ and $\Delta C_8=-\frac12 E'(x_1,x_2,x_3)$.

\section{\boldmath$(g-2)_\mu$ anomaly}\label{g2}
\begin{figure}[H]
\begin{center}
\subfloat[]{\label{fig3a}\includegraphics[scale=1,angle=0]{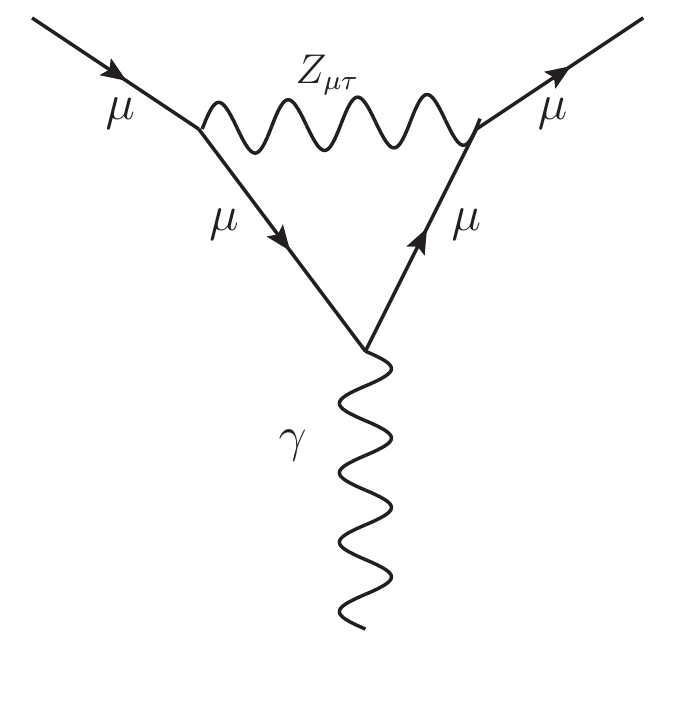}}
\subfloat[]{\label{fig3b}\includegraphics[scale=1,angle=0]{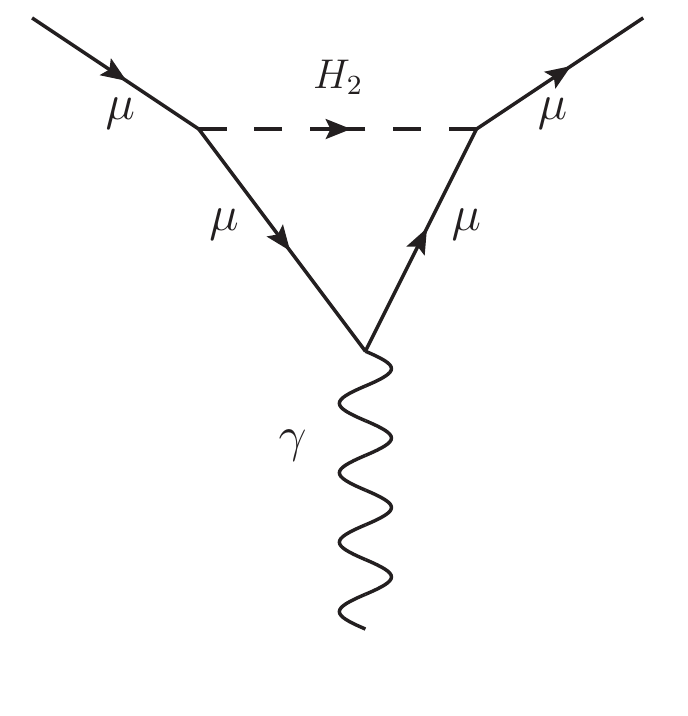}}
\caption{Relevant penguin diagrams that are contributed to the $(g-2)_\mu$ in addition to the SM.}
\label{muong2}
\end{center}
\end{figure}

Using Dirac equation one can define the magnetic moment $\vec{\mathbb{M}}$ of muon in terms of its spin $\vec{\mathbb{S}}$ and gyromagnetic ratio ($g_{\mu}$) in the following way
\begin{eqnarray}
\vec{\mathbb{M}}= g_{\mu} \dfrac{e}{2\,m_\mu} \vec{\mathbb{S}},
\label{mug2}
\end{eqnarray}
which is one of the most accurately measured physical quantities. 
Ideally the value of $g_{\mu}$ is equal to ``2''. 
In SM one can easily calculate the one loop correction to this quantity 
and that gives marginal shift from ``2''. Hence, to measure the deviation of
$g_{\mu}$ from its tree level value one can
define a quantity namely
\begin{eqnarray}
a_{\mu} = \dfrac{g_{\mu}-2}{2}\,.
\end{eqnarray}
This quantity has been precisely measured by the CERN experiments
and later on by the E821 experiment. The current average experimental
value is \cite{Tanabashi:2018oca}
\begin{eqnarray}
a_{\mu}^{\rm exp} = 116592091.0\pm 54\pm 33 \times 10^{-11}\,.
\label{mug2exp}
\end{eqnarray}
On the other hand total theoretical prediction of 
this quantity considering all kinds of source of contributions
in SM is \cite{Tanabashi:2018oca}
\begin{eqnarray}
a_{\mu}^{\rm th} = 116591823.1\pm 34\pm 26 \times 10^{-11}\,.
\label{mug2th}
\end{eqnarray}
It is quite evident from the above Eqs.~\ref{mug2exp} and \ref{mug2th} that both the experimentally measured 
and theoretically predicted values of $a_{\mu}$
are close to each other, however there still exists
some disagreement between these two quantities at the $3.5\sigma$
significance which is \cite{Tanabashi:2018oca},
\begin{eqnarray}
\Delta a_{\mu} = a_{\mu}^{\rm exp} - a_{\mu}^{\rm SM}
=268\pm 63\pm43 \times 10^{-11}\,.
\label{mug2delta}
\end{eqnarray}
Therefore, this anomaly with respect to the SM expectation
requires the interference of BSM theories where one obtains
extra contributions from some NP particles. In the present model\footnote{See
Ref.\,\,\cite{Lindner:2016bgg} for a review on $(g-2)_{\mu}$
in various BSM extensions.}, apart from the SM contribution,
we have two additional one loop diagrams (see Fig.~\ref{muong2}) in which the extra neutral gauge boson $\zmt$
and extra CP-even scalar $H_2$ are involved.
The additional contribution from Fig.~\ref{fig3a} is given by \cite{Gninenko:2001hx, Baek:2001kca},
\begin{equation}\label{g2_Z}
\delta a_\mu^{Z_{\mu\tau}} = \frac{1}{8\pi^2}~
\left(a^2_{Z_{\mu\tau}}F^a_{Z_{\mu\tau}}(R_{Z_{\mu\tau}})-b^2_{Z_{\mu\tau}}F^b_{Z_{\mu\tau}}(R_{Z_{\mu\tau}})\right)\; \\
\end{equation}
with $R_{Z_{\mu\tau}}\equiv M^2_{Z_{\mu\tau}}/m^2_{\mu}$ and
\begin{eqnarray}
a_{Z_{\mu\tau}}&=&\frac{g_2}{4\cos\theta_W}(1-4\sin^2\theta_W)\sin\theta_{\mu\tau}-\bigg(g_{Z_{\mu\tau}}-\frac 34 \frac{g_2\sin\theta_W\epsilon}{\cos\theta_W}\bigg)\cos\theta_{\mu\tau}\;,\\ 
\label{azz} 
b_{Z_{\mu\tau}}&=-&\frac{g_2}{4\cos\theta_W}\bigg(\sin\theta_{\mu\tau}-\sin\theta_W\epsilon\cos\theta_{\mu\tau}\bigg)\;,
\label{bzz} 
\end{eqnarray}
\begin{eqnarray}
F^a_{Z_{\mu\tau}}(R_{Z_{\mu\tau}}) &=& \int_0^1 dx\, \frac{2x(1-x)^2}{(1
 -x)^2+R_{Z_{\mu\tau}} x}
 \;, \\
F^b_{Z_{\mu\tau}}(R_{Z_{\mu\tau}}) &=& \int_0^1 dx\,\frac{2(1-x)(3+x)}{(1
 -x)^2+R_{Z_{\mu\tau}} x}\;.
\end{eqnarray}
Furthermore, the contribution from the extra CP-even scalar $H_2$ is given by \cite{Krawczyk:1996sm, Dedes:2001nx}
\begin{eqnarray}
\delta a_\mu^{H_2} = \frac{G_Fm_\mu^2}{4\pi^2 \sqrt{2}}~
\sin^2{\theta_s} ~R_{H_2}~ F_{H_2}(R_{H_2}) \;, \\
\label{g2_H2}
\end{eqnarray}
with $R_{H_2}\equiv m_\mu^2/M_{H_2}^2$~and
\begin{eqnarray}\label{g2_H2_int}
F_{H_2}(R_{H_2}) = \int_0^1 dx\, \frac{x^2(2-x)}{R_{H_2} x^2-x+1}.
\end{eqnarray}
However, we have checked that the contribution of CP-even scalar $H_2$ is insignificant with respect to 
$\zmt$ in the allowed parameter space.
\begin{figure}[h!]
\centering
\includegraphics[height=8cm,width=10cm,angle=0]{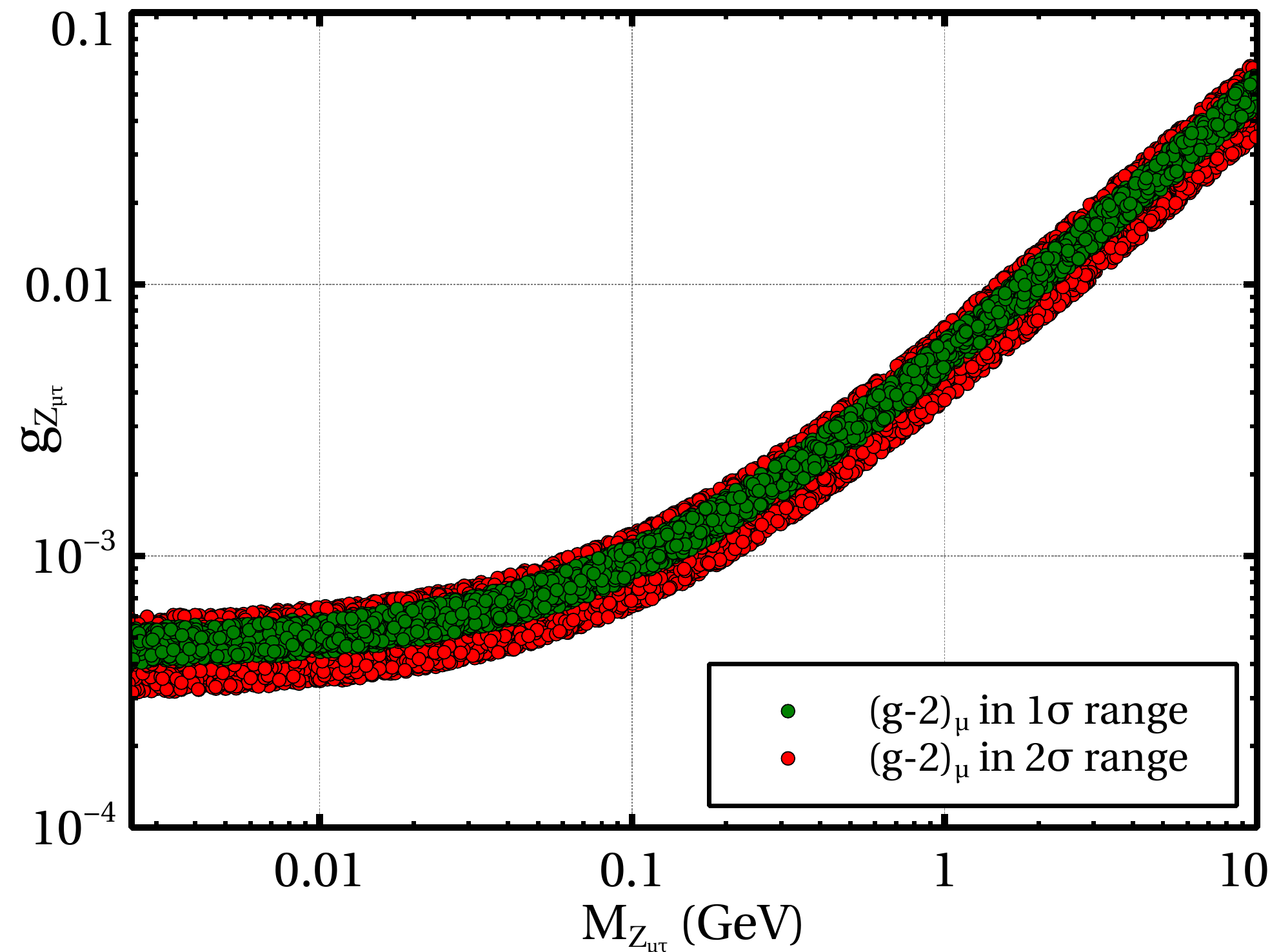}
\caption{Allowed region in $g_{Z_{\mu\tau}}-M_{Z_{\mu\tau}}$
plane which explains the deviation between theoretical (SM)
prediction and experimental result in  $1\sigma$ (green coloured points)
and $2\sigma$ (red coloured points) ranges respectively.}
\label{Fig:muong-2}
\end{figure}

In Fig.\,\ref{Fig:muong-2}, we have shown the allowed region
of $M_{Z_{\mu\tau}}$ and $g_{Z_{\mu\tau}}$ in $g_{Z_{\mu\tau}}-M_{Z_{\mu\tau}}$
plane by red coloured points, which can explain the discrepancy between
theoretical prediction (SM) and experimentally measurable value of
the anomalous magnetic moment of muon in $2\sigma$ range. The corresponding
$1\sigma$ allowed region is also indicated by green coloured points. We will
come back to this parameter space ($g_{Z_{\mu\tau}}-M_{Z_{\mu\tau}}$ plane)
with a detailed analysis, which includes constraints like dark matter relic density,
direct detection, observables related to rare $B$-meson decays ($R_{K^{(*)}}$, Br($B\rightarrow X_s\gamma$))
and also bounds from ongoing and future experiments like CCFR, LHC, DUNE, Borexino
etc. in the next section (see Fig.\,\ref{Fig:mzp-gzp} and related discussions). 
\section{Dark Matter}
\label{dm}
We are in a stage, where we can discuss dark matter phenomenology. 
The scalar sector of the present scenario contains 
two $\mathbb{Z}_2$-odd scalar representations, one of them
is an SU$(2)_{L}$ doublet $\Phi$ having a nonzero $L_{\mu}-L_{\tau}$
charge while the rest is a gauge singlet scalar $S$. As we have seen
earlier in the Section \ref{model}, the term proportional to $\lambda_8$ 
in the scalar potential (Eq.\,\,\ref{Vpot}) enforces a mixing between
the CP-even component $\phi^0$ of the doublet $\Phi$ and the singlet
$S$. Therefore, in the odd sector we have three physical neutral scalars
namely, $\rho_1$, $\rho_2$ and $\rho_3$, out of which $\rho_1$ and
$\rho_2$ are two mutually orthogonal linear combinations of $S$
and $\phi^0$ while $\rho_3$ coincides with the CP-odd component
$a^0$ as the latter does not have any mixing with others. Being
$\mathbb{Z}_2$-odd, the lightest one among the neutral scalars
$\rho_1$, $\rho_2$ and $\rho_3$ is automatically stable and
can be an excellent dark matter candidate of the Universe.
In this work, we consider $\rho_1$ as the potential dark matter candidate
and depending upon the dark sector mixing angle $\theta_D$,
$\rho_1$ will be either ``singlet-like" or ``doublet-like" or
a mixed state. Later in this Section, we will show that although the
combined effects of both dark matter relic density bound and
flavour physics anomalies (including $(g-2)_\mu$) considering
in this work dictates that the dark matter candidate $\rho_1$ to be mostly
a ``single-like" state, its freeze-out process involves extra annihilation
channels involving $L_{\mu}-L_{\tau}$ gauge boson $Z_{\mu\tau}$, making this
scenario significantly different from the case of standard Scalar Singlet
dark matter \cite{McDonald:1993ex, Burgess:2000yq, Biswas:2011td, Cline:2013gha}. 

The viability of the proposed dark matter candidate $\rho_1$ has been
investigated first by computing its relic density\footnote{Here, DM represents the short form of dark matter.}
$\Omega_{\rm DM} h^2$. This requires comoving number density
$Y$ at the present epoch ($T=T_0$, $T_0$ is the present temperature of the Universe),
which is a solution of the Boltzmann equation involving
all relevant annihilation and co-annihilation processes
in the collision term. The Boltzmann equation in terms of
$Y$ is given by \cite{Gondolo:1990dk, Griest:1990kh, Edsjo:1997bg},
\begin{eqnarray}
\dfrac{dY}{dx} = -\left(\dfrac{45\,G}{\pi}\right)^{-\frac{1}{2}}
\dfrac{M_{\rho_1}\,\sqrt{g_{\star}}}{x^2}
\langle{\sigma {\rm v}}\rangle_{\rm eff}\,
(Y^2-(Y^{\rm eq})^2)\,,
\label{eq:BEapprox}
\end{eqnarray}
where $Y=\sum_i Y_i$ with
$Y_i=\dfrac{n_i}{\rm s}$ being the comoving number density of $\mathbb{Z}_2$-odd particle
$i$ having number density $n_i$ and ${\rm s}$ stands for the entropy density
of the Universe. Moreover, $x=\dfrac{M_{\rho_1}}{T}$ is a dimensionless
variable and $G$ is the Newton's gravitational constant. The function
$g_{\star}$ \cite{Gondolo:1990dk} depends on degrees of freedom for entropy and energy densities
of the Universe. The quantity $\langle{\sigma {\rm v}}\rangle_{\rm eff}$ has
been defined as \cite{Griest:1990kh}
\begin{eqnarray}
\langle{\sigma {\rm v}}\rangle_{\rm eff} =
\sum_{i\,j}\langle{\sigma_{i\,j} {\rm v}_{i\,j}}\rangle
\times r_i\,r_j\,,
\label{eq:sigmaveff}
\end{eqnarray}
where, $\langle{\sigma_{i\,j} {\rm v}_{i\,j}}\rangle$
is the thermal averaged annihilation cross section between
particle $i$ and $j$ having relative velocity ${\rm v}_{i\,j}$.
$\langle{\sigma_{i\,j} {\rm v}_{i\,j}}\rangle$ has the following
expression in terms of cross section $\sigma_{i\,j}$,
\begin{eqnarray}
\langle{\sigma_{i\,j} {\rm v}_{i\,j}}\rangle &=&
\frac{1}{2\,M^2_{i}\,M^2_{j}\,T\,{\rm K}_2\left(\dfrac{M_i}{T}\right)\,
{\rm K}_2\left(\dfrac{M_j}{T}\right)} \times \int^{\infty}_{(M_i+M_j)^2}
\sigma_{ij}\,\,p^2_{ij}\,\sqrt{s}\,{\rm K}_1
\left(\frac{\sqrt{s}}{T}\right)\,ds\,, \nonumber \\
p_{ij} &=& \dfrac{\sqrt{s - (M_i+M_j)^2}
\sqrt{s-(M_i-M_j)^2}}{2\,\sqrt{s}}\,,
\label{eq:sigmavij}
\end{eqnarray}
with
\begin{eqnarray}
r_i = \dfrac{Y^{\rm eq}_i}{Y} = \dfrac{n^{\rm eq}_i}{n}=
\dfrac{g_i\left(1+\Delta_i\right)^{3/2}\exp[-\Delta_i\,x]}
{\sum_{i} g_i\left(1+\Delta_i\right)^{3/2}\exp[-\Delta_i\,x]}\,,
\label{eq:ri}
\end{eqnarray}
where, ${\rm K}_i$ is the $i^{\rm th}$ order Modified Bessel function of
second kind and $s$ is the Mandelstam variable. Further, $Y_i^{\rm eq}$ and $n^{\rm eq}_i$ 
are the equilibrium values of $Y_i$ and $n_i$ respectively while $n=\sum_i n_i$ is the
total number density of all the odd sector particles. This
is the most relevant quantity instead of individual $n_i$s,
since all heavier particles, which survive annihilation, will
eventually decay into the LOP ($\rho_1$). This is the actual reason of
expressing the Boltzmann equation in terms of total comoving number
density $Y$ instead of individual $Y_i$s. In the above,
$\Delta_i=\dfrac{M_{i}-M_{\rho_1}}{M_{\rho_1}}$, represents
the mass splitting between LOP and other heavier $\mathbb{Z}_2$-odd
particles. After implementation of the present model in {\tt FeynRules}~\cite{Alloul:2013bka} we have solved Boltzmann equation at $T=T_0$
using \texttt{micrOMEGAs} \cite{Belanger:2013oya}. Finally, we have obtained
$Y(T_0)$ which is related to the relic density
of LOP through the following relation \cite{Edsjo:1997bg}
\begin{eqnarray}
\Omega_{\rm DM} h^2 = 2.755\times 10^8\,\left(\dfrac{M_{\rho_1}}
{\rm GeV}\right)\,Y(T_0)\,.
\label{eq:omega}
\end{eqnarray}
Relic density $\Omega_{\rm DM} h^2$ of dark matter
has been measured precisely by satellite borne experiments
like Planck and WMAP and its present acceptable range
is $0.1172 \leq \Omega_{\rm DM} h^2 \leq 0.1226$
at 67\% confidence level (C.L.) \cite{Ade:2015xua}. 
\begin{figure}[h!]
\includegraphics[scale=0.85]{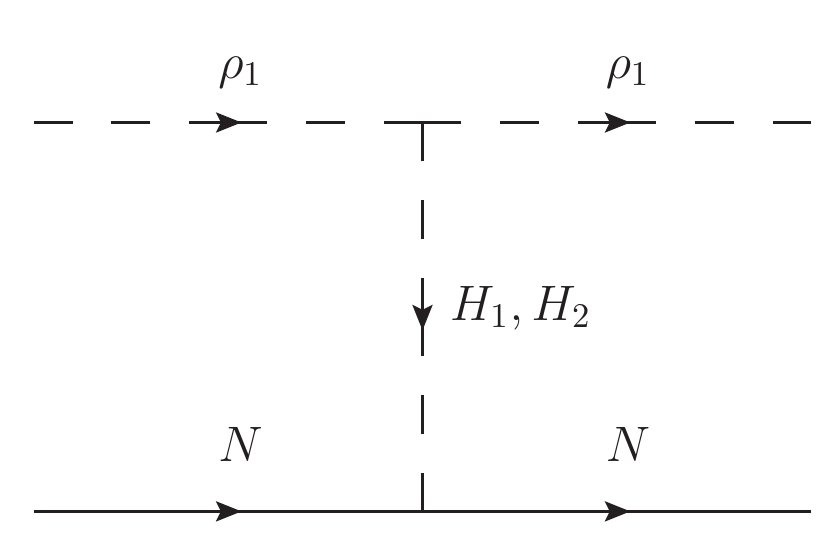}
\caption{Feynman diagram for the elastic scattering of
$\rho_1$ with nucleon $N$ through the exchange of scalar bosons
$H_1$ and $H_2$.}
\label{Fig:DD}
\end{figure}
Apart from this, one has to take into account the latest bound
on dark matter nucleon scattering cross section
from the ``ton-scale" direct detection experiment 
namely XENON1T \cite{Aprile:2018dbl}, which till now provides
the most stringent upper bound on dark matter nucleon
spin independent scattering cross section ($\sigma_{\rm SI}$)
for dark matter mass ranging from 6 GeV to 1 TeV. Since the dark matter
candidate of the present scenario is a scalar, it has only spin independent
scattering with nucleon and such scattering is possible only
though scalar bosons $H_1$ and $H_2$. Feynman diagrams
of such elastic scattering $\rho_1 + N \rightarrow \rho_1 +N$
are shown in Fig. \ref{Fig:DD}. The corresponding expression of $\sigma_{\rm SI}$
is given by
\begin{eqnarray}
\sigma_{\rm SI} = \dfrac{\mu^2_{\rm red}}{4\pi}\left[\dfrac{M_N\,f_N}
{M_{\rho_1}\,v_1}\left(\dfrac{\,g_{H_1\rho_1\rho_1}}{M^2_{H_1}}
+\dfrac{\,g_{H_2\rho_1\rho_1}}{M^2_{H_2}}\right)\right]^2\,,
\label{eq:sigmasi}
\end{eqnarray}
where $g_{H_1(H_2)\rho_1\rho_1}$ is the coupling between $H_1$($H_2$)
and a pair of $\rho_1$. Expressions of these couplings are listed
in Appendix \ref{Dmcouplings}. Moreover, $f_N$ and $M_N$ are nuclear form
factor and nucleon mass respectively. For dark matter
scattering mediated by scalars $f_N\sim 0.3$ \cite{Cline:2013gha}.
We already know that non-observation of any dark matter
signal at direct detection experiments impose severe upper
bound on $\sigma_{\rm SI}$ with respect to dark matter mass.
From, the above expression of $\sigma_{\rm SI}$, it can be
seen clearly that such exclusion limit on $\sigma_{\rm SI}$
in turn puts an upper bound on the involved couplings like
$g_{H_1\rho_1\rho_1}$ and $g_{H_2\rho_1\rho_1}$. 

Moreover, SM Higgs to $\rho_1\rho_1$ coupling for $M_{H_1}> 2\,M_{\rho_1}$
case is also constrained from the maximum allowed limit of
Higgs invisible decay width . At present, the upper limit on
invisible branching fraction of the SM Higgs boson is 0.24 
at 95\% C.L. \cite{Khachatryan:2016whc}.
In the present model, the SM like Higgs boson in addition
to its ``standard decay modes", can also decay into
$Z_{\mu\tau}Z_{\mu\tau}$, $ZZ_{\mu\tau}$, $\rho_1\rho_1$ and
$\rho_1\rho_2$ final states \footnote{In this work, we are
focusing on low mass $Z_{\mu\tau}$ ($\sim 1$ MeV$-$100 MeV)
to address $(g-2)_\mu$ anomaly.}. Decay widths of such processes
are given below,
\begin{eqnarray}
\Gamma_{H_1\rightarrow Z_{\mu\tau}Z_{\mu\tau}} &=& 
\dfrac{g_{H1Z_{\mu\tau}Z_{\mu\tau}}^2\,M^3_{H_1}}{128\,\pi M^4_{Z_{\mu\tau}}}
\left(12 \frac{M^4_{Z_{\mu\tau}}}{M^4_{H_1}} -
4 \frac{M^2_{Z_{\mu\tau}}}{M^2_{H_1}} + 1\right)
\sqrt{1-4\frac{M^2_{Z_{\mu\tau}}}{M^2_{H_1}}}\,,
\end{eqnarray}
\begin{eqnarray}
\Gamma_{H_1\rightarrow ZZ_{\mu\tau}} &=& 
\dfrac{g^2_{H_1ZZ_{\mu\tau}}}{64\,\pi\,M_{H_1}}
\left(8+\dfrac{\left(M^2_{H_1}-M^2_{Z_{\mu\tau}}-M^2_{Z}\right)^2}
{M^2_{Z}M^2_{Z_{\mu\tau}}}\right)
\sqrt{1-\left(\dfrac{M_Z+M_{Z_{\mu\tau}}}{M_{H_1}}\right)^2} \times \nonumber \\
&&\sqrt{1-\left(\dfrac{M_Z-M_{Z_{\mu\tau}}}{M_{H_1}}\right)^2}\,\,,\\
\Gamma_{H_1\rightarrow\rho_1\rho_1} &=& \dfrac{g^2_{H_1\rho_1\rho_1}}{32\,\pi\,M_{H_1}}\,
\sqrt{1-4\frac{M^2_{\rho_1}}{M^2_{H_1}}}\,\,,\\
\Gamma_{H_1\rightarrow\rho_1\rho_2} &=&
\dfrac{g^2_{H_1\rho1\rho2}}{16\,\pi\,M_{H_1}}
\sqrt{1-\left(\dfrac{M_{\rho_1}+M_{\rho_2}}{M_{H_1}}\right)^2}
\sqrt{1-\left(\dfrac{M_{\rho_2}-M_{\rho_1}}{M_{H_1}}\right)^2}\,\,,
\end{eqnarray}
and
\begin{eqnarray}
\Gamma_{H_1}^{\rm Inv} = \Gamma_{H_1\rightarrow Z_{\mu\tau}Z_{\mu\tau}}
+ \Gamma_{H_1\rightarrow ZZ_{\mu\tau}} +
\Gamma_{H_1\rightarrow \rho_1\rho_2} + 
\Gamma_{H_1\rightarrow \rho_1\rho_1}\,. 
\end{eqnarray}
Expressions of all the coupling involved in the above decay widths are
given in Appendix \ref{Dmcouplings}. According to the latest results from LHC,
$\Gamma_{H_1}^{\rm Inv} \leq 0.24\,\,\Gamma^{\rm SM}_{\rm Higgs}$,
where $\Gamma^{\rm SM}_{\rm Higgs} = 4.13$ MeV, is total
decay width of the SM Higgs boson \cite{Denner:2011mq}. 

\begin{figure}[h!]
\hskip -0.2in
\includegraphics[height=2.7cm,width=3.0cm,angle=0]{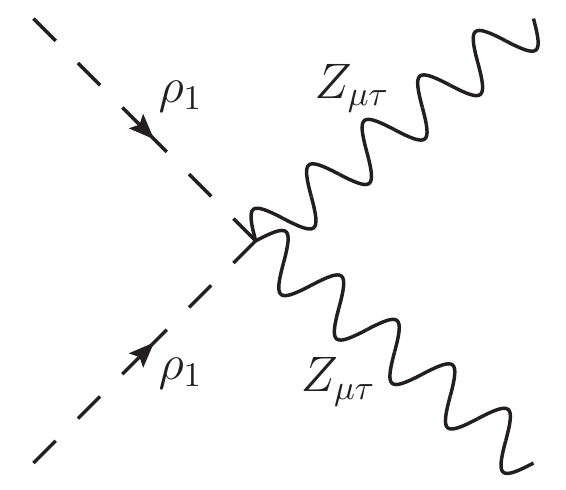}
\hskip 0.1in
\includegraphics[height=2.7cm,width=4.3cm,angle=0]{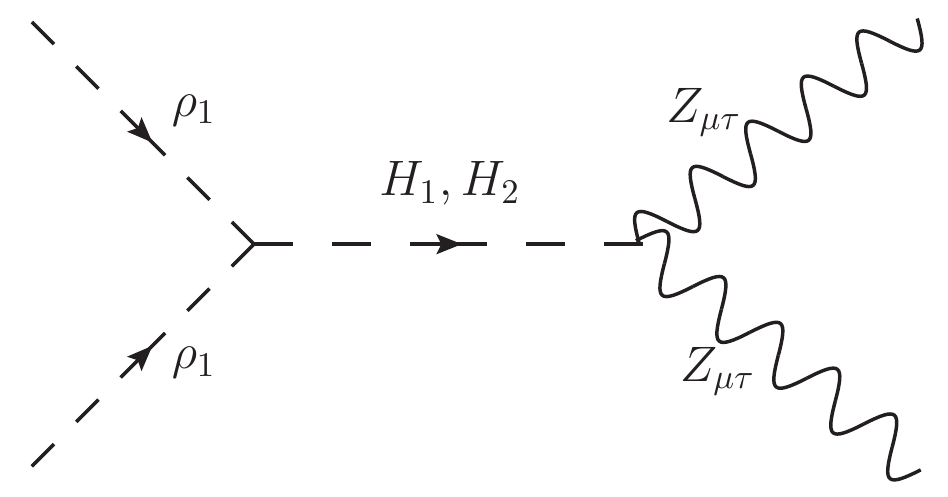}
\hskip 0.1in
\includegraphics[height=2.7cm,width=4.3cm,angle=0]{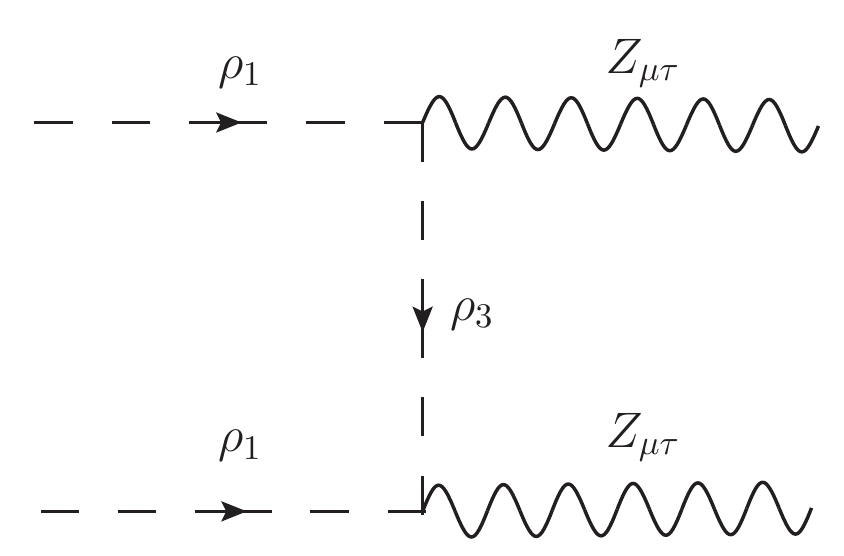}
\hskip 0.1in
\includegraphics[height=2.7cm,width=4.3cm,angle=0]{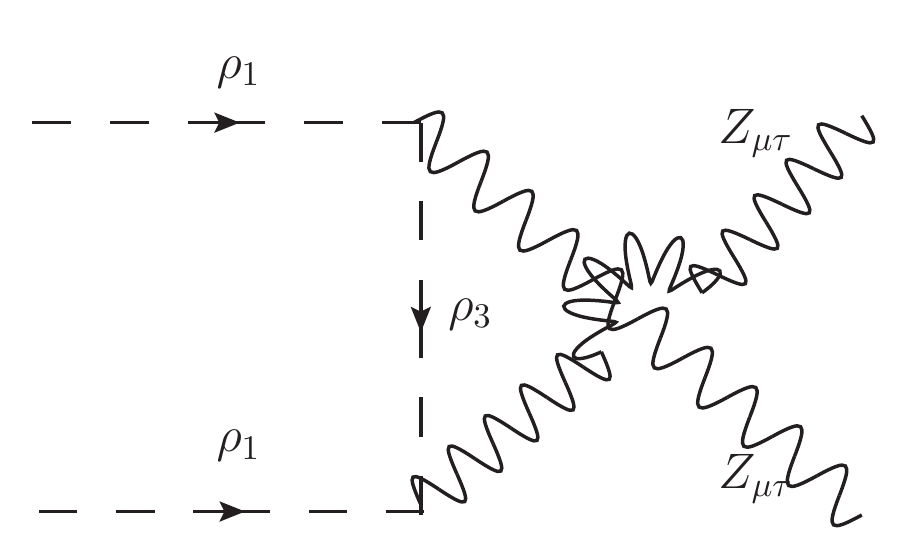}\\
\vskip 0.05in
\hskip -0.2in
\includegraphics[height=2.7cm,width=3.0cm,angle=0]{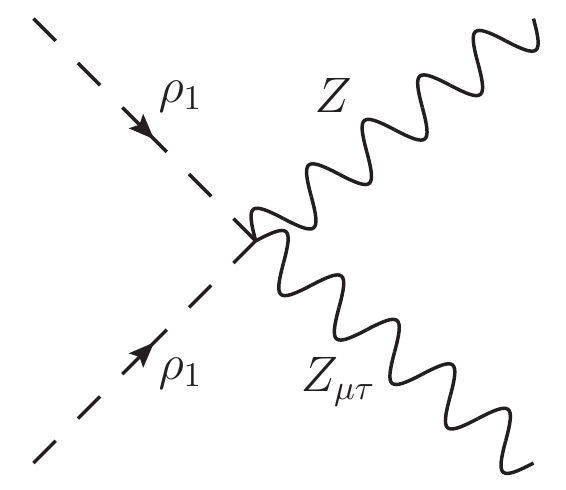}
\hskip 0.1in
\includegraphics[height=2.7cm,width=4.3cm,angle=0]{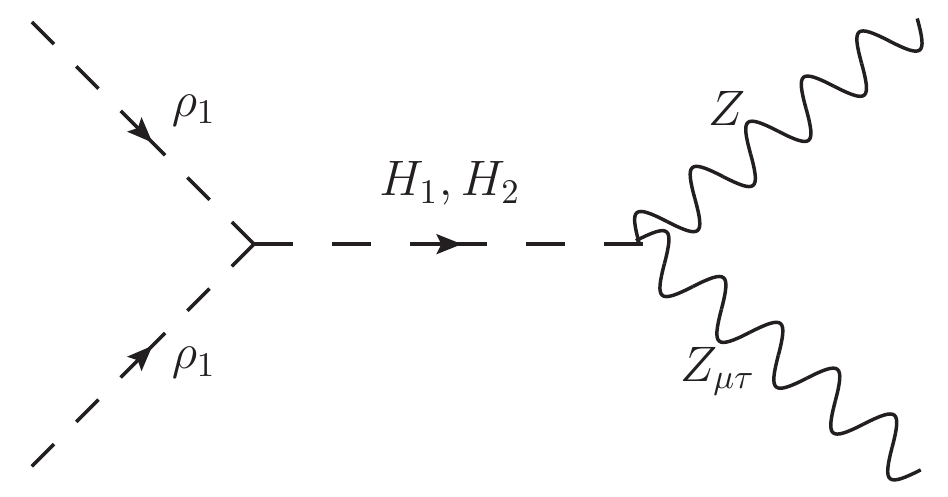}
\hskip 0.1in
\includegraphics[height=2.7cm,width=4.3cm,angle=0]{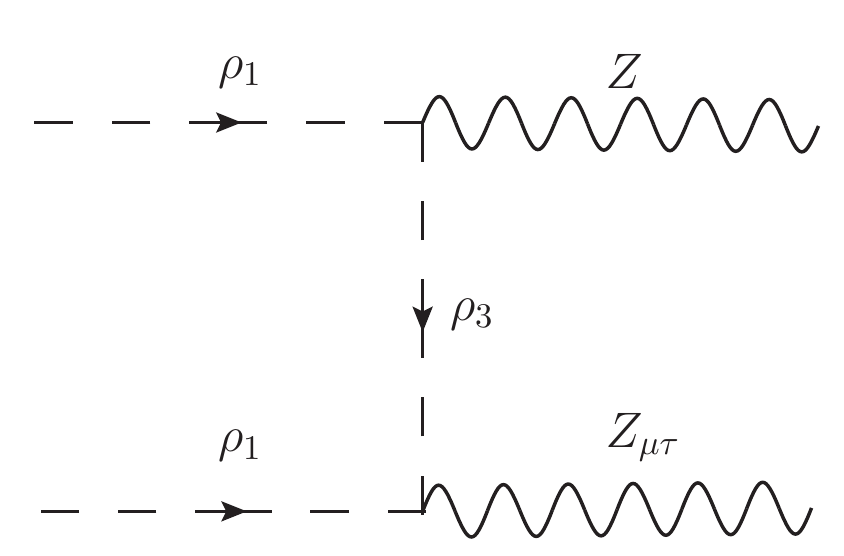}
\hskip 0.1in
\includegraphics[height=2.7cm,width=4.3cm,angle=0]{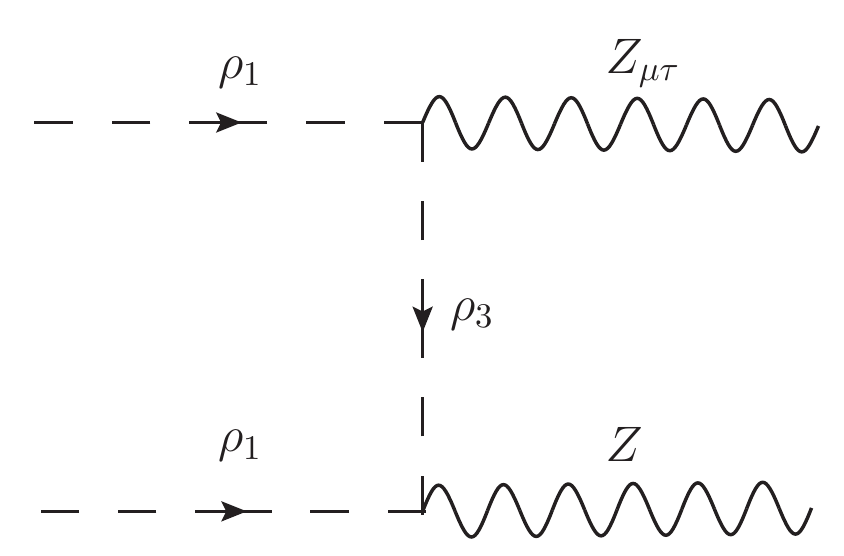} \\
\vskip 0.05in
\hskip -0.2in
\includegraphics[height=2.7cm,width=3.0cm,angle=0]{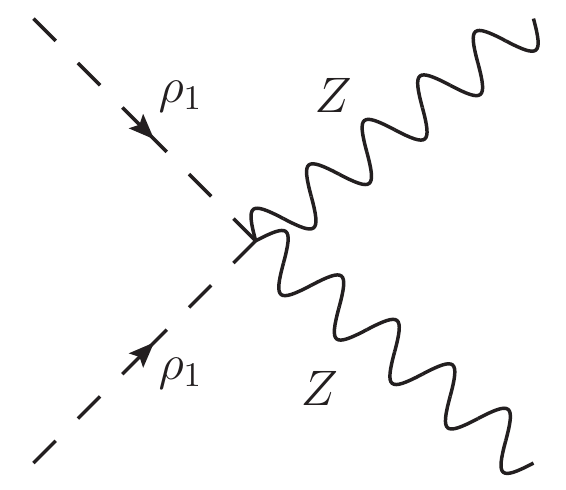} 
\hskip 0.1in
\includegraphics[height=2.7cm,width=4.3cm,angle=0]{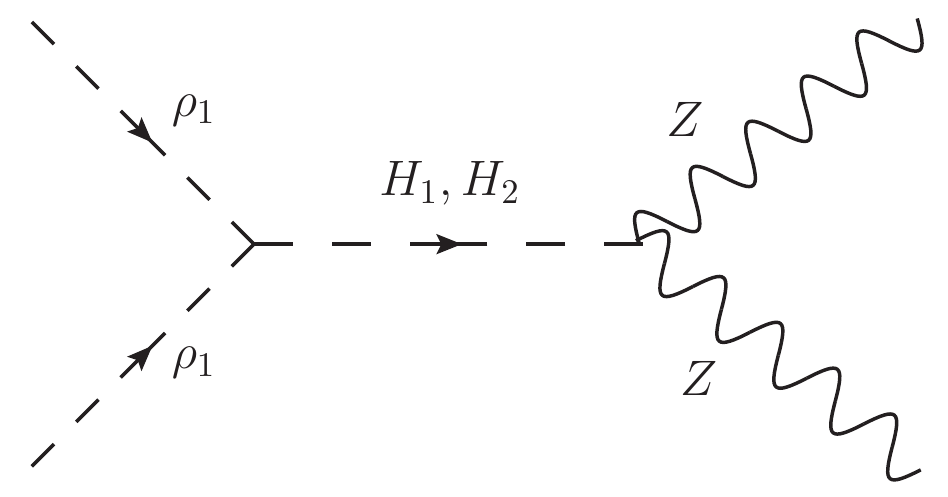}
\hskip 0.1in
\includegraphics[height=2.7cm,width=4.3cm,angle=0]{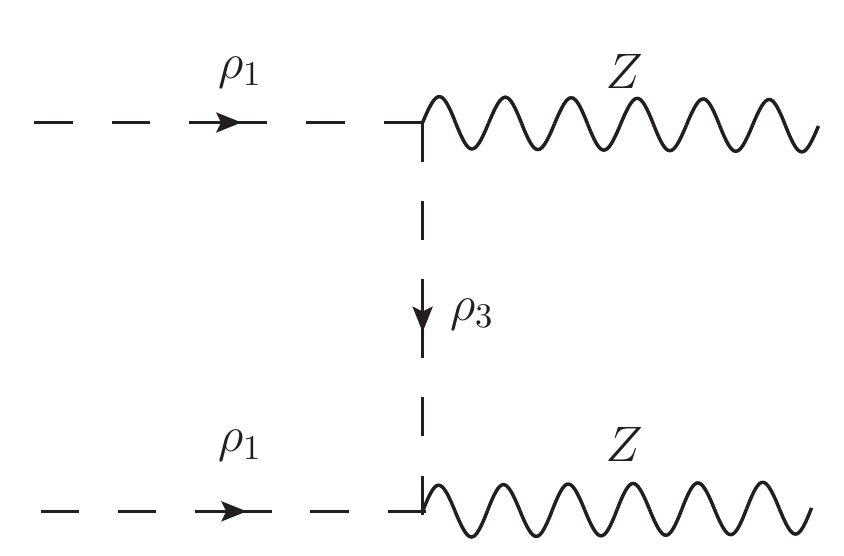}
\hskip 0.1in
\includegraphics[height=2.7cm,width=4.3cm,angle=0]{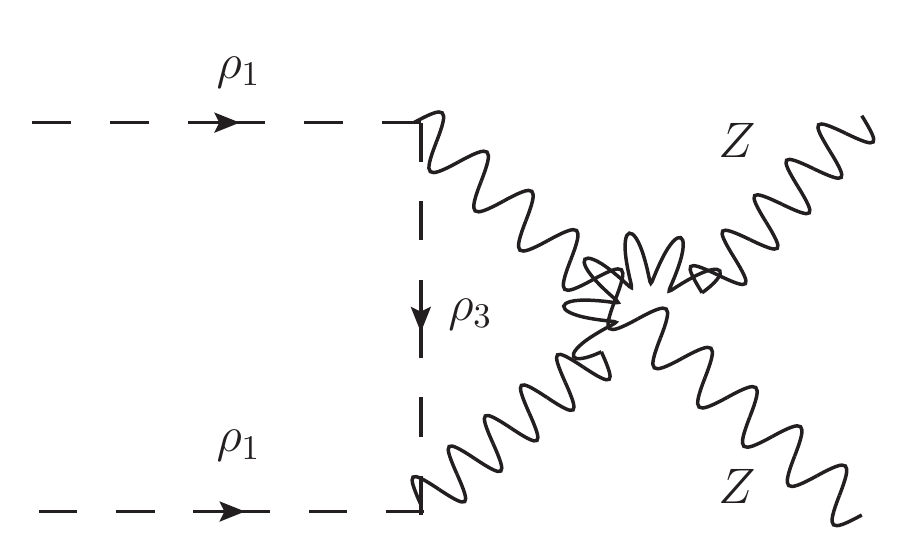} \\
\vskip 0.05in
\hskip -0.2in
\includegraphics[height=2.7cm,width=3.0cm,angle=0]{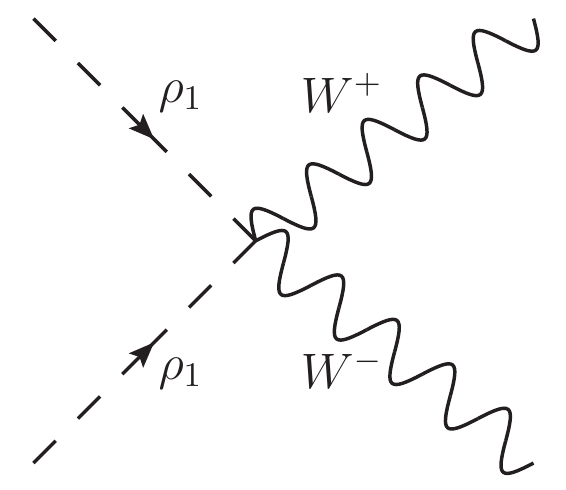}
\hskip 0.1in
\includegraphics[height=2.7cm,width=4.3cm,angle=0]{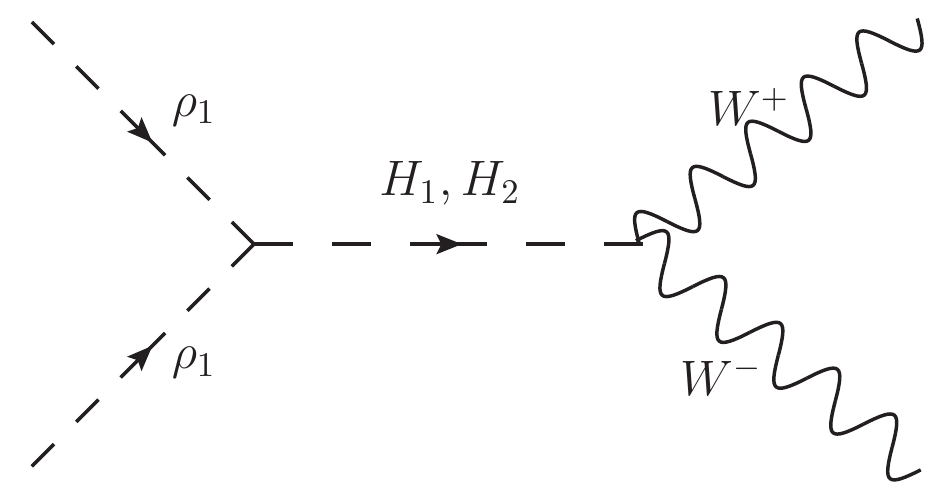}
\hskip 0.1in 
\includegraphics[height=2.7cm,width=4.3cm,angle=0]{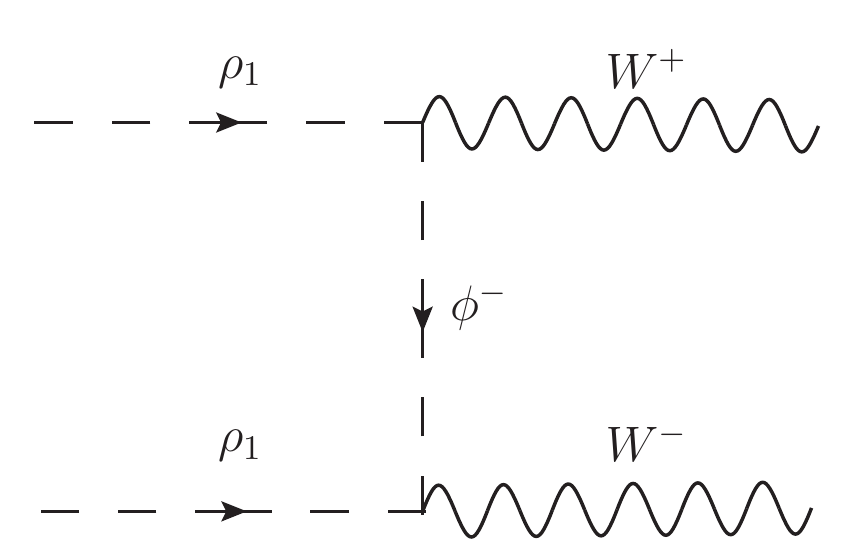}
\hskip 0.1in
\includegraphics[height=2.7cm,width=4.3cm,angle=0]{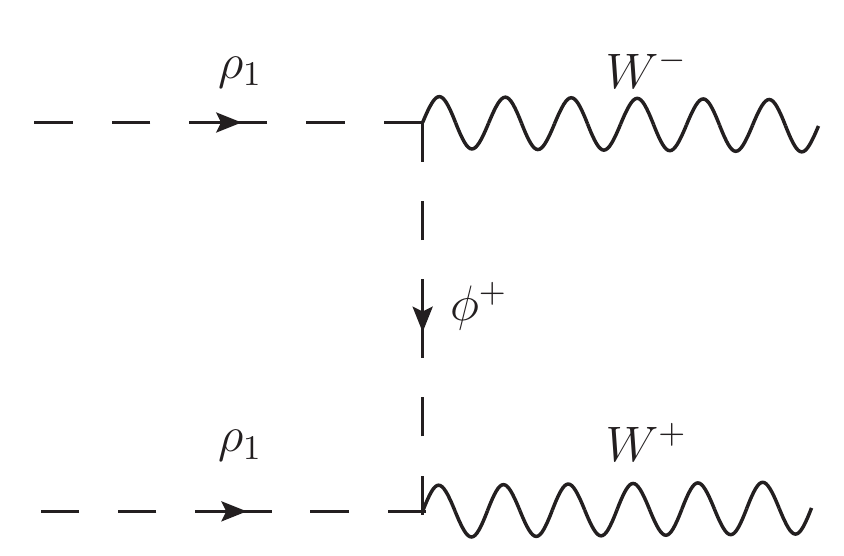} \\
\vskip 0.05in
\hskip -0.2in
\includegraphics[height=2.7cm,width=3.0cm,angle=0]{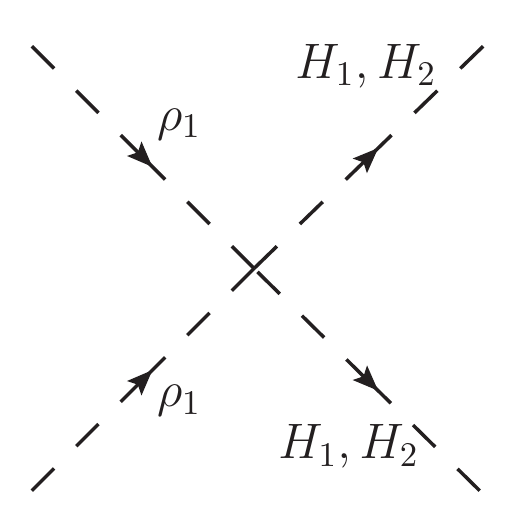} 
\hskip 0.1in
\includegraphics[height=2.7cm,width=4.3cm,angle=0]{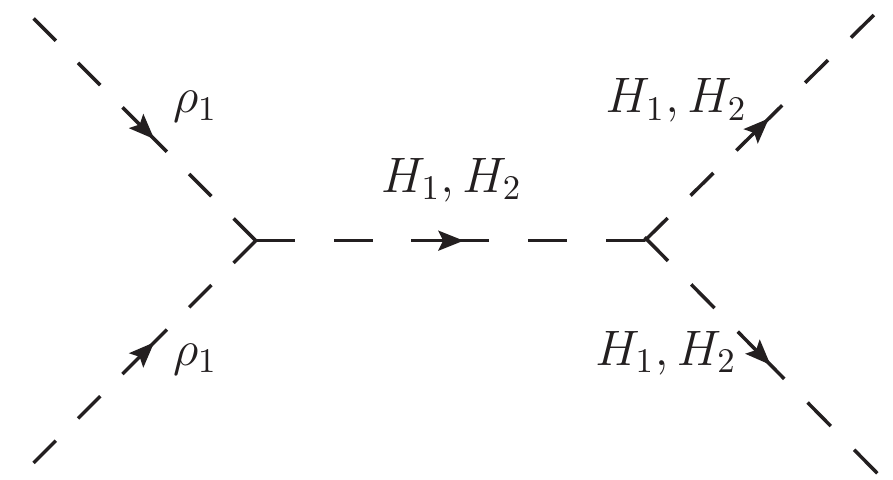}
\hskip 0.1in
\includegraphics[height=2.7cm,width=4.3cm,angle=0]{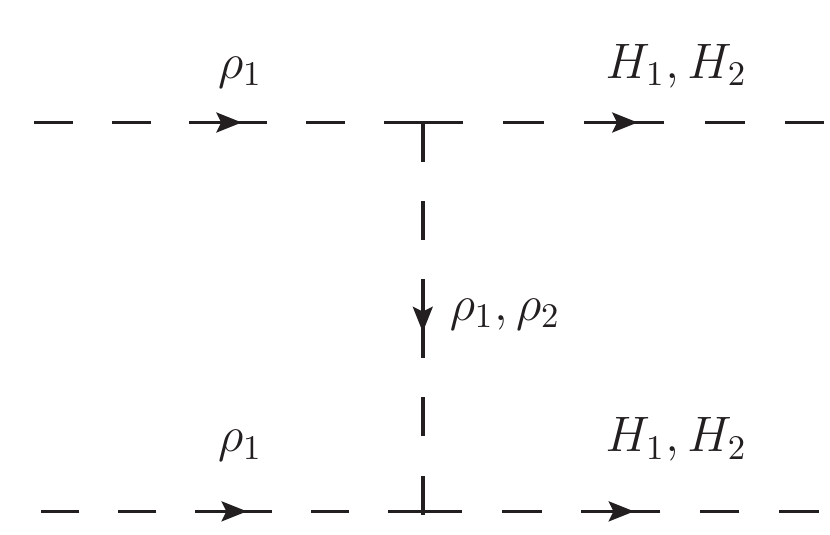}
\hskip 0.1in
\includegraphics[height=2.7cm,width=4.3cm,angle=0]{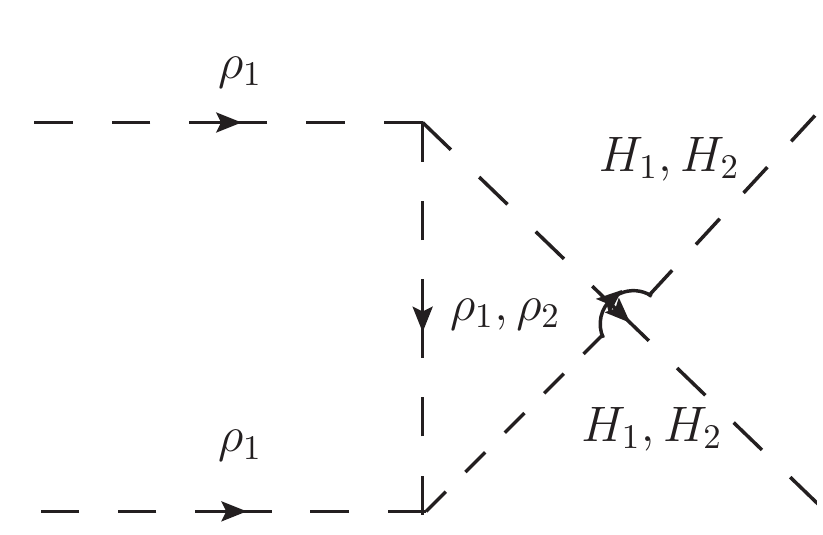} \\
\vskip 0.05in
\hskip -0.2in
\includegraphics[height=2.7cm,width=3.0cm,angle=0]{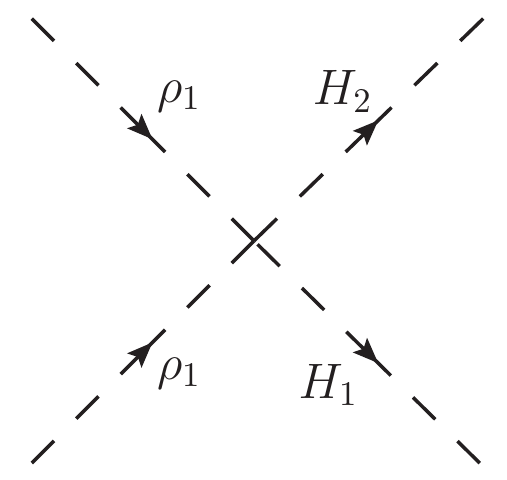}
\hskip 0.1in
\includegraphics[height=2.7cm,width=4.3cm,angle=0]{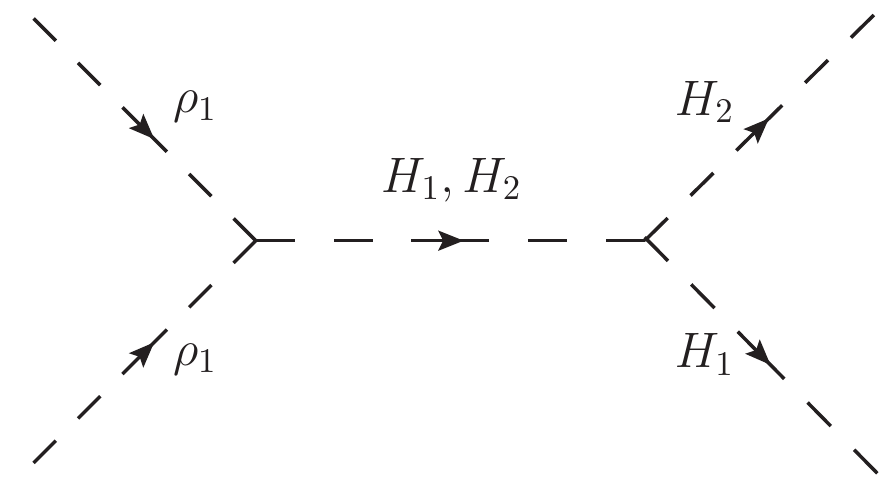}
\hskip 0.1in
\includegraphics[height=2.7cm,width=4.3cm,angle=0]{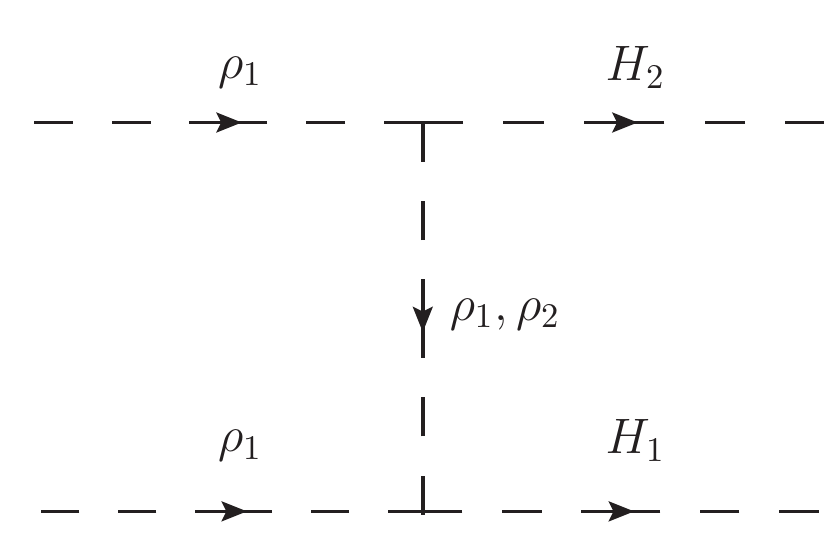}
\hskip 0.1in
\includegraphics[height=2.7cm,width=4.3cm,angle=0]{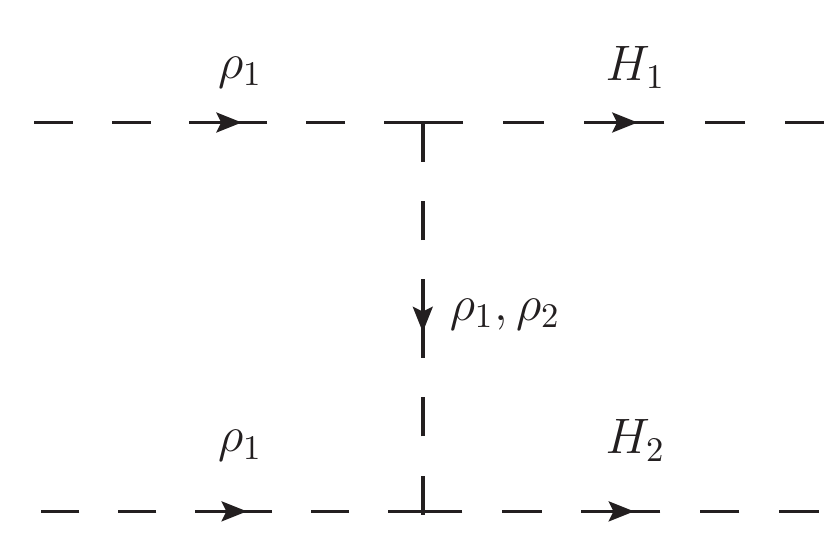}\\
\vskip 0.05in
\includegraphics[height=2.7cm,width=2.7cm,angle=0]{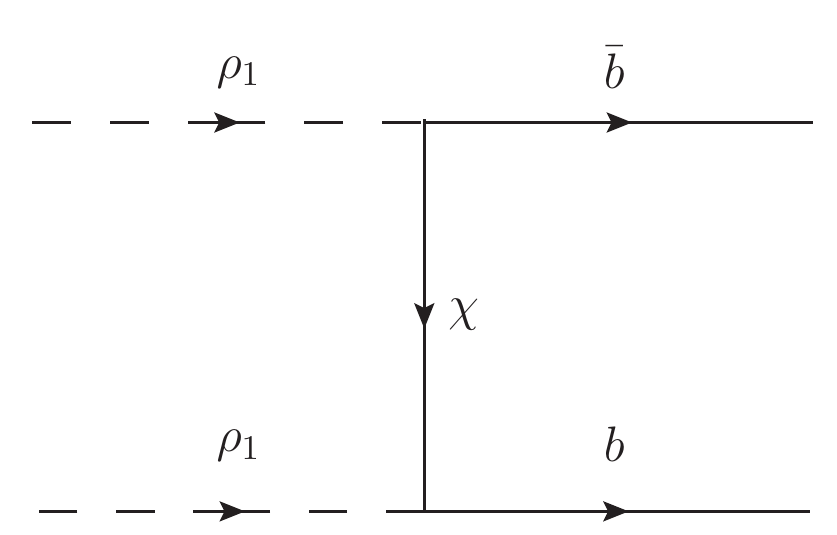}
\includegraphics[height=2.7cm,width=2.7cm,angle=0]{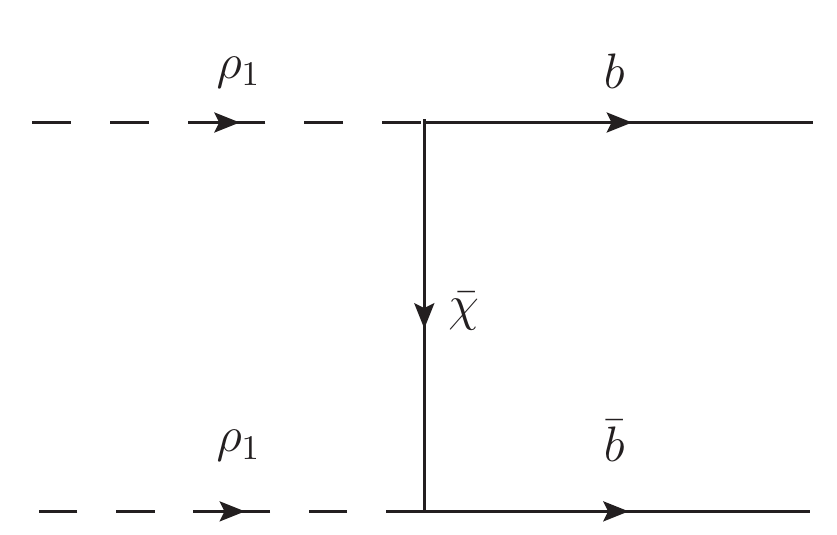}
\includegraphics[height=2.7cm,width=2.7cm,angle=0]{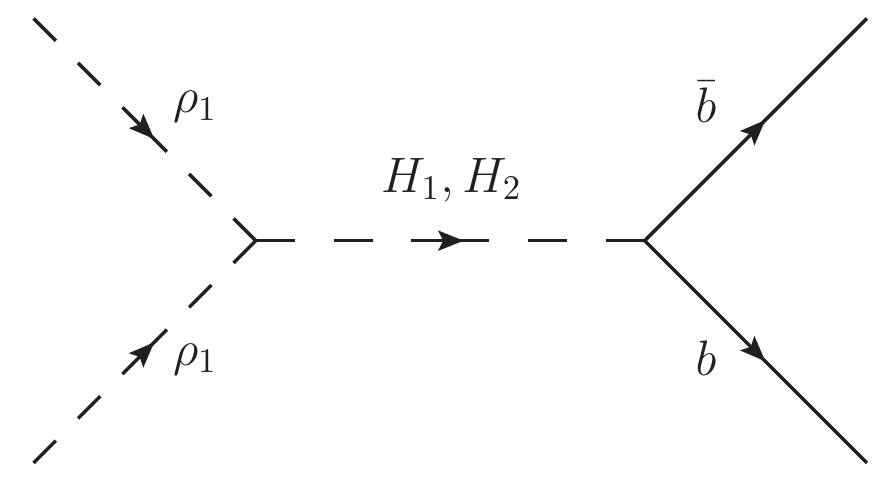}
\includegraphics[height=2.7cm,width=2.7cm,angle=0]{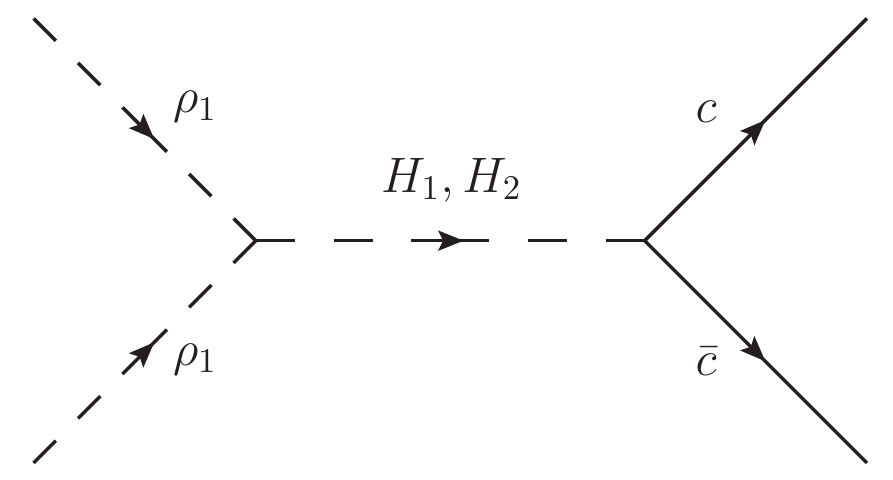}
\includegraphics[height=2.7cm,width=2.7cm,angle=0]{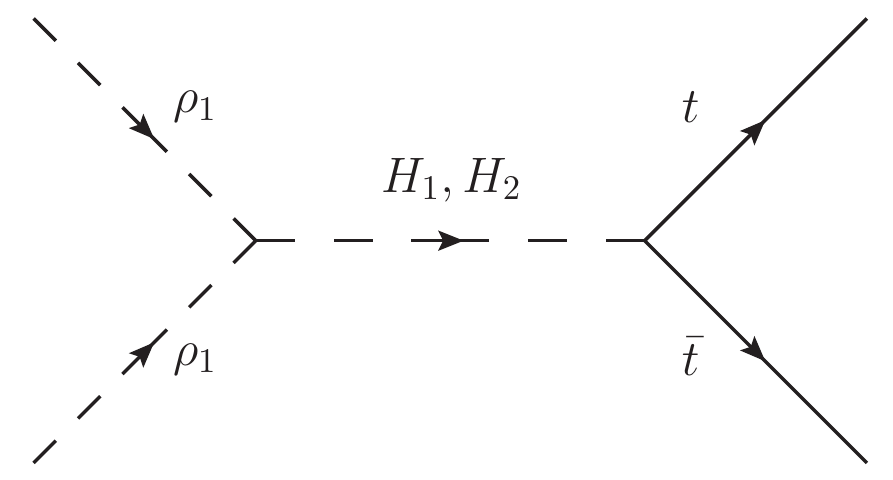} 
\includegraphics[height=2.7cm,width=2.7cm,angle=0]{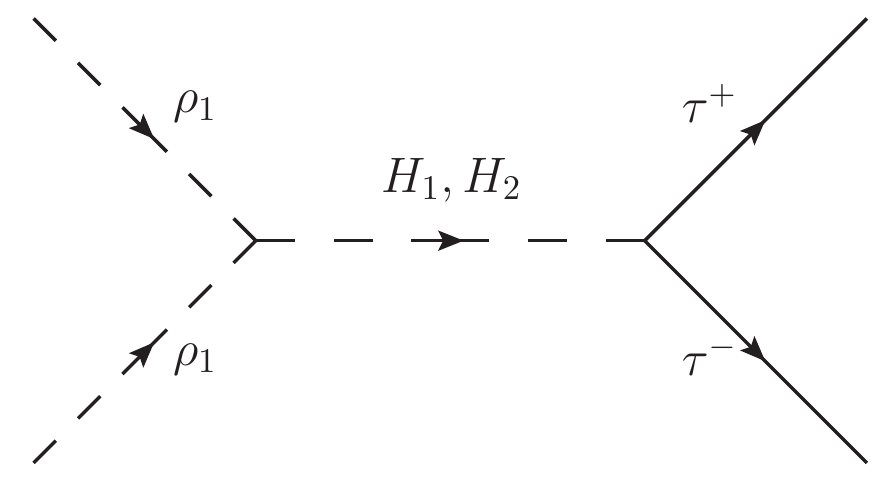} 
\caption{Feynman diagrams of dark matter annihilation channels
contributing significantly to the freeze-out process.}
\label{Fig:FD_anni}
\end{figure}

\begin{figure}[h!]
\hskip -0.2in
\includegraphics[height=2.7cm,width=3.0cm,angle=0]{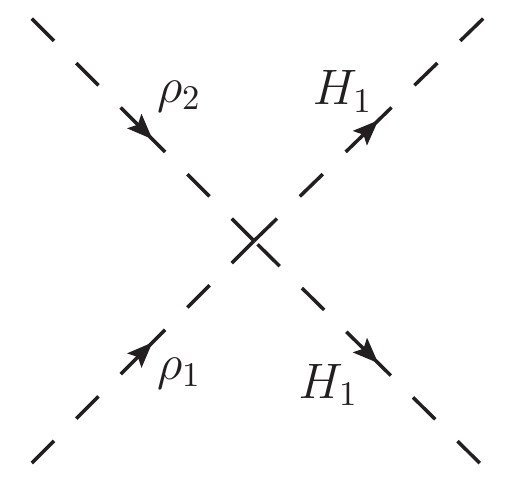}
\hskip 0.1in
\includegraphics[height=2.7cm,width=4.3cm,angle=0]{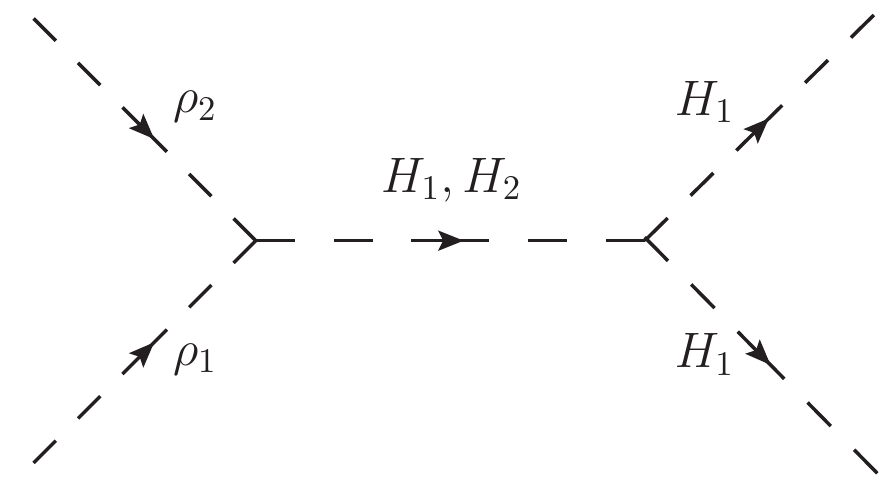}
\hskip 0.1in
\includegraphics[height=2.7cm,width=4.3cm,angle=0]{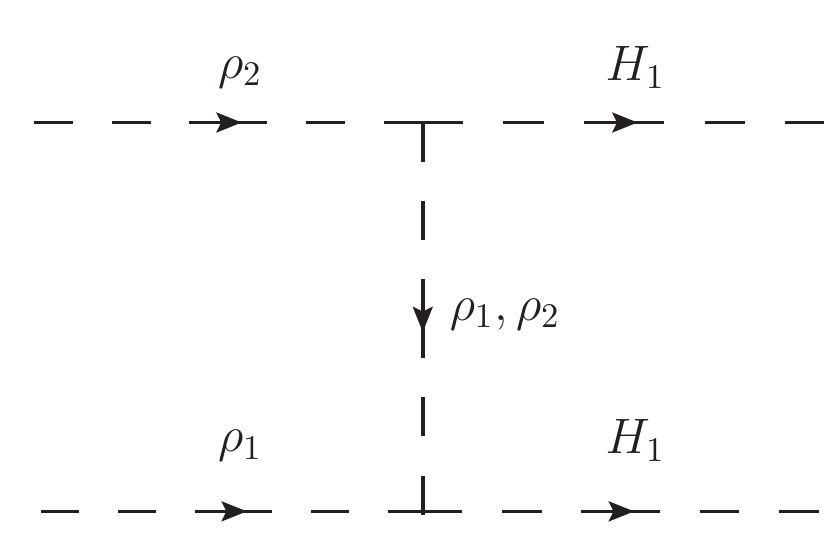}
\hskip 0.1in
\includegraphics[height=2.7cm,width=4.3cm,angle=0]{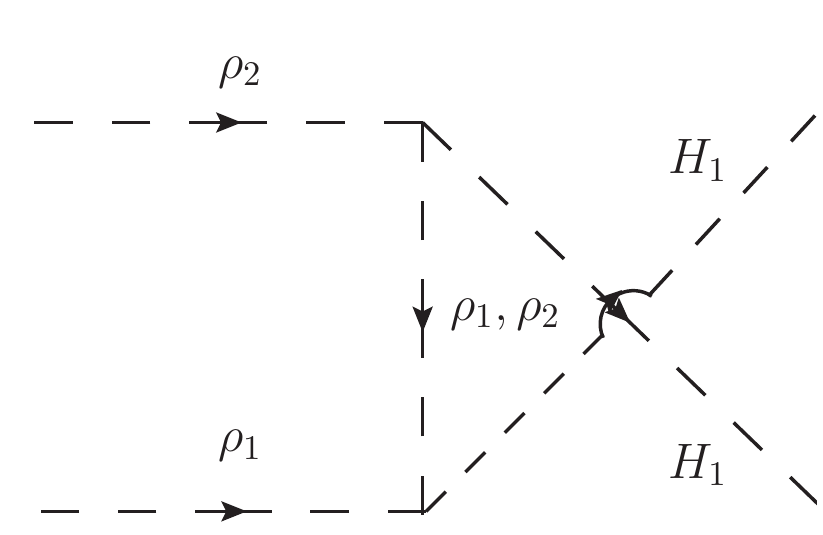}
\vskip 0.05in
\hskip -0.2in
\includegraphics[height=2.7cm,width=3.0cm,angle=0]{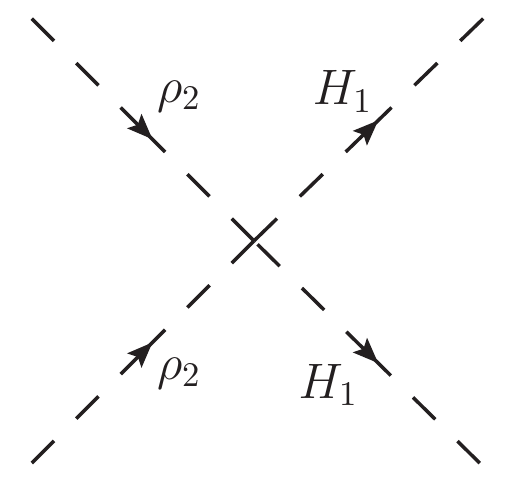}
\hskip 0.1in
\includegraphics[height=2.7cm,width=4.3cm,angle=0]{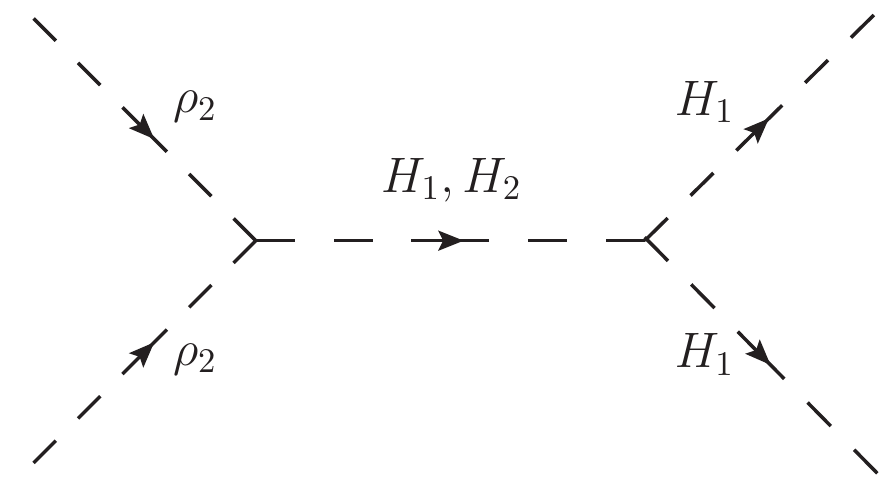}
\hskip 0.1in
\includegraphics[height=2.7cm,width=4.3cm,angle=0]{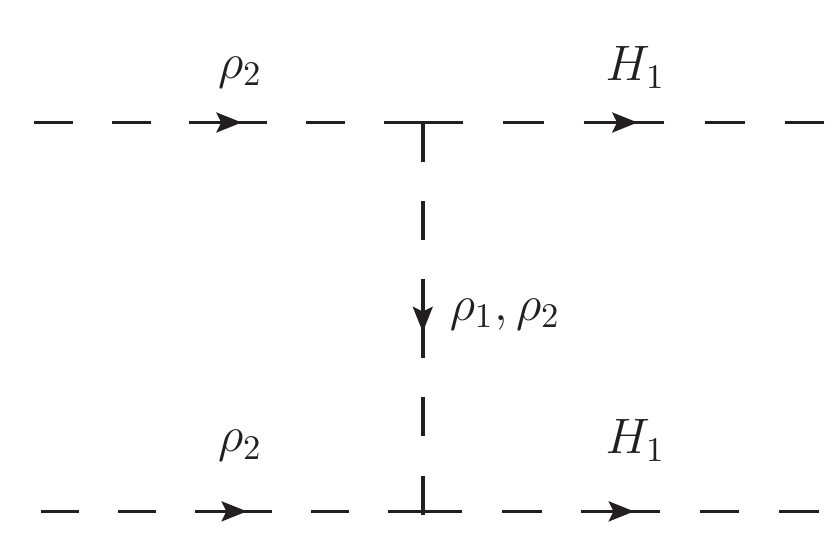}
\hskip 0.1in
\includegraphics[height=2.7cm,width=4.3cm,angle=0]{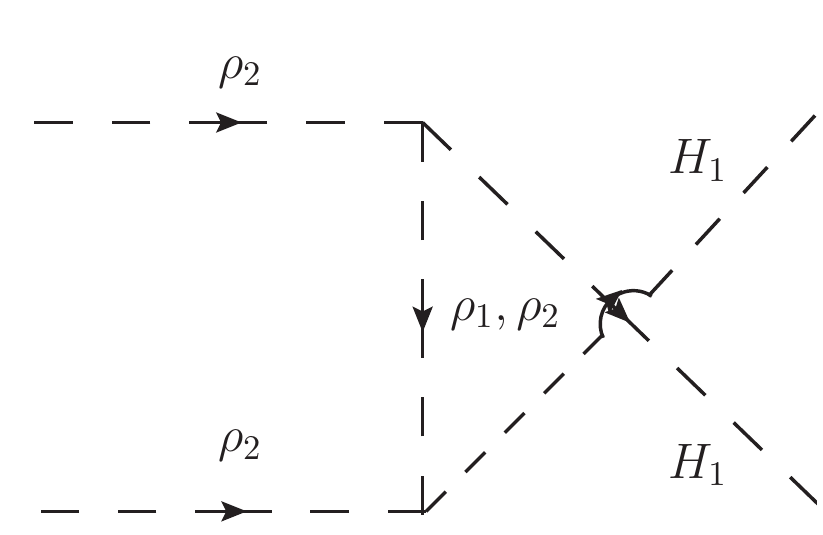}
\vskip 0.05in
\hskip -0.2in
\includegraphics[height=2.7cm,width=3.0cm,angle=0]{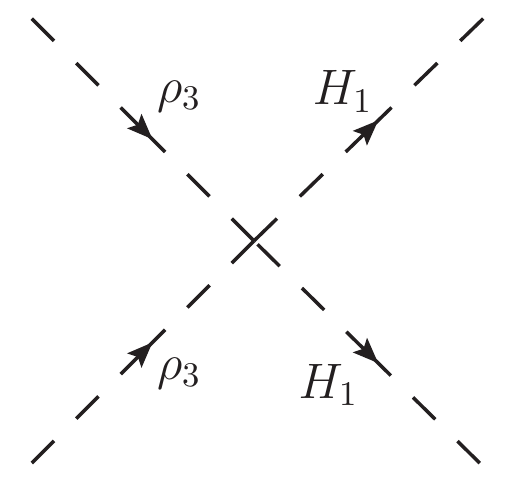}
\hskip 0.1in
\includegraphics[height=2.7cm,width=4.3cm,angle=0]{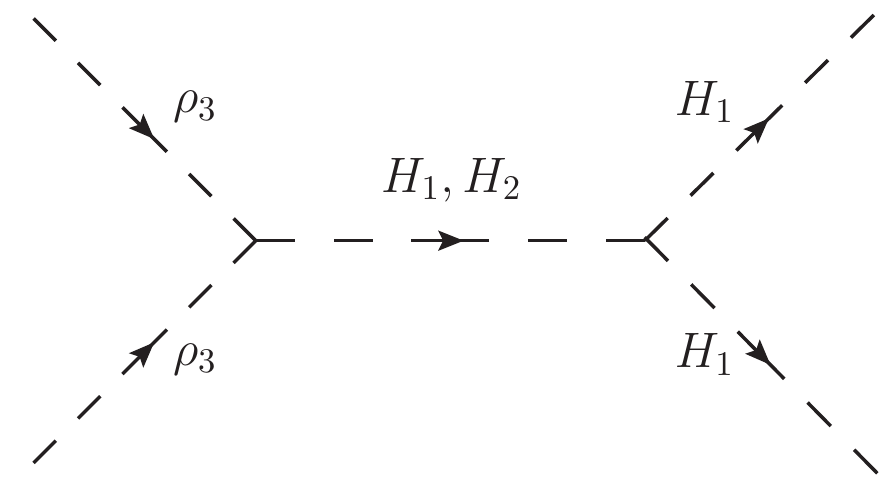}
\hskip 0.1in
\includegraphics[height=2.7cm,width=4.3cm,angle=0]{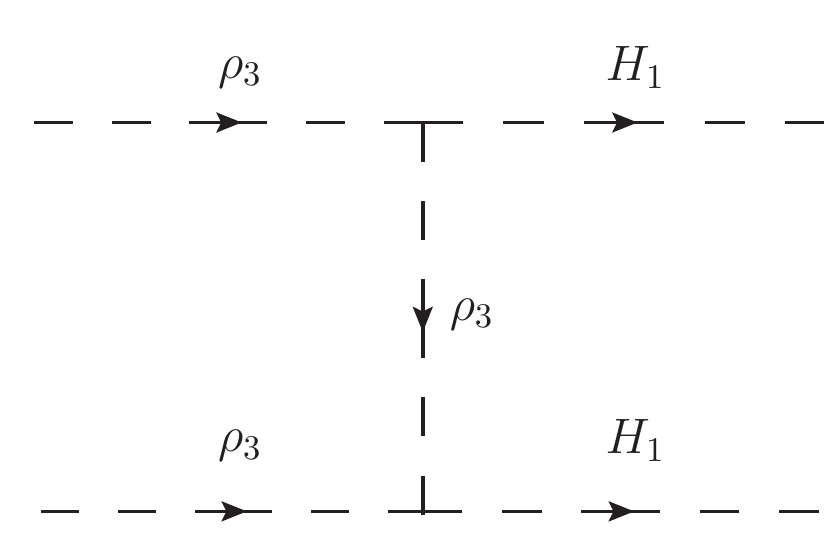}
\hskip 0.1in
\includegraphics[height=2.7cm,width=4.3cm,angle=0]{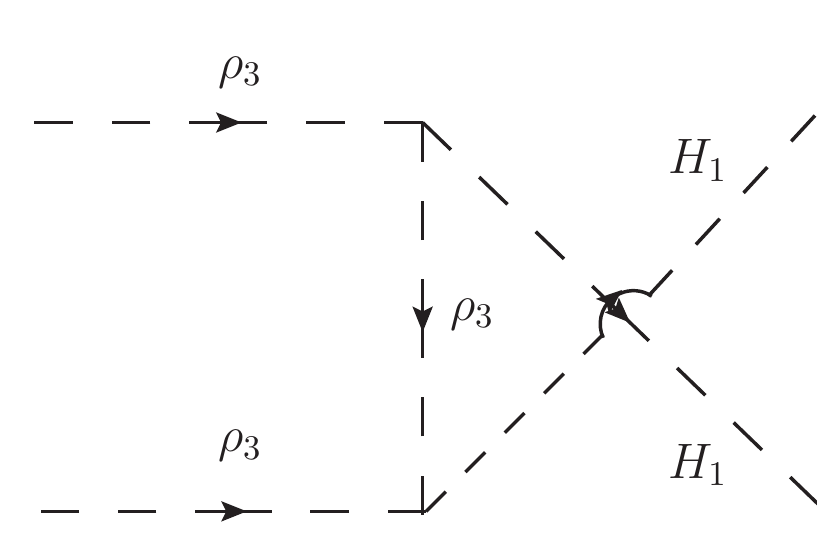}
\vskip 0.05in
\hskip -0.2in
\includegraphics[height=2.7cm,width=3.0cm,angle=0]{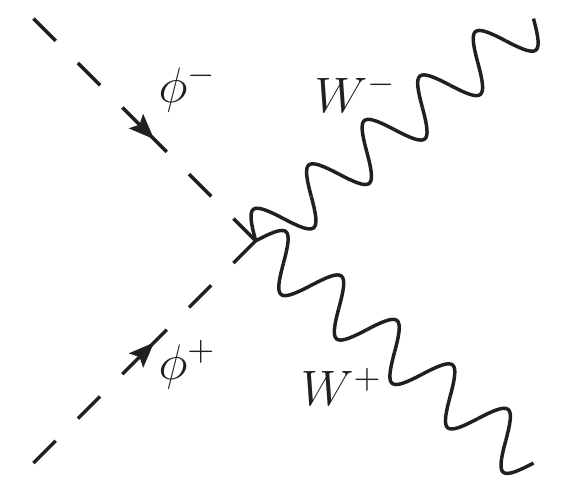}
\hskip 0.1in
\includegraphics[height=2.7cm,width=4.3cm,angle=0]{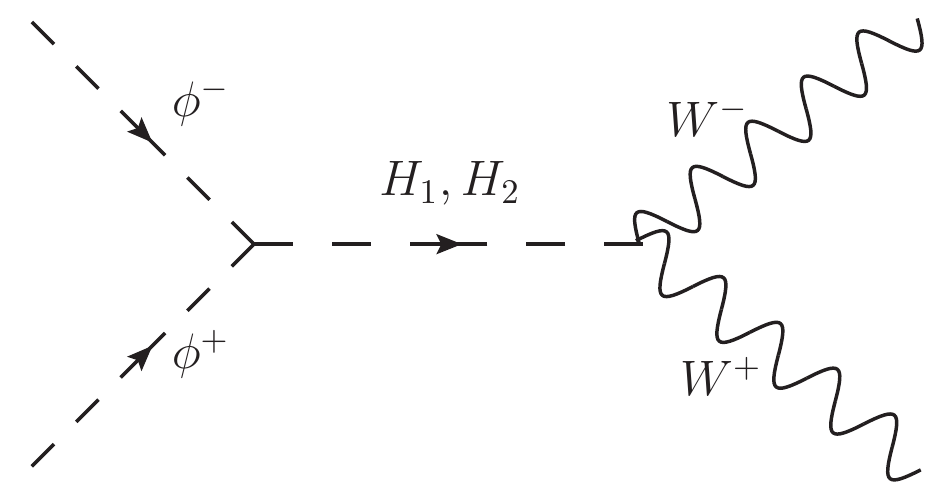}
\hskip 0.1in
\includegraphics[height=2.7cm,width=4.3cm,angle=0]{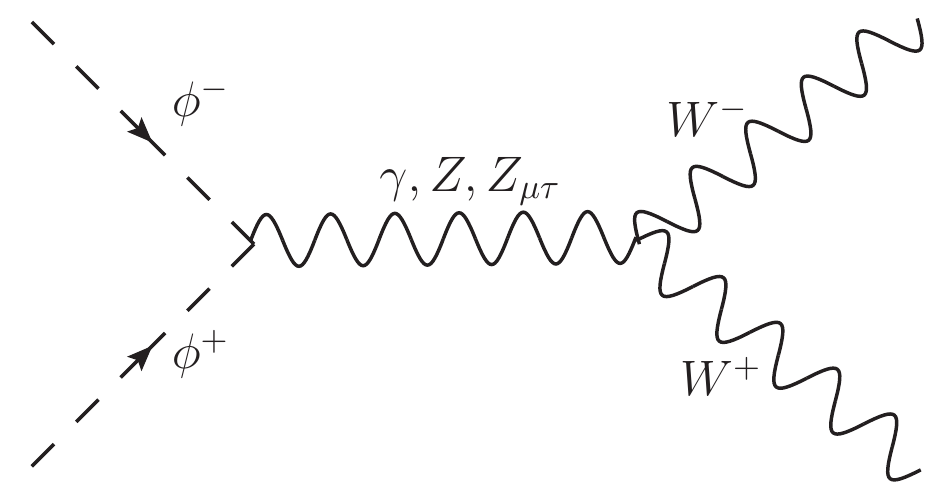}
\hskip 0.1in
\includegraphics[height=2.7cm,width=4.3cm,angle=0]{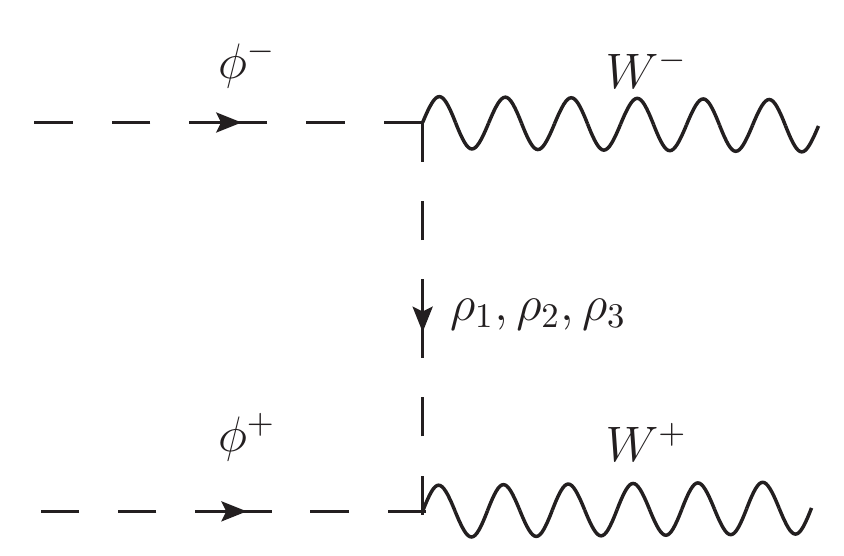}
\vskip 0.05in
\hskip -0.2in
\includegraphics[height=2.7cm,width=3.0cm,angle=0]{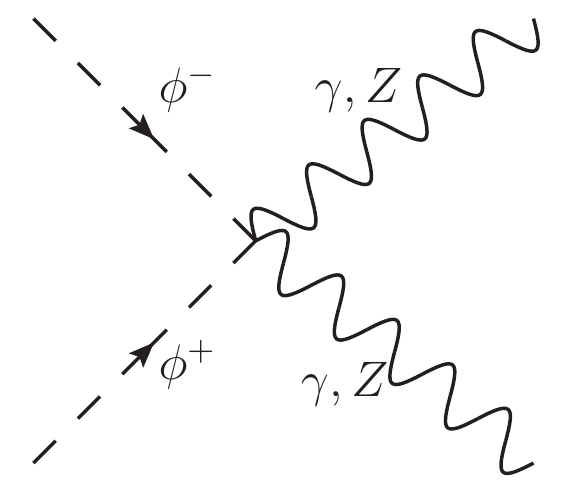}
\hskip 0.1in
\includegraphics[height=2.7cm,width=4.3cm,angle=0]{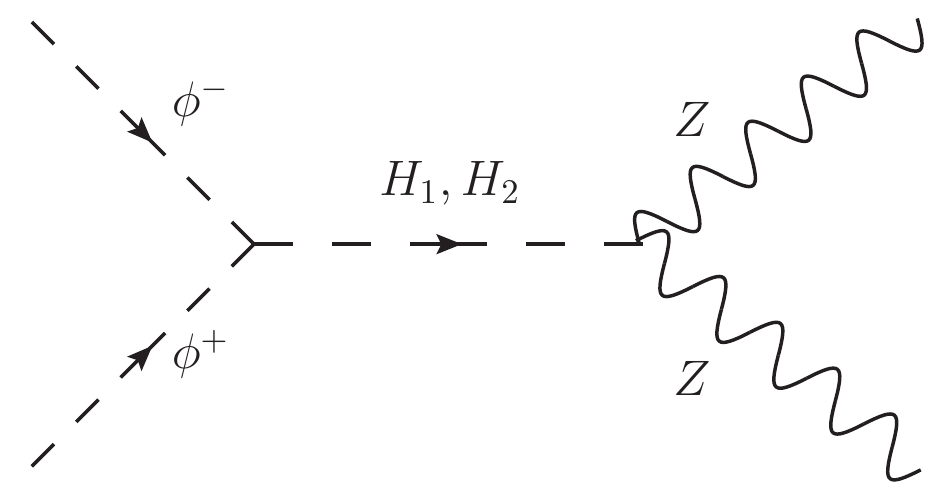}
\hskip 0.1in
\includegraphics[height=2.7cm,width=4.3cm,angle=0]{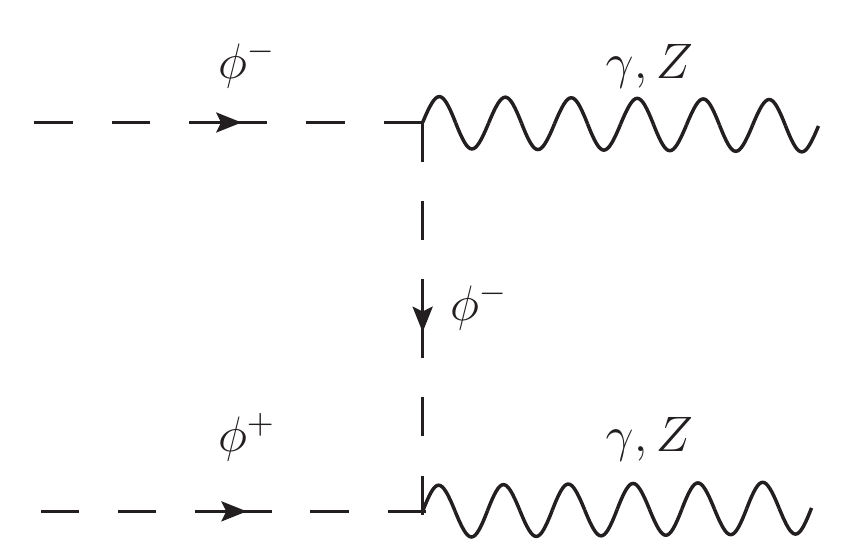}
\hskip 0.1in
\includegraphics[height=2.7cm,width=4.3cm,angle=0]{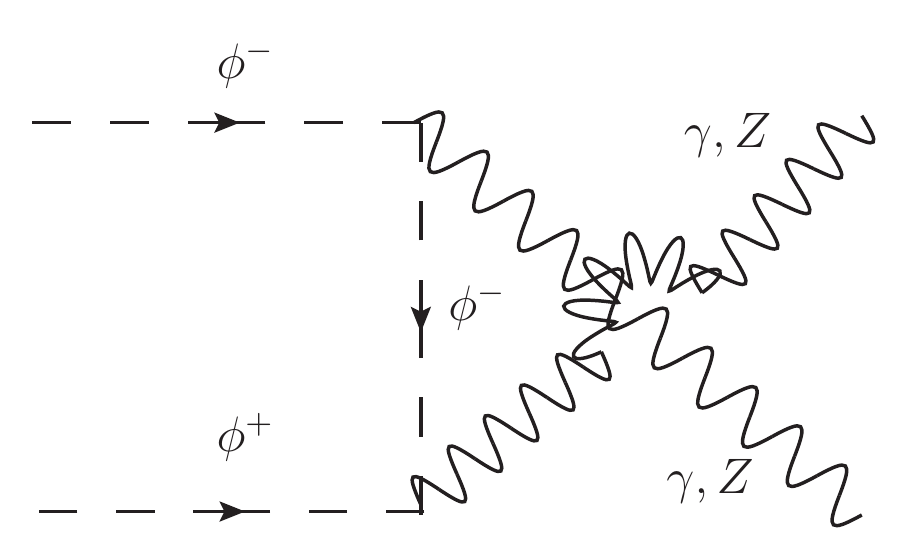}
\vskip 0.05in
\hskip -0.2in
\includegraphics[height=2.7cm,width=3.0cm,angle=0]{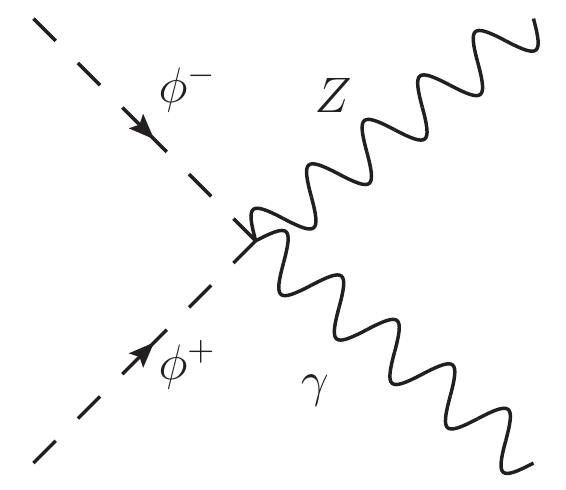}
\hskip 0.1in
\hskip 0.1in
\includegraphics[height=2.7cm,width=4.3cm,angle=0]{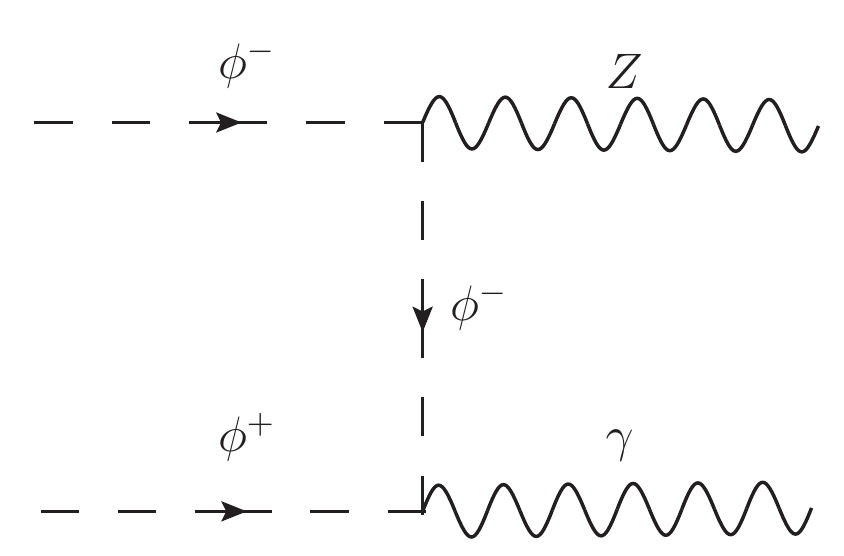}
\hskip 0.1in
\includegraphics[height=2.7cm,width=4.3cm,angle=0]{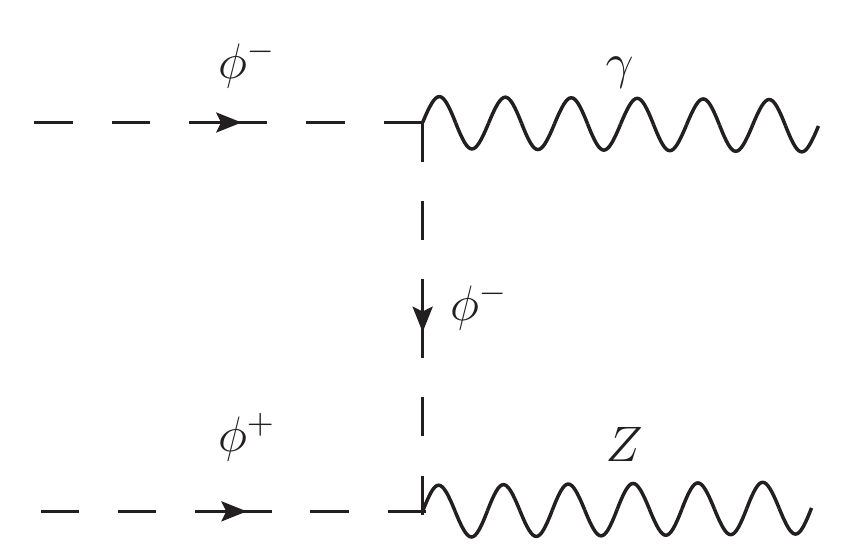}
\caption{Co-annihilation channels contributing to the relic density of
$\rho_1$ in the high mass region.}
\label{Fig:FD_Coanni}
\end{figure}
The dark matter candidate ($\rho_1$) of the present scenario is
a thermal WIMP, which remains in equilibrium with the
thermal bath until its freeze-out though annihilations and co-annihilations into
various final states allowed by the symmetries of the model. 
In this work, we have considered $M_{\rho_1}$ between 10 GeV
to 1 TeV. For low dark matter masses (i.e. $M_{\rho_1}<100$ GeV),
$\rho_1$ predominantly annihilates into a pair of $Z_{\mu\tau}$. In some
cases, depending upon the relevant couplings, $b\bar{b}$, $c\bar{c}$
and $\tau\bar{\tau}$ final states are also possible. Moreover, co-annihilations among
the $\mathbb{Z}_2$-odd particles in the low mass regime are insignificant as
we have considered all heavier $\mathbb{Z}_2$-odd particles masses
larger than 100 GeV throughout this analysis to evade experimental
bounds \cite{Lundstrom:2008ai}. Alternatively, for the heavier mass range
of $\rho_1$, there are various possibilities. First of all depending
upon the mass splitting between $\rho_1$ and other $\mathbb{Z}_2$-odd
particles (parametrised by a quantity $\Delta_i$, defined earlier)
there can either be annihilation or co-annihilations. In the former
case, depending on the values of the associated couplings $\rho_1 \rho_1 \rightarrow
Z_{\mu\tau}Z_{\mu\tau},\,ZZ_{\mu\tau},\,H_2H_2,\,H_1H_1,\,H_1H_2,\,W^+W^-,\,ZZ$
and $t\bar{t}$ final states can be important. On the other hand, co-annihilation
plays a pivotal role during the freeze-out of $\rho_1$ when $\Delta_i \leq 0.2$
\cite{Griest:1990kh} for any $\mathbb{Z}_2$-odd particle
$i$ ($i=\rho_2$, $\rho_3$, $\phi^{\pm}$).
In this circumstances, various co-annihilations among these dark sector
particles like $\rho_1\rho_2\rightarrow H_1H_1$, $\rho_i\rho_i \rightarrow H_1 H_1\,\,(i=1-3)$,
$\phi^+\phi^- \rightarrow W^+W^-,\,\gamma\gamma,\,\gamma Z,\,ZZ$ etc. become
predominant. Feynman diagrams of all significant annihilation and
co-annihilation channels are shown in Figs.\,\ref{Fig:FD_anni}
and \ref{Fig:FD_Coanni} respectively.

\begin{figure}[h!]
\includegraphics[height=9cm,width=12cm,angle=0]{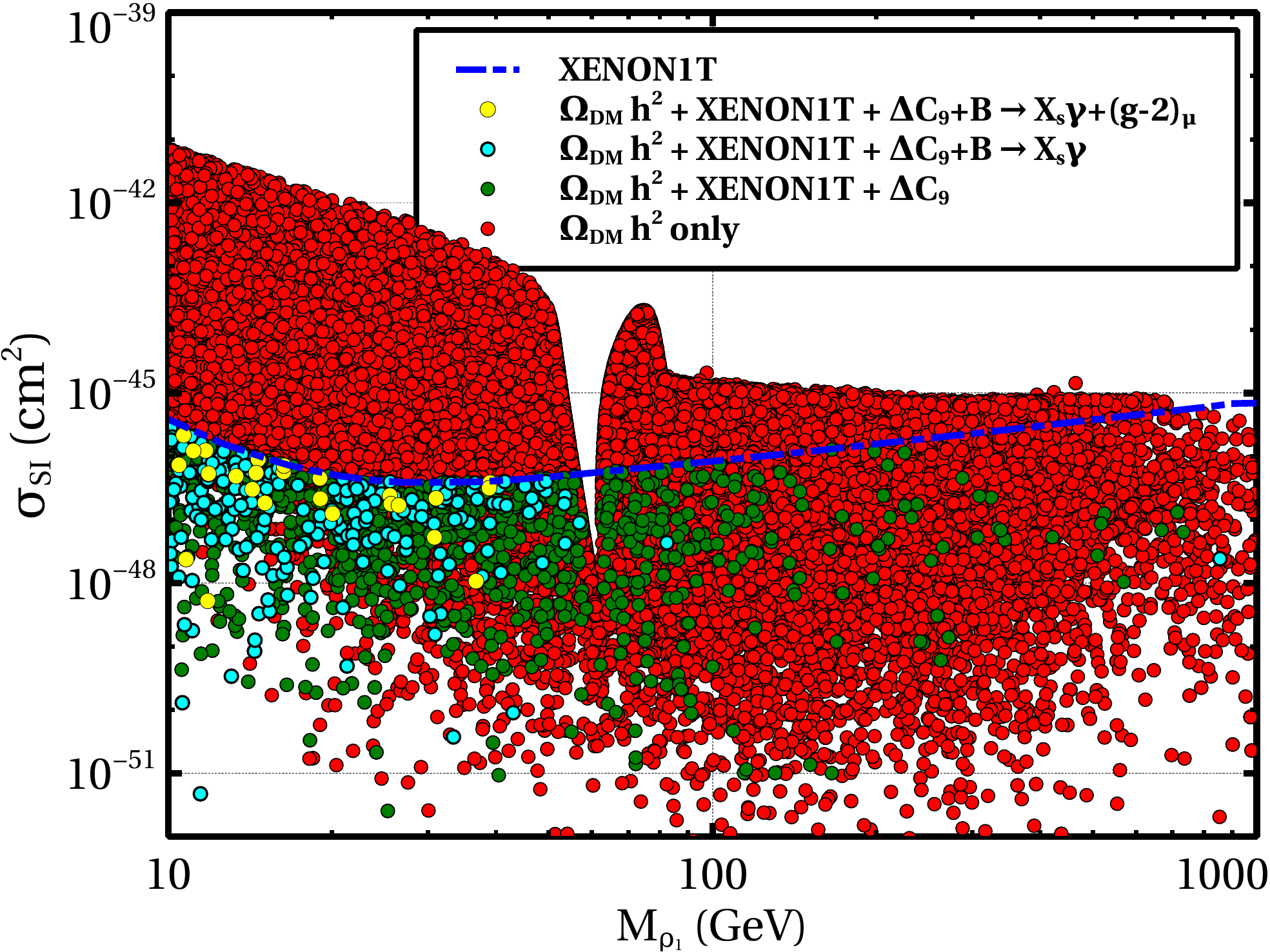}
\caption{Allowed parameter space in $\sigma_{\rm SI}$ vs $M_{\rho_1}$ plane
subject to various experiments bounds indicated in the legend.}
\label{Fig:mdm-sigmaSI}
\end{figure}

In Fig.\ref{Fig:mdm-sigmaSI}, we have plotted spin independent scattering
cross section $\sigma_{\rm SI}$ of $\rho_1$ with its mass $M_{\rho_{1}}$,
varying between 10 GeV to 1 TeV. In this plot all red coloured points
satisfy relic density constraint i.e., $0.1172\leq\Omega_{\rm DM}h^2\leq0.1226$
and bound related to Higgs invisible decay modes as well.  
The blue dashed-dot line represents the latest bound on $\sigma_{\rm SI}$
from XENON1T experiment. All the parameter space below the
blue dashed-dot line are still allowed and can be probed in near future
by ``multi-ton-scale'' direct detection experiments like XENONnT. Therefore, if we consider
direct constrains like relic density, direct detection and Higgs
invisible decay only, there are still enough parameter space
left (although few portion mostly in the low mass dark matter regime
have already been ruled-out) for the entire considered mass range
of $\rho_1$. However, the situation does not remain same when one
tries to explain flavour physics anomalies and $(g-2)_\mu$ anomaly within this framework. The allowed parameter space in
$\sigma_{\rm SI}-M_{\rho_1}$ plane gets severely restricted
when we impose bound on the NP contribution
to the WC $C_9$ (i.e. $-1.26\leq\Delta{C_9}\leq-0.63$
in $2\sigma$ range \cite{Aebischer:2019mlg}) to explain $R_{K^{(*)}}$ anomalies.
This has been indicated by green coloured points in the above plot where
one can notice that the low dark matter mass regime (i.e. $M_{\rho_1}\la 100$ GeV)
is the most favourable to address $R_{K^{(*)}}$ anomalies. This can be understood
from the behaviour of $\Delta{C_9}$ (Eq.\,(\ref{dc9}))
with respect to the mass of ${\rho_1}$ as illustrated in
Fig.\,\ref{Fig:dc9-vs-mrho1}, where the magnitude
of $\Delta{C_9}$ sharply decreases with the increase of
$M_{\rho_1}$. Furthermore, in this framework, we have also tried to explain 
both Br($B\rightarrow X_s \gamma$) and $(g-2)_{\mu}$ anomaly, the two
long-standing anomalies of the SM from their experimental
counterparts. These are indicated by cyan and yellow coloured
points respectively in $\sigma_{\rm SI}-M_{\rho_1}$ plane.
For the branching ratio of $B\rightarrow X_s \gamma$, we have
used $3\sigma$  
($2.84 \leq {\rm Br}(B\rightarrow X_s \gamma)\times 10^4
\leq 3.80$ \cite{Amhis:2016xyh}) while the $2\sigma$ band
i.e. $115.44 \leq \Delta{a_{\mu}} \times 10^9 \leq 420.56$ \cite{Tanabashi:2018oca}
for $(g-2)_{\mu}$ has been taken into account\footnote{Here we would like to mention that, another constraint e.g., $B^0_s-\bar{B^0_s}$ mixing which could be relevant for the present scenario. However, NP contributions to the $B^0_s-\bar{B^0_s}$ mixing arise from the present scenario via box diagrams and these are negligibly small. The reason is that, apart from the dark matter particle, all non-standard particles which generate box diagrams are sufficiently massive (especially the non-standard fermion $\chi$, whose mass that we have taken $\geq 1$ TeV throughout the analysis). At this point it is relevant to mention that, from the recent 13 TeV LHC data \cite{Aaboud:2018pii}, a down-type quark ($\mathcal{B}$) with charge (-1/3) is excluded for masses below 1.22 TeV for the decay channels $\mathcal{B}\to Z b/Wt/ {\rm SM~Higgs}~b$. However, this bound is not applicable in our case, as in our model the field $\chi$ is odd under $\mathbb{Z}_2$ symmetry, therefore such decays are restricted by the $\mathbb{Z}_2$ symmetry. Although, for the sake of conservative approach we use $M_\chi \geq 1$ TeV in our analysis. Hence, the loop functions are substantially suppressed. Consequently, the NP contribution to $B^0_s-\bar{B^0_s}$ mixing would not put any stringent constraint in our scenario.}.
We have checked that in the low dark matter mass
region ($M_{\rho_1}\leq 100$ GeV), $\rho_1$
predominantly annihilates into the $Z_{\mu\tau}$ pair. This
actually makes dark matter physics strongly correlated with the
physics of rare $B$-decays and anomalous magnetic moment of $\mu$,
where the role of new gauge boson $Z_{\mu\tau}$ is extremely crucial.
Moreover, it also helps us to evade the strong bound
coming from the experiments of direct detection \cite{Aprile:2018dbl},
indirect detection \cite{Ahnen:2016qkx} and also from 
the collider on Higgs invisible branching \cite{Khachatryan:2016whc}
for the low mass scalar dark matter \cite{Athron:2017kgt, Casas:2017jjg, Biswas:2017dxt},
where $b\bar{b}$ final state is the principal annihilation channel.
Therefore, in spite of being a gauge singlet $\mathbb{Z}_2$-odd
scalar field, the mixing with another $\mathbb{Z}_2$-odd field
(part of an ${\rm SU(2)}_{L}$ doublet) having nonzero $L_{\mu}-L_{\tau}$
charge, makes the entire dynamics of our dark matter
candidate $\rho_1$ strikingly different from the standard
Scalar Singlet dark matter scenario \cite{McDonald:1993ex,
Burgess:2000yq, Biswas:2011td, Cline:2013gha}.     
Finally, for the completeness we would like to mention here that the yellow
coloured points in $\sigma_{\rm SI}-M_{\rho_1}$ plane are those
which satisfy all the experimental results we have considered in
this work.

\begin{figure}[t!]
\includegraphics[height=7.5cm,width=8.2cm,angle=0]{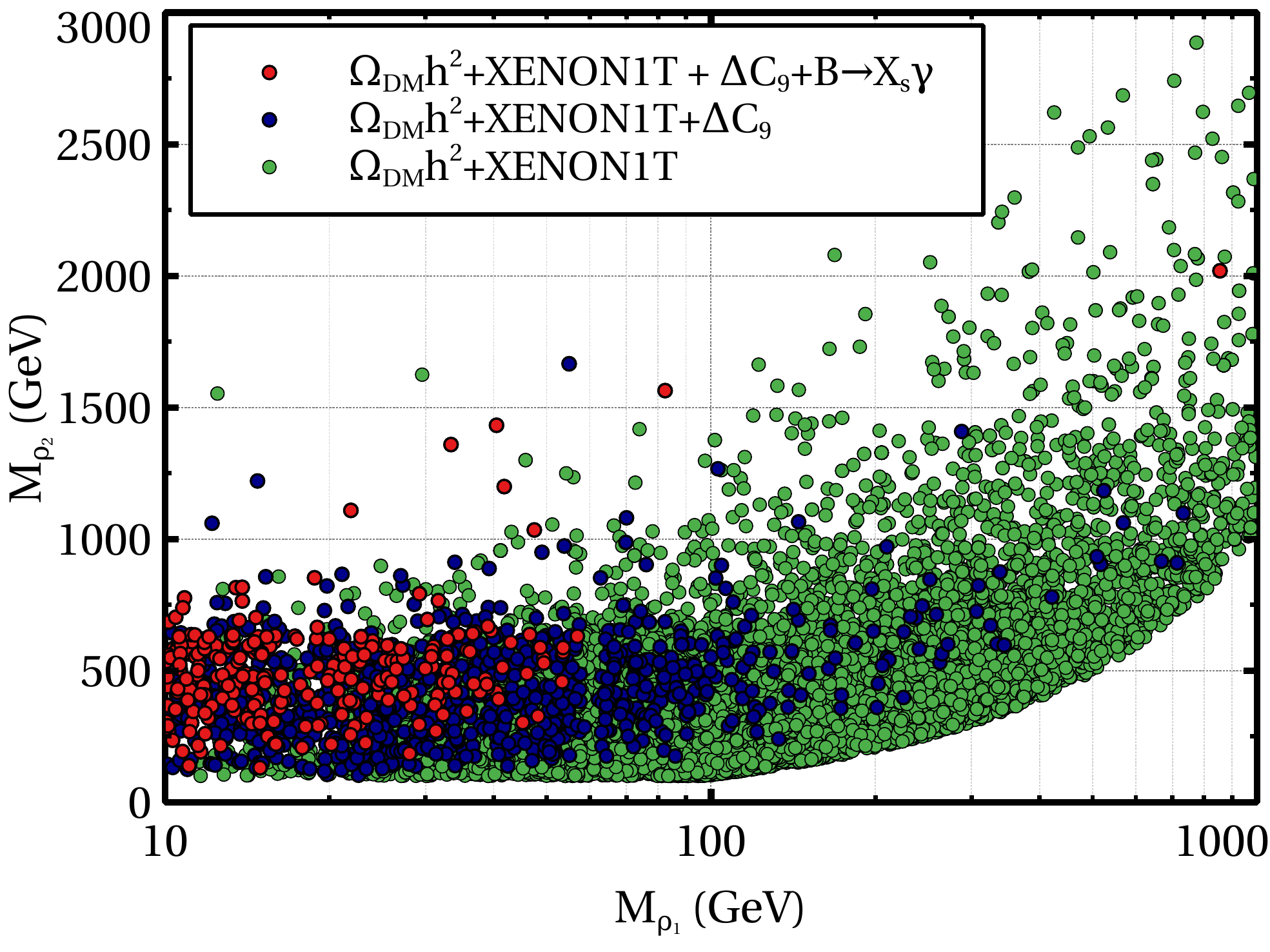}
\includegraphics[height=7.5cm,width=8.2cm,angle=0]{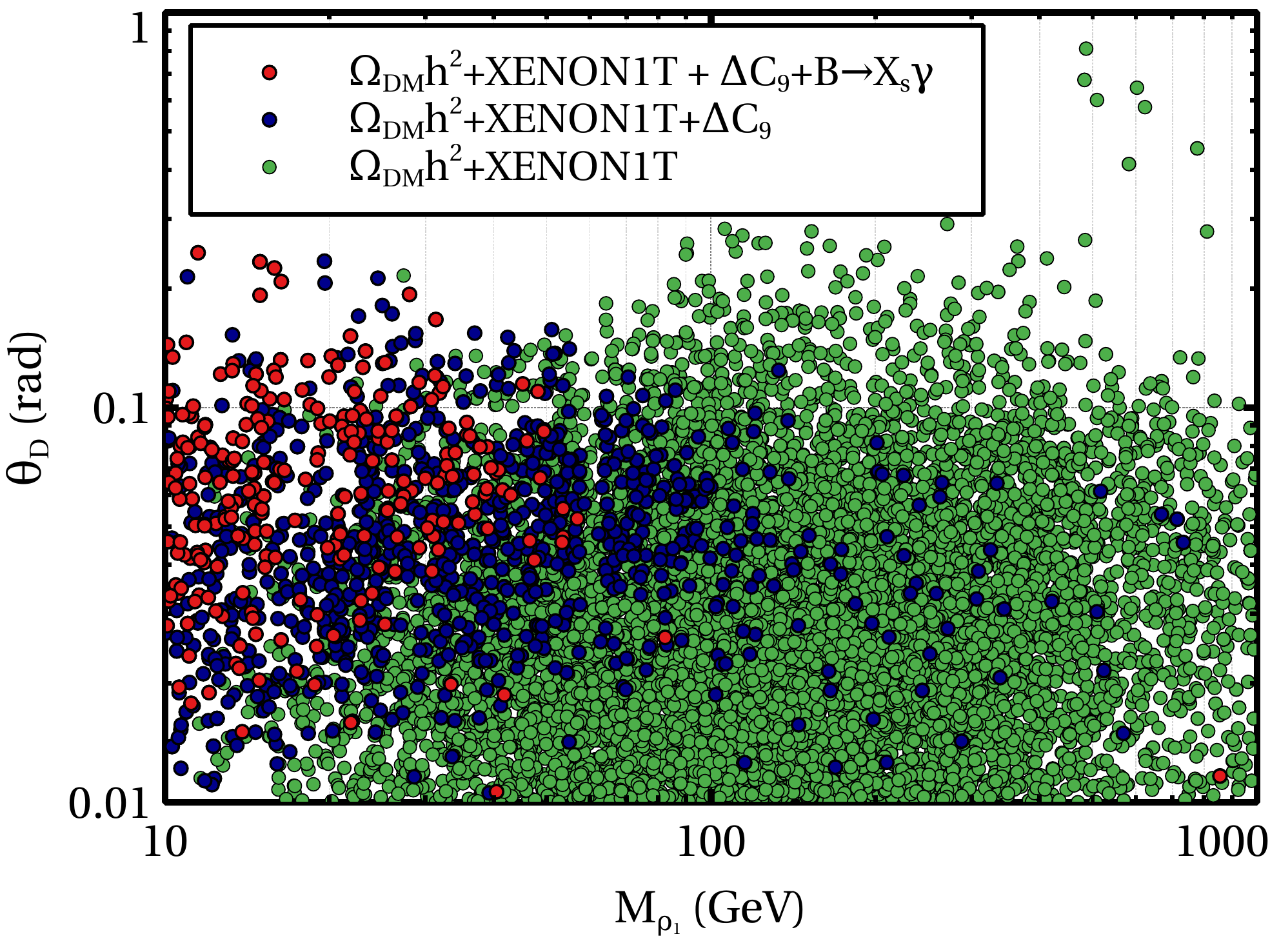}
\caption{Left(Right) panel: Allowed region in
$M_{\rho_2}-M_{\rho_1}$($\theta_D-M_{\rho_1}$) plane from
various experimental results considered in this work.}
\label{Fig:mrho1-mrho2}
\end{figure}

In the left panel of Fig.\,\ref{Fig:mrho1-mrho2}, we have
shown ranges of $M_{\rho_1}$ and $M_{\rho_2}$ allowed
by various experimental results. The allowed region in
$M_{\rho_2}-M_{\rho_1}$ plane from both relic
density as well as direct detection bounds are indicated
by the green coloured points. Similar to the previous
plot in Fig.\,\ref{Fig:mdm-sigmaSI}, here also when we have imposed
various flavour physics constraints, the allowed parameter
space shrinks to a smaller region concentrated mainly
in the low mass regime of $\rho_1$. The parameter
space which reproduces $\Delta{C_9}$
in $2\sigma$ range for explaining $R_{K^{(*)}}$ anomalies has been
shown by the blue coloured points. On the other hand, the red coloured
points are indicating those values of $M_{\rho_1}$ and $M_{\rho_2}$
which in addition to above mentioned experimental results also
satisfy Br($B\rightarrow X_s\gamma$) in $3\sigma$ range. Moreover,
as we have already known that the dark matter candidate $\rho_1$ is an admixture
of a real scalar singlet $S$ and a CP-even neutral component ($\phi^0$) of a
doublet $\Phi$. While both $S$ and $\Phi$ are $\mathbb{Z}_2$-odd
but only $\Phi$ has nonzero $L_{\mu}-L_{\tau}$ charge. Therefore, the
interaction of $\rho_1$ with $L_{\mu}-L_{\tau}$ gauge boson $Z_{\mu\tau}$
(e.g. annihilation of $\rho_1$ into a pair of $Z_{\mu\tau}$) is governed
by the mixing angle $\theta_D$. Larger the mixing angle, larger is the annihilation
rate into $Z_{\mu\tau}Z_{\mu\tau}$ final state, making $\rho_1$ less abundant
at the present epoch. Therefore, the relic density bound puts an upper limit
on the maximum allowed value of $\theta_D$, which is more stringent
in the low dark matter mass region where $Z_{\mu\tau}Z_{\mu\tau}$
is the principal annihilation mode. This feature is clearly visible
in the right panel of Fig.\,\ref{Fig:mrho1-mrho2}, where we
have shown the allowed range of $\theta_D$ with respect to $M_{\rho_1}$.
However, in the high mass regime ($M_{\rho_1}\geq500$ GeV), large values of $\theta_D\ga0.3$ rad
are still allowed because for such large $\theta_D$, $\rho_1$ is mostly an ${\rm SU(2)}_L$
doublet like state (similar to the Inert Doublet dark matter in high
mass range \cite{Hambye:2009pw, Chakrabarty:2015yia, Biswas:2017dxt})
which attains the present abundance of dark matter through co-annihilations with other
$\mathbb{Z}_2$-odd fields into various bosonic final states (both vector and scalar).
Moreover, we have also seen from the Fig.\,\ref{Fig:mdm-sigmaSI} 
that the magnitude of $\Delta{C_9}$ (Eq.\,(\ref{dc9})) decreases with the increasing values of masses
of the particles $\rho_1$, $\rho_2$ and $\rho_3$ involving
within $b\rightarrow s$ transition loops. Now, although the masses of
$\rho_1$ and $\rho_2$ are indeed free parameters of the
present model, the mass of the remaining scalar $\rho_3$
becomes fixed for a particular choice of $M_{\rho_1}$
$M_{\rho_2}$ and $\theta_D$ via Eq.\,(\ref{mho3-relation}).
Here, $M_{\rho_3}$ actually oscillates between
$M_{\rho_2}$ and $M_{\rho_1}$ as we vary $\theta_D$
from $0$ to $\pi/2$. As we are working
in the limit $M_{\rho_1}<M_{\rho_2}$ (since $\rho_1$ is our
dark matter candidate), large $\theta_D$
ensures low mass for $\rho_3$ (using Eq.\,(\ref{mho3-relation}))
and hence enlarge loop contribution to $\Delta{C_9}$.
Thus, $R_{K^{(*)}}$ anomalies prefer larger values of $\theta_D$,
which is a contrasting situation compared to the low mass regime
of $\rho_1$, where relic density bound favours relatively smaller
values of mixing angle to suppress large
annihilation into $Z_{\mu\tau}Z_{\mu\tau}$. As a result, both dark matter
relic density bound and $R_{K^{(*)}}$ anomalies are simultaneously
addressable for $0.01< \theta_D\,({\rm rad})< 0.3$, when
$M_{\rho_1}$ is mostly concentrated below 100 GeV range. This has been
demonstrated by the blue coloured points in $\theta_D-M_{\rho_1}$ plane.
Similar to the left panel, here also red colour points represent the portion in the parameter space which has been satisfied by the constraint of Br($B\rightarrow X_s\gamma$) as well.

\begin{figure}[h!]
\includegraphics[height=8cm,width=10cm,angle=0]{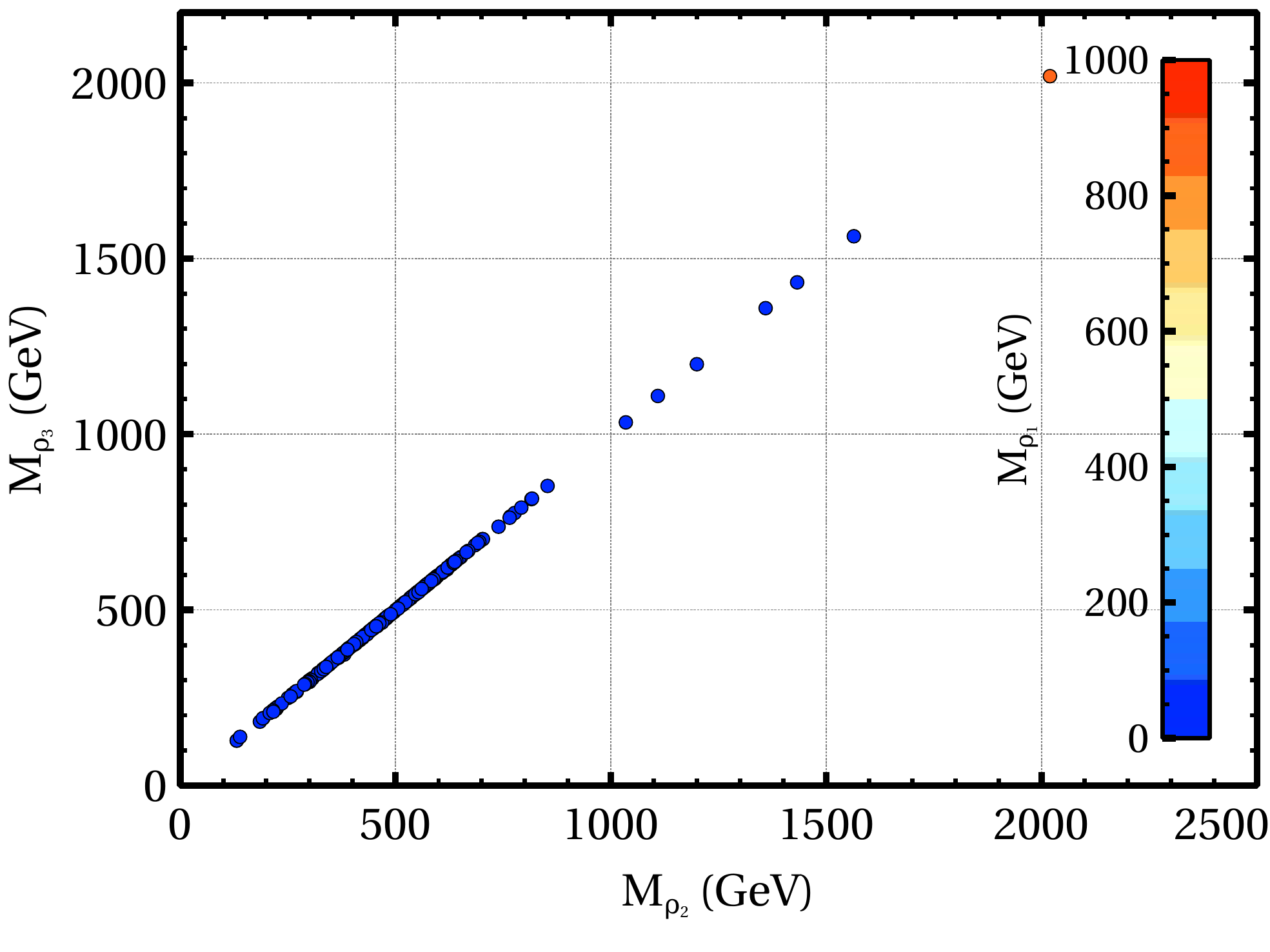}
\caption{Allowed values of $M_{\rho_2}$ and $M_{\rho_3}$
satisfying all the considered experimental constraints. Degeneracy
between $M_{\rho_2}$ and $M_{\rho_3}$ is indicating the fact that
low values of $\theta_D$ ($0.01< \theta_D\,({\rm rad})
< 0.3$) are only allowed while dark matter mass
$M_{\rho_1}$ lies below 100 GeV (shown in colour code).}
\label{Fig:Mrho2-Mrho3}
\end{figure}
Since, the allowed values of $\theta_D$ which satisfy all
the experimental results considered in this work fall in the
range $0.01< \theta_D\,({\rm rad})< 0.3$ (red coloured points
in the right panel of Fig.\,\ref{Fig:mrho1-mrho2}), this makes $M_{\rho_2}$
and $M_{\rho_3}$ almost degenerate and this has been demonstrated
in Fig.\,\ref{Fig:Mrho2-Mrho3}, where the colour bar is indicating
corresponding values of mass of the dark matter candidate $\rho_1$.

\begin{figure}[h!]
\includegraphics[height=10cm,width=12cm,angle=0]{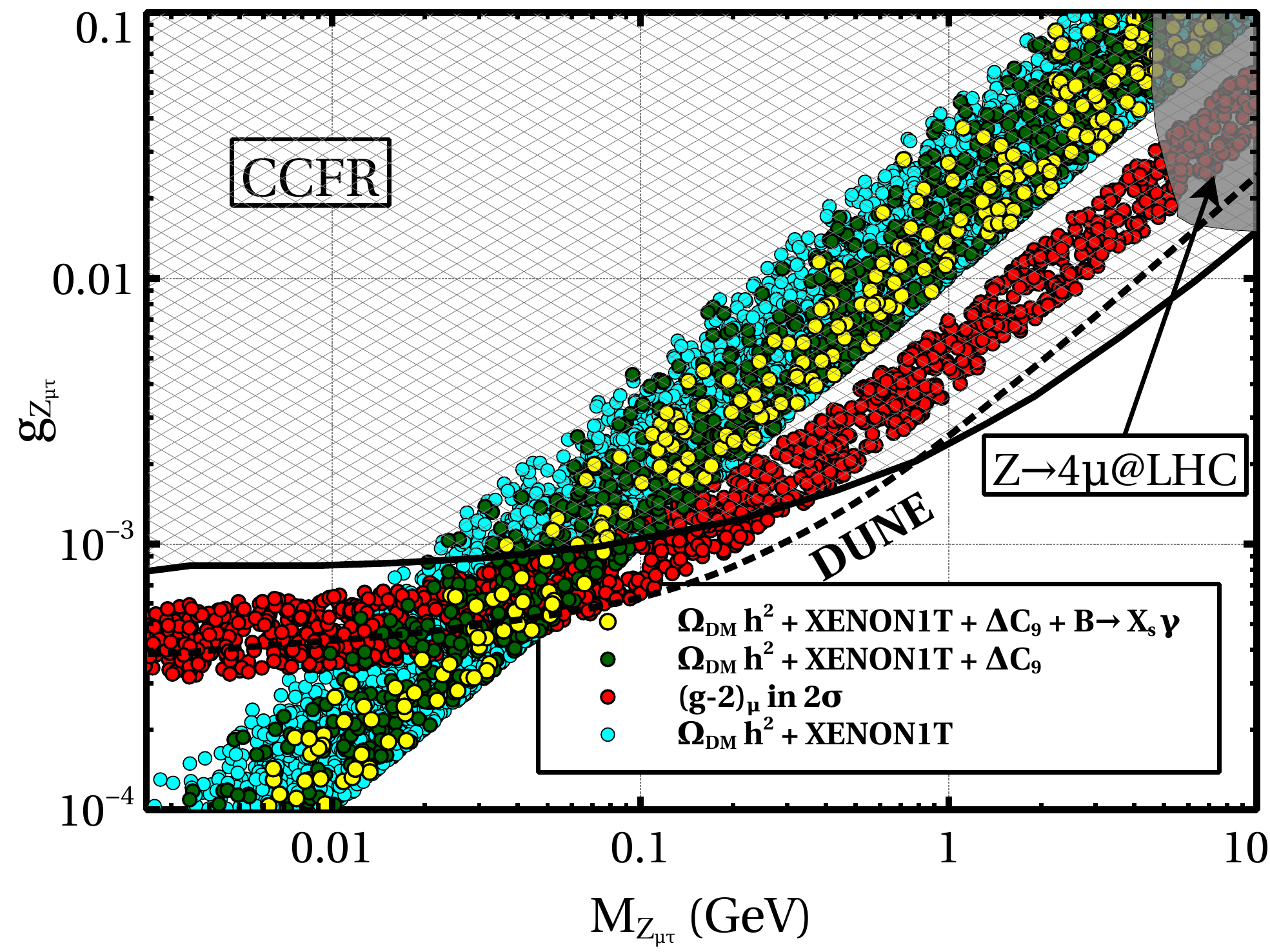}
\caption{Current status of the $g_{Z_{\mu\tau}}-M_{Z_{\mu\tau}}$
plane in the light of various experimental results. In this plane
we have shown the allowed regions satisfying bounds from Planck + XENON1T (cyan coloured
points), Planck + XENON1T + $R_{K^{(*)}}$ anomalies (green coloured points) and
Planck + XENON1T + $R_{K^{(*)}}$ anomalies + Br($B \rightarrow X_s \gamma$)
(yellow coloured points) respectively. Moreover, red coloured points are
indicating those values of $M_{Z_{\mu\tau}}$ and
$g_{Z_{\mu\tau}}$ which address $(g-2)_\mu$ in
$2\sigma$ range. In this plot, we have also taken into account
the invisible decay branching constraint of the SM-like
Higgs boson $H_1$.}
\label{Fig:mzp-gzp}
\end{figure}

Finally, in Fig.\,\ref{Fig:mzp-gzp} we have shown our results in $g_{\mu\tau}-M_{Z_{\mu\tau}}$ plane, which is at the present moment extremely constrained by
various experimental results. In this figure (Fig.\,\ref{Fig:mzp-gzp}), the red coloured points represent those values of $g_{\mu\tau}$ and $M_{Z_{\mu\tau}}$ which
explain $(g-2)_\mu$ in $2\sigma$ range. Here, in
the $g_{\mu\tau}-M_{\mu\tau}$ plane, most strongest constraint till now
comes from neutrino trident production. Neutrino trident production is a
process of producing $\mu^+\mu^-$ pair via neutrino scattering in the Coulomb
field of a target nucleus ($N$), i.e. $\nu_{\mu} (\overline{\nu_{\mu}}) + N \rightarrow
\nu_{\mu} (\overline{\nu_{\mu}}) + \mu^+ \mu^- + N$. In the SM, this process
is possible via $W^\pm$ and $Z$ bosons only. Moreover, if there exists any new
neutral gauge boson (similar to $Z_{\mu\tau}$ in the present work) which couples
to both muons and muon-neutrinos then that gauge boson can also contribute significantly to
the trident production cross section. However, all the experimental collaborations
namely, CCFR \cite{Mishra:1991bv}, CHARM-II \cite{Geiregat:1990gz} and NuTeV \cite{Adams:1999mn}
have measured neutrino trident events and their measured cross sections are in good agreement with that
of the SM prediction i.e. $\dfrac{\sigma_{\rm CCFR}}{\sigma_{\rm SM}} = 0.82\pm 0.28$,
$\dfrac{\sigma_{\rm CHARM-II}}{\sigma_{\rm SM}} = 1.58\pm 0.57$ 
and $\dfrac{\sigma_{\rm NuTeV}}{\sigma_{\rm SM}} = 0.72^{+1.73}_{-0.72}$ respectively.
These results therefore put strong constraint in the mass-coupling plane of the 
new gauge boson. In Fig.\,\ref{Fig:mzp-gzp}, the crossed region
above the black dashed line represents 95\% C.L. upper bound \cite{Altmannshofer:2014pba}
on $g_{\mu\tau}$ as a function of $M_{Z_{\mu\tau}}$
using neutrino trident cross section measured by
the CCFR collaboration\footnote{Furthermore, it is clearly evident from the Fig.\,\ref{Fig:mzp-gzp}, that, due to the consideration of CCFR experimental data we naturally incorporate the constraint of the branching ratio of $\tau\to\mu\nu_\tau\bar{\nu}_\mu$. The reason is that the parameter space (in $g_{\mu\tau}-M_{Z_{\mu\tau}}$ plane) which describes all the concerned observables simultaneously does not overlap with the portion that has already been ruled out from the branching ratio of $\tau\to\mu\nu_\tau\bar{\nu}_\mu$ \cite{Altmannshofer:2014cfa}. Moreover, we have explicitly checked that the NP contribution for the decay $\tau\to\mu\nu_\tau\bar{\nu}_\mu$ due to $Z_{\mu\tau}$ is practically vanishing in nature in the allowed parameter space of the present scenario.}. Consequently, all the crossed regions above black dashed line are excluded by neutrino trident production. Besides, there is a further constraint from the measurement of the SM $Z$ boson decay to $4\mu$ final state at the LHC. This has also been indicated by the
grey region at the topmost corner of right side of this plot.
Cyan coloured points represent those values of $g_{Z_{\mu\tau}}$ and $M_{Z_{\mu\tau}}$
which satisfy bounds related to dark matter physics namely, relic density,
direct detection and Higgs invisible branching ratio. On top of 
the existing dark matter constraints, the effects of flavour physics
observables like $R_{K^{(*)}}$ anomalies ($2\sigma$ bound on $\Delta C_9$)
and $R_{K^{(*)}}$ + Br($B \rightarrow X_s \gamma$) on the mass as well 
as the coupling of $Z_{\mu\tau}$ have been shown by green and
yellow coloured points respectively. Therefore, from this plot
it can be easily seen that although maximum portions of $g_{\mu\tau}-M_{Z_{\mu\tau}}$
plane have already been excluded by the results of CCFR collaboration,
there is still a small but interesting region left in this parameter
space which is $0.01 \leq M_{Z_{\mu\tau}}\,({\rm GeV})\leq 0.1$ and
$3\times 10^{-4}\leq g_{\mu\tau} \leq 10^{-3}$. This region of the parameter space of the present model can address dark matter, $(g-2)_{\mu}$ anomaly,
$R_{K^{(*)}}$ anomalies and Br($B \rightarrow X_s \gamma$) simultaneously
and more exciting thing is that this parameter space can be probed
within a few years by the DUNE experiment \cite{Acciarri:2015uup} measuring neutrino
trident events (shown by black dashed line) \cite{Altmannshofer:2019zhy}.
This will surely be the test of our model, at least the benchmark
points (if not the full model) in the low mass dark matter region
which are compatible to both dark matter and
flavour physics issues. For completeness in Table \ref{tab:BP}, we present three plausible benchmark points (BP1, BP2 and BP3) and corresponding numerical values of several physical quantities of the present scenario.
\begin{table}[h!]
\begin{center}
\vskip 0.5cm
\begin{tabular} {|c|c|c|c|}
\hline
Parameters/& BP1& BP2 & BP3\\
Observables & & &\\
\hline
\hline
$M_{\rho_1}$ (GeV) & 14.499& 26.515 & 36.767 \\
$M_{\rho_2}$ (GeV) & 478.254& 506.009 & 450.276  \\
$M_{\rho_3}$ (GeV) & 475.201& 503.742 & 449.255\\
$M_{\phi^{\pm}}$ (GeV) & 160.591& 121.443 & 101.748\\
$M_{H_2}$ (GeV) & 353.418& 401.503 & 352.41\\
$M_{\chi}$ (GeV) &1107.840 & 1300.660 & 1087.52\\
$M_{Z_{\mu\tau}}$ (GeV) & $5.052\times10^{-2}$ & $7.577\times10^{-2}$
& $3.167\times10^{-2}$\\
$v_2$ (GeV) & 76.328& 81.151 & 71.229\\
$g_{Z_{\mu\tau}}$ & $6.619\times10^{-4}$& $9.339\times10^{-4}$&
$4.447\times10^{-4}$\\
$\tan \theta_{\mu\tau}$ & $2.752\times10^{-6}$&$1.637\times10^{-5}$&
$5.337\times10^{-6}$ \\
$\tan \theta_{D}$ & 0.1135 &  $9.511\times10^{-2}$&
$6.769\times10^{-2}$ \\
$\tan \theta_{s}$ & $3.203\times10^{-4}$& $9.893\times10^{-4}$ &
$4.643\times10^{-3}$ \\
$\lambda_{\Phi}$  & 0.1 & 0.1&0.1\\
$\lambda_{S}$  & 0.1&  0.1 &0.1\\
$\lambda_{2}$  & $9.520\times10^{-3}$ & $7.935\times10^{-2}$ &
$2.360\times10^{-4}$ \\
$\lambda_{4}$  & $2.499\times10^{-3}$& $9.691\times10^{-2}$ &
$1.128\times10^{-3}$ \\
$\lambda_{5}$  & $9.236\times10^{-3}$ & $1.994\times10^{-4}$ &
$1.066\times10^{-3}$ \\
$\lambda_{6}$  & $5.364\times10^{-4}$ & $1.835\times10^{-2}$ &
$8.175\times10^{-4}$ \\
$\lambda_{7}$  & $4.724\times10^{-2}$ & $2.243\times10^{-3}$ &
$2.599\times10^{-4}$ \\
$f_2 \times f_3$ & 1.657 &2.533& 3.228\\
$\Omega_{\rm DM}h^2$ & 0.1218 & 0.1206 & 0.1213\\
$\sigma_{\rm SI}$ (cm$^2$)& $5.480\times10^{-47}$& $1.688\times10^{-47}$ &
$1.076\times10^{-48}$ \\
${\rm Br}(\Gamma^{\rm Inv}_{H_1})$ & $1.639\times10^{-4}$& $1.954\times 10^{-3}$&
$2.094\times10^{-2}$\\
$\Delta{C_9}$ & -0.973&  -0.7578 & -0.684\\
${\rm Br}(B \rightarrow X_s \gamma)$ &$3.196\times 10^{-4}$
& $3.173\times10^{-4}$ & $2.974\times10^{-4}$\\
$\Delta{a_{\mu}}$ & $218.495\times10^{-11}$ & $311.557\times10^{-11}$
& $129.438\times10^{-11}$\\
\hline
\hline
\end{tabular}
\end{center}
\caption{Viable benchmark points (BP1, BP2 and BP3) and corresponding numerical values of several physical quantities of the present scenario.}
\label{tab:BP}
\end{table} 

\section{Neutrino masses and mixings}\label{neu}
In this section, we will discuss briefly about neutrino masses
and mixings. It has now been firmly established from the phenomena of
neutrino oscillations that there exist two tiny mass square differences
between three neutrino mass eigenstates i.e. $\Delta{m^2_{21}}
=7.39^{+0.21}_{-0.20} \times 10^{-5}$ eV$^{2}$, \footnote{$\Delta{m^2_{ij}}$
is defined as $m^2_i-m^2_j$.} and $\Delta{m^2_{31}} = 2.525^{+0.033}_{-0.032}
(-2.512^{+0.034}_{-0.032}) \times 10^{-3}$ eV$^{2}$
for the normal(inverted) hierarchy \cite{Esteban:2018azc} in $3\sigma$
range. This also indicates that to explain solar, atmospheric
and rector neutrino anomalies though three flavour neutrino
oscillation we need at least two neutrino mass eigenstates
having nonzero masses corresponding to mass squared differences
as mentioned above. Moreover, there are also precise measurements
of three intergenerational mixing angles namely the atmospheric
mixing angle ($40.3^{\degree}(40.6^{\degree})\leq\theta_{23}
\leq 52.4^{\degree}(52.5^{\degree})$)\footnote{Where numbers without(within)
brackets are for the normal(inverted) hierarchical scenario.},
the solar mixing angle ($31.61^{\degree}\leq\theta_{12}\leq 36.27^{\degree}$)
and the reactor mixing angle ($8.22^{\degree}(8.27^{\degree})\leq\theta_{13}
\leq 8.99^{\degree}(9.03^{\degree})$) \cite{Esteban:2018azc}. The latter
one is the most recent entry in that list. In the present model, although
we do not need any extra fermionic degrees of freedom to cancel
$L_{\mu}-L_{\tau}$ anomaly which actually cancels between $\mu$
and $\tau$ generations of charged leptons and corresponding neutrinos,
one can still introduce three right handed neutrinos $N_{Ri}$ ($i=e\,,\mu,\,,\tau$)
in an anomaly free manner, in the model, to address neutrino masses and mixings
via Type I seesaw mechanism. The Lagrangian for right handed neutrinos
are given in Eq.\,(\ref{lagN}). The light neutrino mass matrix $m_{\nu}$ after
spontaneous breaking of both SU(2)$_{L}\times {\rm U(1)}_Y$ and U(1)$_{L_{\mu}-L_{\tau}}$
symmetries has the following structure
\begin{eqnarray}
m_{\nu} &=& -{M_D}\,\mathcal{M_R}^{-1}\,M_D^T\,, \nonumber \\
 &=& \dfrac{1}{2\,p} \left(\begin{array}{ccc}
y_{e}^{2}M_{\mu \tau}\,v^2_1\,e^{i\xi}  &
-\dfrac{y_{e}\,y_{\mu}\,y_{e \tau} v^2_1\,v_2}{\sqrt{2}}\, &
-\dfrac{y_{e}\,y_{\tau}\,y_{e \mu} v^2_1\,v_2}{\sqrt{2}}\\
-\dfrac{y_{e}\,y_{\mu}\,y_{e \tau} v^2_1\,v_2}{\sqrt{2}}\, &
\dfrac{y_{\mu}^{2}\,y_{e\tau}^2\,v_1^2\,v_2^2\,e^{-i\xi}}{2\,M_{\mu\tau}} &
\dfrac{y_{\mu}\,y_{\tau}\,v_1^2}{2\,M_{\mu\tau}}(M_{ee}\,M_{\mu\tau}-p\,e^{-i\xi})\\
-\dfrac{y_{e}\,y_{\tau}\,y_{e \mu} v^2_1\,v_2}{\sqrt{2}}\, &
\dfrac{y_{\mu}\,y_{\tau}\,v_1^2}{2\,M_{\mu\tau}}(M_{ee}\,M_{\mu\tau}-p\,e^{-i\xi})  &
\dfrac{y_{\tau}^{2}\,y_{e\mu}^2\,v^2_1\,v_2^2\,e^{-i\xi}}{2\,M_{\mu\tau}} \\
\end{array}\right) \,\,,
\label{mass-matrix}
\end{eqnarray}
while the mass matrix for the heavy neutrinos coincides with $\mathcal{M}_R$. 
In the above, $p=y_{e\mu}\,y_{e\tau}\,v_2^2-M_{ee}\,M_{\mu\tau}\,e^{i\xi}$. 
Majorana mass matrix $\mathcal{M}_R$ and Dirac mass matrix
$M_D$ are given by,
\begin{eqnarray}
\mathcal{M}_{R} = \left(\begin{array}{ccc}
M_{ee} ~~&~~ \dfrac{v_2}{\sqrt{2}} y_{e \mu}
~~&~~\dfrac{v_2}{\sqrt{2}} y_{e \tau} \\
~~&~~\\
\dfrac{v_2}{\sqrt{2}} y_{e \mu} ~~&~~ 0
~~&~~ M_{\mu \tau} \,e^{i\xi}\\
~~&~~\\
\dfrac{v_2}{\sqrt{2}} y_{e \tau} ~~&
~~ M_{\mu \tau}\,e^{i\xi} ~~&~~ 0 \\
\end{array}\right) \,,\,\,\,\,
M_{D} = \dfrac{v_1}{\sqrt{2}}\left(\begin{array}{ccc}
y_e ~~&~~ 0 ~~&~~ 0 \\
~~&~~\\
0 ~~&~~ y_{\mu} ~~&~~ 0 \\
~~&~~\\
0 ~~&~~ 0 ~~&~~ y_{\tau} \\
\end{array}\right) \,.
\label{MR-md}
\end{eqnarray}
In the present case, due to $L_{\mu}-L_{\tau}$ flavour
symmetry, the Dirac mass matrix is exactly diagonal while
before U(1)$_{L_{\mu}-L_{\tau}}$ symmetry breaking only
three elements (only two are independent) are there in the
Majorana mass matrix $\mathcal{M}_{R}$. Only after symmetry
breaking we get additional elements proportional to $v_2$. Therefore,
$L_{\mu}-L_{\tau}$ symmetry breaking plays a crucial role here to get
desire structure of $m_{\nu}$ matrix. Also, looking at
both $M_D$ and $\mathcal{M}_R$ matrices, one can easily notice
that there can only be one complex element. Phases of other elements
can be absorbed by redefining both SM leptons and right handed
neutrinos. Now, one can calculate mass eigenvalues and mixing angles of light
neutrinos by diagonalising this $m_{\nu}$ matrix, which is a
complex symmetric matrix, indicating the Majorana nature
of the light neutrinos. If we consider, $v_2\sim 10^2$ GeV
(in the right ballpark to produce desired contribution to $(g-2)_{\mu}$),
$0.1 \la y_{e\mu},\,y_{e\tau} \la 1.0$ and $10\,{\rm GeV} \la M_{ee},\,M_{\mu\tau} \la$1 TeV
(100 GeV to TeV scale right handed neutrinos)
then we need Dirac couplings $10^{-7} \la y_{e},\,y_{\mu},\,y_{\tau}\la 10^{-5}$
to reproduce neutrino oscillation parameters.
Detail analysis of mass matrix diagonalisation and
comparison with latest 3$\sigma$ range of oscillation
parameters have been done in Ref. \cite{Biswas:2016yan}.
Moreover, we would like to mention here that although only two right handed neutrinos ($N^{\mu}_R$
and $N^{\tau}_R$) are sufficient to make the present model anomaly free,
such scenario is unable to reproduce all neutrino oscillation
parameters due to special flavour structure of the Dirac mass matrix.

\section{Constraint from di-lepton resonance search at 13 TeV LHC}\label{cldr}
Depending on the mass ranges, the {\it non-standard} $Z$ boson (which we designate as $Z_{\mu\tau}$ in this article) confronts constraints from collider searches. For example, if the mass of $Z_{\mu\tau}$ is less then SM $Z$ boson then there exists some viable parameter region for the favorable kind among the various NP models that exist in the literature. Furthermore, as the $Z_{\mu\tau}$ has no direct coupling with electron\footnote{Only possible via $Z$-$Z_{\mu\tau}$ mixing. Therefore, the interaction strength is insignificant.}, hence LEP searches cannot provide direct constraint on the light $Z_{\mu\tau}$. On the other hand, the Tevatron~\cite{Abazov:2010ti, Aaltonen:2011gp} and LHC~\cite{Khachatryan:2016zqb, Aaboud:2017buh, ATLAS:2019vcr} searches for $Z_{\mu\tau}$ to di-lepton final state only apply, however in this case $M_{Z_{\mu\tau}}>100~$GeV. Moreover, only relevant limit to the light $Z_{\mu\tau}$ case obtained from the LHC searches at $ p p \to Z\to 4\mu$\cite{Altmannshofer:2014pba}.  At this point we remark in passing that, in our present article even though in the low mass limit of $Z_{\mu\tau}$ we have obtained certain
region of parameter space (depicted in Fig.\,\ref{Fig:mzp-gzp}) which has been satisfied by
some flavour physics data, dark matter constraints and $(g-2)_\mu$ anomaly, however, cross section for a process like $pp \to Z_{\mu\tau} \to \ell^+ \ell^-$ in that region of
parameter space is extremely tiny at the 13 TeV LHC. 

On the other hand in the high mass region of $Z_{\mu\tau}$, the LHC searches put the tightest bound on its mass ($3\sim5$ TeV ~\cite{Khachatryan:2016zqb, Aaboud:2017buh, ATLAS:2019vcr}) in the di-muon final states. Thus, in the present article we use the exclusion data obtained by ATLAS collaboration \cite{ATLAS:2019vcr} for a di-lepton resonance search at the LHC experiment to constraint parameter space of the present scenario. In order to embed this limit in the present scenario, we first implement the model using {\tt FeynRules}~\cite{Alloul:2013bka}. Then we generate the cross section for the process $pp \to Z_{\mu\tau} \to \ell^+ \ell^-$ using {\tt Madgraph5}~\cite{Alwall:2014hca} with the default parton distribution functions {\tt NNPDF3.0}~\cite{Ball:2014uwa} at 13 TeV LHC\footnote{Production of $Z_{\mu\tau}$ at the LHC in the present model is possible due to the couplings of $q_i \bar{q_i}Z_{\mu\tau}$ which are generated via $Z$-$Z_{\mu\tau}$ mixing.}. Here $\ell(\equiv e, \mu)$, however, significant contribution has been generated from $\mu^+\mu^-$ final state. Finally, for a specific combination of coupling $g_{Z_{\mu\tau}}$ and $Z$-$Z_{\mu\tau}$ mixing angle $\theta_{\mu\tau}$ we compare the theoretical prediction of cross section for any particular value of mass (confined within the range [0.5, 5] TeV) of $Z_{\mu\tau}$  with the corresponding experimental data given by ATLAS collaboration \cite{ATLAS:2019vcr}.

\begin{figure}[t]
\begin{center}
\includegraphics[height=9cm,width=12cm,angle=0]{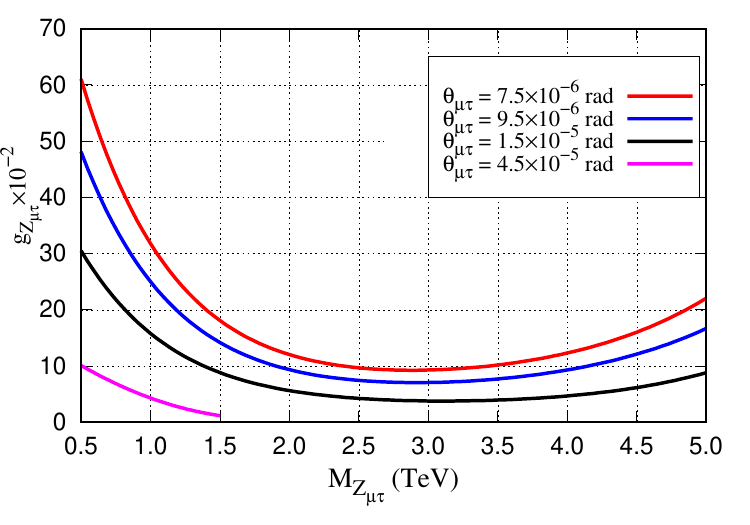}
\caption{Using the non-observation of a resonant $\ell^{+}\ell^{-}$ signal at the LHC running at 13 TeV, we have depicted the exclusion plots at 95\% C.L. in the  $M_{Z_{\mu\tau}}-g_{Z_{\mu\tau}}$ plane for four different values of $Z$-$Z_{\mu\tau}$ mixing angle $\theta_{\mu\tau}$. The region above a particular curve has been ruled out from the non-observation of a resonant $\ell^{+}\ell^{-}$ signal in the 13 TeV run of LHC by latest ATLAS data \cite{ATLAS:2019vcr} considering mass range [0.5, 5] TeV.}
\label{mz_gz}
\end{center}
\end{figure}

In Fig.~\ref{mz_gz} we show the exclusion curves at 95\% C.L. in the $M_{Z_{\mu\tau}}-g_{Z_{\mu\tau}}$ plane for four different values of $Z$-$Z_{\mu\tau}$ mixing angle $\theta_{\mu\tau}$ using the ATLAS data \cite{ATLAS:2019vcr} for non-observation of a resonant $\ell^{+}\ell^{-}$ signal at the LHC running at 13 TeV with integrated luminosity 139 ${\rm fb}^{-1}$. In this case the region above a particular curve has been ruled out at 95\% C.L. from the non-observation of a resonant $\ell^{+}\ell^{-}$ signal in the 13 TeV run of LHC by ATLAS data \cite{ATLAS:2019vcr}. If we focus on a particular curve fixed by a particular value of mixing angle $\theta_{\mu\tau}$ then we observe that for the lower values of mass the coupling $g_{Z_{\mu\tau}}$ rapidly falls with the increasing values of mass $M_{Z_{\mu\tau}}$. This phenomena can be explained in the following way. In the lower mass range if we vary the mass then the cross section does not fall rapidly as desired by the ATLAS data. Hence, to acquire the proper cross section for a particular mass one should decrease the value of the coupling $g_{Z_{\mu\tau}}$. Once the lower mass range is over then with the increasing values of mass the curve exactly replicates the exclusion plot as given in \cite{ATLAS:2019vcr}. At this point, we would like to mention another notable feature of the exclusion curves (which is true for all over the mass range) that for a fixed value of mass if the mixing angle increases then to satisfy ATLAS data \cite{ATLAS:2019vcr} one requires decreasing values of coupling constant $g_{Z_{\mu\tau}}$. Furthermore, it is clearly evident form the Fig.~\ref{mz_gz} that as the mixing angle $\theta_{\mu\tau}$ increases large amount of area in the $M_{Z_{\mu\tau}}-g_{Z_{\mu\tau}}$ plane has been ruled out by the ATLAS data. Both of the features can be explained, if we analyse the structure of the coupling\footnote{The relevant couplings have been given in Appendix~\ref{Dmcouplings}.} between ${Z_{\mu\tau}}$ and $\ell^+ \ell^-$. If we decompose the coupling then we can find that there is one vectorial part and other is axial vectorial in nature. The latter one has no significant role in the concerned process but totally controlled by the vectorial part. We have also checked that, one can control the coupling (which in turn the vectorial part of the coupling) that satisfy the exclusion data with lower values of mixing angle $\theta_{\mu\tau}$. However, as the mixing increases then one looses the control over the coupling, i.e., there is no variation of coupling with larger mixing angle. Therefore, with larger mixing angle one can not vary the cross section properly, hence one can not have the required cross section for a particular mass. For example, if the mixing is set at $4.5\times 10^{-4}$ rad, then one can not go beyond 1500 GeV mass of $M_{Z_{\mu\tau}}$. Since in this situation after 1500 GeV mass we can not have the desired cross section by changing the value of $g_{Z_{\mu\tau}}$. Therefore, in order to translate the exclusion limit obtained by ATLAS data \cite{ATLAS:2019vcr} for non-observation of di-lepton resonance search at the LHC experiment in our model we have restricted ourselves within the relatively smaller values of mixings angle $\theta_{\mu\tau}$.

\section{Conclusions}\label{con}
In order to simultaneously resolving $R_{K^{(*)}}$ anomalies and
dark matter enigma, we have proposed a unified scenario by introducing an
extra local ${\rm U(1)}_{L_{\mu}-L_{\tau}}$ gauge symmetry to the Standard Model.
This ${\rm U(1)}_{L_{\mu}-L_{\tau}}$ gauge symmetry provides a neutral non-standard
gauge boson $Z_{\mu\tau}$ which has versatile effects on different phenomenological
aspects that have been considered in this article. For the purpose of breaking of
the ${\rm U(1)}_{L_{\mu}-L_{\tau}}$ symmetry spontaneously a complex scalar field
$\eta$ has been invoked to the scalar sector in addition to the usual Standard Model
Higgs doublet $H$. Three singlet right handed neutrinos have also been introduced
in order to explain the observed oscillation data by incorporating neutrino masses
and mixings via Type-I seesaw mechanism. Furthermore, for the proper establishment
of correlation between $R_{K^{(*)}}$ anomalies and dark matter puzzle, a bottom quark
like coloured fermion field $\chi$ has been included in this scenario. This non-standard
fermion field $\chi$ is transformed vectorially under the ${\rm U(1)}_{L_{\mu}-L_{\tau}}$
symmetry and further it is odd under the $\mathbb{Z}_2$ parity. Apart from these,
an SU(2)$_L$ scalar doublet $\Phi$ with nonzero ${\rm U(1)}_{L_{\mu}-L_{\tau}}$
charge and a real scalar singlet $S$ have also been incorporated in the present
scenario. Both of these non-standard scalar fields are odd under $\mathbb{Z}_2$ symmetry.
The mixing (which is parametrised by a mixing angle $\theta_D$) between these two
$\mathbb{Z}_2$-odd scalar fields gives a potential dark matter candidate $\rho_1$
and also two heavier $\mathbb{Z}_2$-odd physical particles $\rho_2$ and $\rho_3$.
All of these three scalar fields provide significant contributions not only in dark
matter phenomenology but also in rare $B$-meson decay processes. 

Existence of lepton flavour universality violation in neutral current sector has
been measured by $R_{K^{(*)}}$ in which $b\to s \ell^+\ell^-$ ($\ell\equiv e, \mu$)
transition is involved. This type of flavour changing neutral current is highly
suppressed in the Standard Model and therefore, even for a small deviation between
the experimental data and the Standard Model could play significant role for finding
of new physics effects. In this work, the introduced new physics particles have played
crucial role in the concerned $b\to s$ transition processes which are in general loop
induced\footnote{Apart from leptoquark scenarios where $b\to s$ transition is possible at tree level.}. Particularly, the dark matter particle $\rho_1$ with two heavier $\mathbb{Z}_2$-odd
neutral scalar fields $\rho_2$, $\rho_3$ and the non-standard fermion $\chi$
generate extra loop contributions. Furthermore, the extra non-standard gauge
boson $Z_{\mu\tau}$ behaves as a propagator (in addition to the SM $Z$ boson)
for the process $b\to s \ell^+\ell^-$. Now, due to the very basic structure of
our model, the process $b\to s \mu^+\mu^-$ is more favourable with respect to
$b\to s e^+e^-$, consequently one obtains the significant non-standard
contribution to the Wilson coefficients $C^{\rm NP}_{9}$ for ``$\mu$" but not for ``$e$".
Therefore, in our work, we have easily satisfied the current fit result for
$C^{\rm NP,\mu}_{9} \in [-1.26, -0.63]$ in $2\sigma$ interval to explain
the $R_{K^{(*)}}$ anomalies and thereby we have constrained the parameter
space of the proposed scenario. On top of that, we have also calculated
another rare decay process $B\to X_s\gamma$ which has also been a class of
processes that characterised by $b\to s$ transition. We have estimated the
branching ratio for $B\to X_s\gamma$ process, and have used the corresponding
experimental data within $3\sigma$ interval as one of the constraints in
our analysis. Moreover, we have calculated the contribution of non-standard
gauge boson $Z_{\mu\tau}$ to the $(g-2)_\mu$ and considering the recent experimental
data with some error bars ($1\sigma$ and $2\sigma$) we have further constrained the
parameter space allowed by dark matter and flavour physics observables.

In the present scenario, we have extensively studied the dark matter phenomenology
by choosing $\rho_1$ as a WIMP type dark matter candidate. This $\rho_1$ is an admixture
of a real scalar singlet $S$ and the CP-even neutral component ($\phi^0$) of the
doublet $\Phi$. In our work, first we have calculated dark matter relic abundance 
by considering all possible annihilation and co-annihilation channels for a
wide range (10 GeV $\leq$ 1 TeV) of the mass of $\rho_1$. Thereafter, we have
imposed necessary constraints like Planck limit on relic density
($0.1172\leq\Omega_{\rm DM} h^2\leq 0.1226$), latest direct detection bounds on
$\sigma_{\rm SI}$ from XENON1T and also the bound on Higgs invisible branching
ratio from LHC to find the allowed parameter space.\,\,We have found that in the
case of low mass region ($M_{\rho_1}< 100$ GeV), our dark matter candidate $\rho_1$
predominantly annihilates into $Z_{\mu\tau}$ pair while co-annihilations among other
$\mathbb{Z}_2$-odd particles are insignificant as we have considered all heavier
$\mathbb{Z}_2$-odd particles masses larger than 100 GeV throughout this analysis
to respect the experimental bounds form LEP collider. Due to this primary
annihilation channel ($\rho_1\rho_1\to Z_{\mu\tau}Z_{\mu\tau}$), in spite of
being a gauge singlet $\mathbb{Z}_2$-odd scalar field, the mixing with another
$\mathbb{Z}_2$-odd field (part of an ${\rm SU(2)}_{L}$ doublet) having nonzero
$L_{\mu}-L_{\tau}$ charge, makes the entire dynamics of our dark matter
candidate $\rho_1$ remarkably different from the standard Scalar Singlet dark matter
scenario where $b\bar{b}$ final state is in general the principal annihilation channel
and low mass region has already been ruled out by direct detection, indirect detection
and also by the upper limit on Higgs invisible decay branching ratio.
On the other hand for the higher values of $M_{\rho_1}$, depending upon the mass
splitting between $\rho_1$ and other $\mathbb{Z}_2$-odd particles several annihilation
or co-annihilation channels may appear and have contributed significantly to
the relic density. Since, one of our prime motivations of this article
is to correlate dark matter puzzle with some specific flavour physics anomalies
associated with FCNC processes, therefore, we have used experimental data of some flavour
physics observables (e.g., $R_{K^{(*)}}$ anomalies and Br($B\to X_s\gamma$)) as
further constraints on the parameter space which is already allowed by experiments
related to dark matter physics. As a consequence, both the effects of $R_{K^{(*)}}$
anomalies and dark matter phenomenology allow only a very restrictive values
of dark sector mixing angle $\theta_D$ which remains confined within a certain
range (0.01$<$ $\theta_D$ (rad) $<$ 0.3) when $M_{\rho_1}\leq 100$ GeV. This is
a unique feature of our proposed model. 

Additionally, we have used some other constraints which have been relevant
to our present scenario. For example, we have imposed constraint from neutrino
trident production and for that purpose we have used the CCFR experimental data
which is currently the most stringent one for the neutrino trident production
process. Furthermore, we have imposed constraint from the measurement of the
Standard Model $Z$ boson decay to 4$\mu$ final state at the LHC. As a consequence
there is a substantial amount of reduction in the parameters space due to the inclusion
of such constraints. However, there still exists a few portion of the parameter space
of the present model which can address dark matter, $R_{K^{(*)}}$ anomalies, $(g-2)_{\mu}$ and
Br($B \rightarrow X_s \gamma$) simultaneously. Most importantly our predicted parameter
space and hence our model can be tested within a few years by neutrino trident processes
at DUNE. Therefore, in view of the above discussion
we can readily conclude that our proposed scenario can reasonably connect the dark matter
puzzle with some of the flavour physics anomalies. Besides, within the scope of our proposed model,
we have also briefly discussed the origin of neutrino masses and mixing angles via
Type-I seesaw mechanism, which is a common feature of most of the ${L_{\mu}-L_{\tau}}$
models. 

Finally, for the purpose of constraining the parameter space of the present scenario from the LHC, we have used the latest ATLAS data of non-observation of a resonant $\ell^{+}\ell^{-}$ signal at the LHC running at 13 TeV with an integrated luminosity 139 ${\rm fb}^{-1}$. For this purpose we have estimated the cross section for the process $pp \to Z_{\mu\tau} \to \ell^+ \ell^-$ at the 13 TeV LHC for the mass range $M_{Z_{\mu\tau}}\in [0.5, 5]$ TeV in the present scenario. By comparing the theoretical predictions of the cross section with corresponding ATLAS data of cross section for non-observation of a resonant $\ell^{+}\ell^{-}$ signal at the 13 TeV LHC one yields some specific combination of coupling $g_{Z_{\mu\tau}}$ and $Z$-$Z_{\mu\tau}$ mixing angle $\theta_{\mu\tau}$. Consequently, with those combinations we have excluded some portion of the parameter space of the present scenario at 95\% C.L. From our analysis it has been observed that, for a larger values of mixing angle one can exclude larger region of parameter space in the $M_{Z_{\mu\tau}}-g_{Z_{\mu\tau}}$ plane. For example if the mixing angle is $4.5\times 10^{-5}$ rad then one can maximally exclude the region of parameter space in the $M_{Z_{\mu\tau}}-g_{Z_{\mu\tau}}$ plane.

\noindent{\bf Acknowledgments}
A.S. would like to thank Heerak Banerjee for useful discussions.
A.B. would like to acknowledge the cluster computing facility
(http://www.hri.res.in/cluster/) of Harish-Chandra Research Institute, Allahabad.
He also thanks Alexander Pukhov for a few email conversation
regarding the package micrOMEGAs. Moreover, A.B. acknowledges all the
members of Particle Group Meeting of IACS, especially Sourov Roy,
Satyanarayan Mukhopadhyay, Heerak Banerjee, Sougata Ganguly,
Ananya Tapadar and Disha Bhatia for a useful discussion
on kinetic mixing between two U(1) gauge groups.
\begin{appendices}
\renewcommand{\thesection}{\Alph{section}}
\renewcommand{\theequation}{\thesection-\arabic{equation}} 
\setcounter{equation}{0}

\section{Multiplicative factors and functions that are involved in flavour physics}\label{flav_app}
\begin{eqnarray}
\mathscr{L}^9_{Z(Z_{\mu\tau})}&=&\frac{g_2}{4\cos\theta_W}\bigg(1-4\sin^2\theta_W\bigg)\cos(\sin)\theta_{\mu\tau}\pm \bigg(g_{Z_{\mu\tau}}-\frac 34 \frac{g_2\sin\theta_W\epsilon}{\cos\theta_W}\bigg)\sin(\cos)\theta_{\mu\tau}\;,\\\label{lzz9} 
\mathscr{L}^{10}_{Z(Z_{\mu\tau})}&=-&\frac{g_2}{4\cos\theta_W}\bigg(\cos(\sin)\theta_{\mu\tau}\pm\epsilon\sin\theta_W\sin(\cos)\theta_{\mu\tau}\bigg)\;,\\
\label{lzz10} 
\mathcal{G}_{Z(Z_{\mu\tau})}&=&\frac{g_2}{3\cos\theta_W}\sin^2\theta_W\bigg(\cos(\sin)\theta_{\mu\tau}\pm\frac{\epsilon}{\sin\theta_W}\sin(\cos)\theta_{\mu\tau}\bigg)\pm g_{Z_{\mu\tau}}\sin(\cos)\theta_{\mu\tau}\;,\\
\label{gzz}  
\mathcal{C}_{Z(Z_{\mu\tau})}&=&\frac{g_2}{\cos\theta_W}\cos(\sin)\theta_{\mu\tau}\pm \bigg(2g_{Z_{\mu\tau}}+\frac{g_2\sin\theta_W\epsilon}{\cos\theta_W}\bigg)\sin(\cos)\theta_{\mu\tau}\;,\\  
\label{czz}
\mathcal{S}_{Z(Z_{\mu\tau})}&=&\frac{g_2}{\cos\theta_W}\bigg(\left(\frac 12 -\frac{\sin^2\theta_W}{3}\right)\cos(\sin)\theta_{\mu\tau}\pm\epsilon\sin\theta_W\sin(\cos)\theta_{\mu\tau}\bigg)\;.
\label{szz}
\end{eqnarray}

\begin{eqnarray}
h_q(x)&=&\frac{1}{1-x}+\frac{\ln(x)}{(1-x)^2}\;,\\
\label{hq}
h_w(x,r)&=&\frac 32 -\frac{(1+r)^2\ln(1+r)}{r(1+r-x)}-\frac{x^2\ln(x)}{(1-x)(1+r-x)}\;,\\
\label{hw}
h_s(x)&=&\frac 12\left(\frac{1-3x}{1-x}-\frac{2x^2\ln(x)}{(1-x)^2}\right)\;,\\
\label{hs}
h_b(x)&=&-\frac{x^2-5x-2}{12(1-x)^3}+\frac{z\ln(x)}{6(1-x)^4}\;.
\label{hb}
\end{eqnarray}

\section{Couplings required for dark matter phenomenology, flvour physics observables and LHC analysis}\label{Dmcouplings}
$\bullet$ \underline {Trilinear couplings of different SM fermions with $Z(Z_{\mu\tau})$ gauge fields:}
\begin{eqnarray}
\bar{u}_i u_i Z^\alpha&:&i\frac{g_2\gamma^\alpha}{12\cos\theta_W}\Bigg[\Bigg(\bigg(-3+8\sin^2\theta_W\bigg)\cos\theta_{\mu\tau}+5\epsilon\sin\theta_W\sin\theta_{\mu\tau}\Bigg) \\ \nonumber 
&&+\Bigg(3\cos\theta_{\mu\tau}+3\epsilon\sin\theta_W\sin\theta_{\mu\tau}\Bigg)\gamma^5\Bigg]
\label{uuz}
\end{eqnarray}
\begin{eqnarray}
\bar{u}_i u_i Z^\alpha_{\mu\tau}&:&i\frac{g_2\gamma^\alpha}{12\cos\theta_W}\Bigg[\Bigg(\bigg(-3+8\sin^2\theta_W\bigg)\sin\theta_{\mu\tau}-5\epsilon\sin\theta_W\cos\theta_{\mu\tau}\Bigg) \\ \nonumber 
&&+\Bigg(3\sin\theta_{\mu\tau}-3\epsilon\sin\theta_W\cos\theta_{\mu\tau}\Bigg)\gamma^5\Bigg]
\label{uuz1}
\end{eqnarray}
\begin{eqnarray}
\bar{d}_i d_i Z^\alpha&:&-i\frac{g_2\gamma^\alpha}{12\cos\theta_W}\Bigg[\Bigg(\bigg(-3+4\sin^2\theta_W\bigg)\cos\theta_{\mu\tau}+\epsilon\sin\theta_W\sin\theta_{\mu\tau}\Bigg) \\ \nonumber 
&&+\Bigg(3\cos\theta_{\mu\tau}+3\epsilon\sin\theta_W\sin\theta_{\mu\tau}\Bigg)\gamma^5\Bigg]
\label{ddz}
\end{eqnarray}
\begin{eqnarray}
\bar{d}_i d_i Z^\alpha_{\mu\tau}&:&-i\frac{g_2\gamma^\alpha}{12\cos\theta_W}\Bigg[\Bigg(\bigg(-3+4\sin^2\theta_W\bigg)\sin\theta_{\mu\tau}-\epsilon\sin\theta_W\cos\theta_{\mu\tau}\Bigg) \\ \nonumber 
&&+\Bigg(3\sin\theta_{\mu\tau}-3\epsilon\sin\theta_W\cos\theta_{\mu\tau}\Bigg)\gamma^5\Bigg]
\label{ddz1}
\end{eqnarray}
In the above, $i=1,2,3$.
\begin{eqnarray}
\bar{e}e Z^\alpha&:&i\gamma^\alpha\Bigg[\Bigg(\frac{g_2}{4\cos\theta_W}\bigg(1-4\sin^2\theta_W\bigg)\cos\theta_{\mu\tau}-\frac 34 \frac{g_2\sin\theta_W\epsilon}{\cos\theta_W}\sin\theta_{\mu\tau}\Bigg)\\ \nonumber
&&-\Bigg(\frac{g_2}{4\cos\theta_W}\bigg(\cos\theta_{\mu\tau}+\epsilon\sin\theta_W\sin\theta_{\mu\tau}\bigg)\Bigg)\gamma^5\Bigg]
\label{eez}
\end{eqnarray}
\begin{eqnarray}
\bar{e}e Z^\alpha_{\mu\tau}&:&i\gamma^\alpha\Bigg[\Bigg(\frac{g_2}{4\cos\theta_W}\bigg(1-4\sin^2\theta_W\bigg)\sin\theta_{\mu\tau}+\frac 34 \frac{g_2\sin\theta_W\epsilon}{\cos\theta_W}\cos\theta_{\mu\tau}\Bigg)\\ \nonumber
&&-\Bigg(\frac{g_2}{4\cos\theta_W}\bigg(\sin\theta_{\mu\tau}-\epsilon\sin\theta_W\cos\theta_{\mu\tau}\bigg)\Bigg)\gamma^5\Bigg]
\label{eez1}
\end{eqnarray}
\begin{eqnarray}
\bar{\mu}\mu Z^\alpha&:&i\gamma^\alpha\Bigg[\Bigg(\frac{g_2}{4\cos\theta_W}\bigg(1-4\sin^2\theta_W\bigg)\cos\theta_{\mu\tau}+ \bigg(g_{Z_{\mu\tau}}-\frac 34 \frac{g_2\sin\theta_W\epsilon}{\cos\theta_W}\bigg)\sin\theta_{\mu\tau}\Bigg)\\ \nonumber
&&-\Bigg(\frac{g_2}{4\cos\theta_W}\bigg(\cos\theta_{\mu\tau}+\epsilon\sin\theta_W\sin\theta_{\mu\tau}\bigg)\Bigg)\gamma^5\Bigg]
\label{mumuz}
\end{eqnarray}
\begin{eqnarray}
\bar{\mu}\mu Z^\alpha_{\mu\tau}&:&i\gamma^\alpha\Bigg[\Bigg(\frac{g_2}{4\cos\theta_W}\bigg(1-4\sin^2\theta_W\bigg)\sin\theta_{\mu\tau}- \bigg(g_{Z_{\mu\tau}}-\frac 34 \frac{g_2\sin\theta_W\epsilon}{\cos\theta_W}\bigg)\cos\theta_{\mu\tau}\Bigg)\\ \nonumber
&&-\Bigg(\frac{g_2}{4\cos\theta_W}\bigg(\sin\theta_{\mu\tau}-\epsilon\sin\theta_W\cos\theta_{\mu\tau}\bigg)\Bigg)\gamma^5\Bigg]
\label{mumuz1}
\end{eqnarray}
\newpage
$\bullet$ \underline {Trilinear couplings of $\rho_i~(i\equiv 1,2,3)$ with $H_1$ and $H_2$ scalar fields:}

\begin{eqnarray}
\rho_1 \rho_1 H_1 &:& i \Bigg(2\cos^2\theta_D\bigg(v_1\lambda_7\cos\theta_s-v_2\lambda_6\sin\theta_s\bigg)\\ \nonumber
&&+\sqrt{2}\lambda_8\cos\theta_D\sin\theta_D\bigg(v_2\cos\theta_s-v_1\sin\theta_s\bigg)\\ \nonumber
&&+ \sin^2\theta_D\bigg(v_1(\lambda_2+\lambda_3)\cos\theta_s-v_2\lambda_4\sin\theta_s\bigg)\Bigg)\\
\rho_1 \rho_1 H_2 &:& i \Bigg(2\cos^2\theta_D\bigg(v_1\lambda_7\sin\theta_s+v_2\lambda_6\cos\theta_s\bigg)\\ \nonumber
&&+\sqrt{2}\lambda_8\cos\theta_D\sin\theta_D\bigg(v_2\sin\theta_s+v_1\cos\theta_s\bigg)\\ \nonumber
&&+ \sin^2\theta_D\bigg(v_1(\lambda_2+\lambda_3)\sin\theta_s+v_2\lambda_4\cos\theta_s\bigg)\Bigg)
\end{eqnarray}
\begin{eqnarray}
\rho_2 \rho_2 H_1 &:& i \Bigg(2\sin^2\theta_D\bigg(v_1\lambda_7\cos\theta_s-v_2\lambda_6\sin\theta_s\bigg)\\ \nonumber
&&-\sqrt{2}\lambda_8\cos\theta_D\sin\theta_D\bigg(v_2\cos\theta_s-v_1\sin\theta_s\bigg)\\ \nonumber
&&+ \cos^2\theta_D\bigg(v_1(\lambda_2+\lambda_3)\cos\theta_s-v_2\lambda_4\sin\theta_s\bigg)\Bigg)\\
\rho_2 \rho_2 H_2 &:& i\Bigg(2\sin^2\theta_D\bigg(v_1\lambda_7\sin\theta_s+v_2\lambda_6\cos\theta_s\bigg)\\ \nonumber
&&-\sqrt{2}\lambda_8\cos\theta_D\sin\theta_D\bigg(v_2\sin\theta_s+v_1\cos\theta_s\bigg)\\ \nonumber
&&+ \cos^2\theta_D\bigg(v_1(\lambda_2+\lambda_3)\sin\theta_s+v_2\lambda_4\cos\theta_s\bigg)\Bigg)
\end{eqnarray}
\begin{eqnarray}
\rho_1 \rho_2 H_1 &:& \frac{i}{2} \Bigg(\sqrt{2}\cos 2\theta_D\lambda_8\bigg(v_2\cos\theta_s-v_1\sin\theta_s\bigg)\\ \nonumber
&&+ \sin 2\theta_D\bigg(v_1(\lambda_2+\lambda_3-2\lambda_7)\cos\theta_s-v_2(\lambda_4-2\lambda_6)\sin\theta_s\bigg)\Bigg)\\
\rho_1 \rho_2 H_2 &:& \frac{i}{2} \Bigg(\sqrt{2}\cos 2\theta_D\lambda_8\bigg(v_2\sin\theta_s+v_1\cos\theta_s\bigg)\\ \nonumber
&&+ \sin 2\theta_D\bigg(v_1(\lambda_2+\lambda_3-2\lambda_7)\sin\theta_s+v_2(\lambda_4-2\lambda_6)\cos\theta_s\bigg)\Bigg)
\end{eqnarray}
\begin{eqnarray}
\rho_3 \rho_3 H_1 &:& i\Bigg(v_1(\lambda_2+\lambda_3)\cos\theta_s-v_2\lambda_4\sin\theta_s\Bigg) \\
\rho_3 \rho_3 H_2 &:& i\Bigg(v_1(\lambda_2+\lambda_3)\sin\theta_s+v_2\lambda_4\cos\theta_s\Bigg)
\end{eqnarray}
$\bullet$ \underline {Quartic couplings of $\rho_i~(i\equiv 1,2,3)$ with $H_1$ scalar fields:}
\begin{eqnarray}
\rho_1 \rho_1 H_1 H_1 &:&i \Bigg(-2\sqrt{2}\lambda_8\cos\theta_s\sin\theta_s\cos\theta_D\sin\theta_D \\ \nonumber
&&+\cos^2\theta_s\bigg(2\lambda_7\cos^2\theta_D+(\lambda_2+\lambda_3)\sin^2\theta_D\bigg) \\ \nonumber
&&+\sin^2\theta_s\bigg(2\lambda_6\cos^2\theta_D+\lambda_4\sin^2\theta_D\bigg)\Bigg)
\end{eqnarray}
\begin{eqnarray}
\rho_2 \rho_2 H_1 H_1 &:&i \Bigg(2\sqrt{2}\lambda_8\cos\theta_s\sin\theta_s\cos\theta_D\sin\theta_D \\ \nonumber
&&+\cos^2\theta_s\bigg(2\lambda_7\sin^2\theta_D+(\lambda_2+\lambda_3)\cos^2\theta_D\bigg) \\ \nonumber
&&+\sin^2\theta_s\bigg(2\lambda_6\sin^2\theta_D+\lambda_4\cos^2\theta_D\bigg)\Bigg)
\end{eqnarray}
\begin{eqnarray}
\rho_1 \rho_2 H_1 H_1 &:&i \Bigg(-\sqrt{2}\lambda_8\cos\theta_s\sin\theta_s\cos 2\theta_D+\cos\theta_D\sin\theta_D \\ \nonumber
&&\bigg((\lambda_4-2\lambda_6)\sin^2\theta_s+(\lambda_2+\lambda_3-2\lambda_7)\cos^2\theta_s\bigg)\Bigg)
\end{eqnarray}
\begin{eqnarray}
\rho_3 \rho_3 H_1 H_1 &:& i \Bigg((\lambda_2+\lambda_3)\cos^2\theta_s+\lambda_4\sin^2\theta_s\Bigg)
\end{eqnarray}
$\bullet$ \underline {Quartic couplings of $\rho_i~(i\equiv 1,2,3)$ with $H_2$ scalar fields:}
\begin{eqnarray}
\rho_1 \rho_1 H_2 H_2 &:&i \Bigg(2\sqrt{2}\lambda_8\cos\theta_s\sin\theta_s\cos\theta_D\sin\theta_D \\ \nonumber
&&+\sin^2\theta_s\bigg(2\lambda_7\cos^2\theta_D+(\lambda_2+\lambda_3)\sin^2\theta_D\bigg) \\ \nonumber
&&+\cos^2\theta_s\bigg(2\lambda_6\cos^2\theta_D+\lambda_4\sin^2\theta_D\bigg)\Bigg)
\end{eqnarray}
\begin{eqnarray}
\rho_2 \rho_2 H_2 H_2 &:&i \Bigg(-2\sqrt{2}\lambda_8\cos\theta_s\sin\theta_s\cos\theta_D\sin\theta_D \\ \nonumber
&&+\sin^2\theta_s\bigg(2\lambda_7\sin^2\theta_D+(\lambda_2+\lambda_3)\cos^2\theta_D\bigg) \\ \nonumber
&&+\cos^2\theta_s\bigg(2\lambda_6\sin^2\theta_D+\lambda_4\cos^2\theta_D\bigg)\Bigg)
\end{eqnarray}
\begin{eqnarray}
\rho_1 \rho_2 H_2 H_2 &:&i \Bigg(\sqrt{2}\lambda_8\cos\theta_s\sin\theta_s\cos 2\theta_D+\cos\theta_D\sin\theta_D \\ \nonumber
&&\bigg((\lambda_4-2\lambda_6)\cos^2\theta_s+(\lambda_2+\lambda_3-2\lambda_7)\sin^2\theta_s\bigg)\Bigg)
\end{eqnarray}
\begin{eqnarray}
\rho_3 \rho_3 H_2 H_2 &:& i \Bigg((\lambda_2+\lambda_3)\sin^2\theta_s+\lambda_4\cos^2\theta_s\Bigg)
\end{eqnarray}
$\bullet$ \underline {Trilinear couplings between $\mathbb{Z}_2$ odd particles with  gauge fields:}
\begin{eqnarray}
\rho_1\phi^\pm W^{\mp_\alpha} &:& \mp i\frac{e\sin\theta_D}{2\sin\theta_W}(p_1-p_2)^\alpha\\
\rho_2\phi^\pm W^{\mp_\alpha} &:& \mp i\frac{e\cos\theta_D}{2\sin\theta_W}(p_1-p_2)^\alpha\\
\rho_3\phi^\pm W^{\mp_\alpha} &:& -\frac{e}{2\sin\theta_W}(p_1-p_2)^\alpha\\
\rho_1\rho_3 Z^\alpha &:& \frac{\sin\theta_D}{2} \Bigg(\frac{e}{2\sin\theta_W \cos\theta_W}\cos\theta_{\mu\tau}\left(2g_{Z_{\mu\tau}}+\epsilon\frac{e}{\cos\theta_W} \right)\sin\theta_{\mu\tau}\Bigg)(p_1-p_2)^\alpha\\
\rho_1\rho_3 Z^\alpha_{\mu\tau} &:& \frac{\sin\theta_D}{2} \Bigg(\frac{e}{\sin\theta_W \cos\theta_W}\sin\theta_{\mu\tau}-\left(2g_{Z_{\mu\tau}}+\epsilon\frac{e}{\cos\theta_W} \right)\cos\theta_{\mu\tau}\Bigg)(p_1-p_2)^\alpha\\
\rho_2\rho_3 Z^\alpha &:& \frac{\cos\theta_D}{2} \Bigg(\frac{e}{2\sin\theta_W \cos\theta_W}\cos\theta_{\mu\tau}\left(2g_{Z_{\mu\tau}}+\epsilon\frac{e}{\cos\theta_W} \right)\sin\theta_{\mu\tau}\Bigg)(p_1-p_2)^\alpha\\
\rho_2\rho_3 Z^\alpha_{\mu\tau} &:& \frac{\cos\theta_D}{2} \Bigg(\frac{e}{\sin\theta_W \cos\theta_W}\sin\theta_{\mu\tau}-\left(2g_{Z_{\mu\tau}}+\epsilon\frac{e}{\cos\theta_W} \right)\cos\theta_{\mu\tau}\Bigg)(p_1-p_2)^\alpha
\end{eqnarray}
 $\bullet$ \underline {Quartic couplings of dark matter with  gauge fields:}
\begin{eqnarray}
\rho_1 \rho_1 W^{+_\alpha} W^{-_\beta} &:& i\frac{e^2\sin^2\theta_D}{2\sin^2\theta_W}g^{\alpha\beta}
\end{eqnarray}
\begin{eqnarray}
\rho_1 \rho_1 Z^{\alpha}Z^{\beta}  &:& i \frac {\sin^2\theta_D}{2} \Bigg(\bigg(2g_{Z_{\mu\tau}}\sin\theta_{\mu\tau}+\frac{e \cos\theta_{\mu\tau}}{\cos\theta_W\sin\theta_W}\bigg)\\ \nonumber
&&\bigg(2\left(g_{Z_{\mu\tau}}+\epsilon\frac{e}{\cos\theta_W}\right)\sin\theta_{\mu\tau}+\frac{e \cos\theta_{\mu\tau}}{\cos\theta_W\sin\theta_W}\bigg)\Bigg)g^{\alpha\beta}
\end{eqnarray}
\begin{eqnarray}
\rho_1 \rho_1 Z^{\alpha}_{\mu\tau}Z^{\beta}_{\mu\tau}  &:& i \frac {\sin^2\theta_D}{2} \Bigg(\bigg(2g_{Z_{\mu\tau}}\cos\theta_{\mu\tau}-\frac{e \sin\theta_{\mu\tau}}{\cos\theta_W\sin\theta_W}\bigg)\\ \nonumber
&&\bigg(2\left(g_{Z_{\mu\tau}}+\epsilon\frac{e}{\cos\theta_W}\right)\cos\theta_{\mu\tau}-\frac{e \sin\theta_{\mu\tau}}{\cos\theta_W\sin\theta_W}\bigg)\Bigg)g^{\alpha\beta}
\end{eqnarray}
\begin{eqnarray}
\rho_1 \rho_1 Z^{\alpha}_{\mu\tau}Z^{\beta}&:& i \frac {\sin^2\theta_D}{2}\Bigg(\frac{e^2\cos\theta_{\mu\tau}\sin\theta_{\mu\tau}}{\cos^2\theta_W\sin^2\theta_W} \\ \nonumber
&&-\frac{e}{\cos\theta_W \sin\theta_W}\left(2g_{Z_{\mu\tau}}+\epsilon\frac{e}{\cos\theta_W}\right)\cos 2\theta_{\mu\tau} \\ \nonumber
&& -2g_{Z_{\mu\tau}} \left(g_{Z_{\mu\tau}}+\epsilon\frac{e}{\cos\theta_W}\right) \sin 2\theta_{\mu\tau}
\Bigg)g^{\alpha\beta}
\end{eqnarray}

$\bullet$ \underline {Trilinear couplings between $\mathbb{Z}_2$ odd charged particles with $H_1$ and $H_2$ scalar fields:}
\begin{eqnarray}
\phi^+\phi^-H_1&:& i\Bigg(v_1\lambda_2\cos\theta_s-v_2\lambda_4\sin\theta_s\Bigg)\\
\phi^+\phi^-H_2&:& i\Bigg(v_1\lambda_2\sin\theta_s+v_2\lambda_4\cos\theta_s\Bigg)
\end{eqnarray}
$\bullet$ \underline {Trilinear couplings between $\mathbb{Z}_2$ odd charged particles with gauge fields:}
\begin{eqnarray}
\phi^+\phi^-\gamma^\alpha&:&-ie(p_1-p_2)^\alpha \\ 
\phi^+\phi^-Z^\alpha&:&\frac{i}{2}\Bigg(\frac{e\cos 2\theta_W}{\sin\theta_W \cos\theta_W}\cos\theta_{\mu\tau}-\left(2g_{Z_{\mu\tau}}+\epsilon\frac{e}{\cos\theta_W} \right)\sin\theta_{\mu\tau}\Bigg) (p_1-p_2)^\alpha\\
\phi^+\phi^-Z^\alpha_{\mu\tau}&:&\frac{i}{2}\Bigg(\frac{e \cos 2\theta_W}{\sin\theta_W \cos\theta_W}\sin\theta_{\mu\tau}+\left(2g_{Z_{\mu\tau}}+\epsilon\frac{e}{\cos\theta_W} \right)\cos\theta_{\mu\tau}\Bigg) (p_1-p_2)^\alpha
\end{eqnarray}
$\bullet$ \underline {Quartic couplings between $\mathbb{Z}_2$ odd charged particles with gauge fields:}
\begin{eqnarray}
\phi^+\phi^-W^{+\alpha} W^{-\beta} &:&i\frac{e^2}{2\sin^2\theta_W}g^{\alpha\beta}\\
\phi^+\phi^-\gamma^\alpha \gamma^\beta &:&i2e^2g^{\alpha\beta}
\end{eqnarray}
\begin{eqnarray}
\phi^+\phi^-\gamma^\alpha Z^\beta &:&i\frac{e}{2}\Bigg(\frac{e\cos 2\theta_W}{\sin\theta_W \cos\theta_W}\cos\theta_{\mu\tau}-\left(2g_{Z_{\mu\tau}}+\epsilon\frac{e}{\cos\theta_W} \right)\sin\theta_{\mu\tau}\Bigg) g^{\alpha\beta} \\
\phi^+\phi^-Z^\alpha Z^\beta &:&\frac{i}{2}g^{\alpha\beta}\Bigg(\frac{e\cos 2\theta_W}{\sin\theta_W \cos\theta_W}\cos\theta_{\mu\tau}-2g_{Z_{\mu\tau}}\sin\theta_{\mu\tau}\Bigg) \\ \nonumber
&&\Bigg(\frac{e\cos 2\theta_W}{\sin\theta_W \cos\theta_W}\cos\theta_{\mu\tau}-\left(2g_{Z_{\mu\tau}}+\epsilon\frac{e}{\cos\theta_W} \right)\sin\theta_{\mu\tau}\Bigg)
\end{eqnarray}
$\bullet$ \underline {Trilinear couplings between CP-even scalar fields:}
\begin{eqnarray}
H_1 H_1 H_1 &:& i \Bigg(6 v_1\lambda_H \cos^3\theta_s-3\lambda_1\bigg(\cos^2\theta_s\sin\theta_s-\cos\theta_s\sin^2\theta_s\bigg)-6v_2\lambda_\eta\sin^3\theta_s\Bigg)\\
H_2 H_1 H_1 &:& i\Bigg(v_2\lambda_1\cos^3\theta_s+2v_1(3\lambda_H-\lambda_1)\cos^2\theta_s\sin\theta_s\\ \nonumber
&&+2v_2(3\lambda_\eta-\lambda_1)\cos\theta_s\sin^2\theta_s+v_1\lambda_1\sin^3\theta_s\Bigg)
\end{eqnarray}
$\bullet$ \underline {Trilinear couplings of CP-even scalar fields with  gauge fields:}
\begin{eqnarray}
H_1 W^{+\alpha} W^{-\beta} &:& i\frac{e^2v_1}{2\sin^2\theta_W}\cos\theta_s g^{\alpha\beta}\\
H_2 W^{+\alpha} W^{-\beta} &:& i\frac{e^2v_1}{2\sin^2\theta_W}\sin\theta_s g^{\alpha\beta}\\
H_1 Z^{\alpha} Z^{\beta} &:& i\Bigg(\frac{e v_1}{2\sin\theta_W\cos\theta_W}\cos\theta_{\mu\tau}\bigg(\cos\theta_{\mu\tau}\frac{e}{\sin\theta_W\cos\theta_W}\\ \nonumber
&&+2\sin\theta_{\mu\tau}\epsilon\frac{e}{\cos\theta_W}\bigg)\cos\theta_s -2g^2_{Z_{\mu\tau}}v_2\sin^2\theta_{\mu\tau}\sin\theta_s\Bigg)g^{\alpha\beta}\\
H_2 Z^{\alpha} Z^{\beta} &:& i\Bigg(\frac{e v_1}{2\sin\theta_W\cos\theta_W}\cos\theta_{\mu\tau}\bigg(\cos\theta_{\mu\tau}\frac{e}{\sin\theta_W\cos\theta_W}\\ \nonumber
&&+2\sin\theta_{\mu\tau}\epsilon\frac{e}{\cos\theta_W}\bigg)\sin\theta_s +2g^2_{Z_{\mu\tau}}v_2\sin^2\theta_{\mu\tau}\cos\theta_s\Bigg)g^{\alpha\beta}\\
H_1 Z^{\alpha}_{\mu\tau} Z^{\beta}_{\mu\tau} &:& i\Bigg(\frac{e v_1}{2\sin\theta_W\cos\theta_W}\sin\theta_{\mu\tau}\bigg(\sin\theta_{\mu\tau}\frac{e}{\sin\theta_W\cos\theta_W}\\ \nonumber
&&-2\cos\theta_{\mu\tau}\epsilon\frac{e}{\cos\theta_W}\bigg)\cos\theta_s -2g^2_{Z_{\mu\tau}}v_2\cos^2\theta_{\mu\tau}\sin\theta_s\Bigg)g^{\alpha\beta}\\
H_2 Z^{\alpha}_{\mu\tau} Z^{\beta}_{\mu\tau} &:& i\Bigg(\frac{e v_1}{2\sin\theta_W\cos\theta_W}\sin\theta_{\mu\tau}\bigg(\sin\theta_{\mu\tau}\frac{e}{\sin\theta_W\cos\theta_W}\\ \nonumber
&&-2\cos\theta_{\mu\tau}\epsilon\frac{e}{\cos\theta_W}\bigg)\sin\theta_s +2g^2_{Z_{\mu\tau}}v_2\cos^2\theta_{\mu\tau}\cos\theta_s\Bigg)g^{\alpha\beta}
\end{eqnarray}
\begin{eqnarray}
H_1 Z^{\alpha} Z^{\beta}_{\mu\tau} &:& i\Bigg(\frac{e v_1}{2\sin\theta_W\cos\theta_W}\bigg(\sin 2\theta_{\mu\tau}\frac{e}{2\sin\theta_W\cos\theta_W}\\ \nonumber
&&-\cos 2\theta_{\mu\tau}\epsilon\frac{e}{\cos\theta_W}\bigg)\cos\theta_s +g^2_{Z_{\mu\tau}}v_2\sin 2\theta_{\mu\tau}\sin\theta_s\Bigg)g^{\alpha\beta}
\end{eqnarray}
\begin{eqnarray}
H_2 Z^{\alpha} Z^{\beta}_{\mu\tau} &:& i\Bigg(\frac{e v_1}{2\sin\theta_W\cos\theta_W}\bigg(\sin 2\theta_{\mu\tau}\frac{e}{2\sin\theta_W\cos\theta_W}\\ \nonumber
&&-\cos 2\theta_{\mu\tau}\epsilon\frac{e}{\cos\theta_W}\bigg)\sin\theta_s -g^2_{Z_{\mu\tau}}v_2\sin 2\theta_{\mu\tau}\cos\theta_s\Bigg)g^{\alpha\beta}
\end{eqnarray}
$\bullet$ \underline {Trilinear couplings between gauge fields:}
\begin{eqnarray}
\gamma^\sigma W^{+\alpha}W^{-\beta}&:& ie \bigg( g^{\sigma\alpha} (p_2
      -p_1)^\beta + g^{\sigma\beta} (p_1 -p_3)^\alpha +
    g^{\beta\alpha} (p_3 -p_2)^\sigma \bigg) \\
Z^\sigma W^{+\alpha}W^{-\beta}&:& ie \frac{\cos\theta_W \cos\theta_s}{\sin\theta_W}\bigg( g^{\sigma\alpha} (p_2
      -p_1)^\beta + g^{\sigma\beta} (p_1 -p_3)^\alpha +
    g^{\beta\alpha} (p_3 -p_2)^\sigma \bigg) \\
Z^\sigma_{\mu\tau} W^{+\alpha}W^{-\beta}&:& ie \frac{\cos\theta_W \sin\theta_s}{\sin\theta_W}\bigg( g^{\sigma\alpha} (p_2
      -p_1)^\beta + g^{\sigma\beta} (p_1 -p_3)^\alpha +
    g^{\beta\alpha} (p_3 -p_2)^\sigma \bigg) 
\end{eqnarray}
$\bullet$ \underline {Trilinear couplings of CP-even fields scalar with different SM fermion fields:}
\begin{eqnarray}
H_1 c \bar{c} &:& -i\frac{e\; m_c}{\sqrt{2}\sin\theta_W M_W}\cos\theta_s\\
H_2 c \bar{c} &:& -i\frac{e\; m_c}{\sqrt{2}\sin\theta_W M_W}\sin\theta_s\\
H_1 t \bar{t} &:& -i\frac{e\; m_t}{\sqrt{2}\sin\theta_W M_W}\cos\theta_s\\
H_2 t \bar{t} &:& -i\frac{e\; m_t}{\sqrt{2}\sin\theta_W M_W}\sin\theta_s\\
H_1 b \bar{b} &:& -i\frac{e\; m_b}{\sqrt{2}\sin\theta_W M_W}\cos\theta_s\\
H_2 b \bar{b} &:& -i\frac{e\; m_b}{\sqrt{2}\sin\theta_W M_W}\sin\theta_s\\
H_1 \tau^+\tau^- &:& -i\frac{e \;m_\tau}{\sqrt{2}\sin\theta_W M_W}\cos\theta_s\\
H_2 \tau^+\tau^- &:& -i\frac{e\; m_\tau}{\sqrt{2}\sin\theta_W M_W}\sin\theta_s
\end{eqnarray}
$\bullet$ \underline {Trilinear couplings of $\chi$ with SM down-type quarks and $\rho_i~(i\equiv 1,2,3)$ field:}
\begin{eqnarray}
\bar{\chi} \rho_1 b_j: -i\frac{f_j}{2\sqrt{2}}(1-\gamma^5)\sin\theta_D,\;\;\;\;\bar{b}_j \rho_1\chi: -i\frac{f_j}{2\sqrt{2}}(1+\gamma^5)\sin\theta_D
\label{rho1xd}
\end{eqnarray}
\begin{eqnarray}
\bar{\chi} \rho_2 b_j: -i\frac{f_j}{2\sqrt{2}}(1-\gamma^5)\cos\theta_D,\;\;\;\;\bar{b}_j \rho_2\chi: -i\frac{f_j}{2\sqrt{2}}(1+\gamma^5)\cos\theta_D
\label{rho2xd}
\end{eqnarray}
\begin{eqnarray}
\bar{\chi} \rho_3 b_j: -\frac{f_j}{2\sqrt{2}}(1-\gamma^5),\;\;\;\;\bar{b}_j \rho_3\chi: -\frac{f_j}{2\sqrt{2}}(1+\gamma^5)
\label{rho3xd}
\end{eqnarray}
$\bullet$ \underline {Trilinear couplings of $\chi$ with $Z(Z_{\mu\tau})$ gauge field:}
\begin{eqnarray}
\bar{\chi}\chi Z^\alpha&=&-i\gamma^\alpha\Bigg[\frac{g_2}{3\cos\theta_W}\sin^2\theta_W\bigg(\cos\theta_{\mu\tau}+\frac{\epsilon}{\sin\theta_W}\sin\theta_{\mu\tau}\bigg)+ g_{Z_{\mu\tau}}\sin\theta_{\mu\tau}\Bigg]\\
\label{chichiz}
\bar{\chi}\chi Z^\alpha_{\mu\tau}&=&-i\gamma^\alpha\Bigg[\frac{g_2}{3\cos\theta_W}\sin^2\theta_W\bigg(\sin\theta_{\mu\tau}-\frac{\epsilon}{\sin\theta_W}\cos\theta_{\mu\tau}\bigg)- g_{Z_{\mu\tau}}\cos\theta_{\mu\tau}\Bigg]
\label{chichiz1}
\end{eqnarray}
\end{appendices}

\bibliographystyle{jhep}
\bibliography{Lmu_Ltau}

\end{document}